\documentclass[floatfix,amsmath,amssymb,longbibliography]{revtex4}
\usepackage{amsmath,amssymb,color,latexsym,natbib,graphicx,bm}
\usepackage{lpic} \usepackage{subfigure} \usepackage[normalem]{ulem}
\usepackage{booktabs}

\bibpunct{[}{]}{,}{n}{,}{,} 

\def\DEL#1{{\textcolor{green}{}}} 

\newcommand{\vomega}{\mbox{\boldmath $\omega$}}

\newcommand{\be}{\begin{equation}}  \newcommand{\ee}{\end{equation}}

\begin{document}
\title{Coupling large eddies and waves in turbulence: \\
Case study of magnetic helicity at the ion inertial scale}
\author{Annick Pouquet $^{1,2}$, Julia E. Stawarz$^3$ and Duane Rosenberg $^4$}

\affiliation{
$^{1}$  Laboratory for Atmospheric and Space Physics, University of Colorado, Boulder, CO 80309, USA \\
$^{2}$ National Center for Atmospheric Research, P.O.~Box 3000, Boulder, CO 80307, USA \\ 
$^{3}$ Department of Physics, Imperial College London, London, United Kingdom \\
$^{4}$ 288 Harper St., Louisville, CO 80027, USA
}
\begin{abstract}
In turbulence, for neutral or conducting fluids, a large ratio of scales is excited because of the possible occurrence of inverse cascades to large, global scales together with direct cascades to small, dissipative scales, as observed in the atmosphere and oceans, or in the solar environment. In this context, using direct numerical simulations with forcing, we analyze scale dynamics in the presence of magnetic fields with a generalized Ohm's law including a Hall current. The ion inertial length $\epsilon_H$ serves as the control parameter at fixed Reynolds number.
Both the magnetic and generalized helicity -- invariants in the ideal case -- grow linearly with time, as expected from classical arguments. The cross-correlation between the velocity and magnetic field grows as well, more so in relative terms for a stronger Hall current.  We find that the helical growth rates vary exponentially with 
$\epsilon_H$, provided the ion inertial scale resides within the inverse cascade range. These exponential variations are recovered phenomenologically using simple scaling arguments. They are directly linked to the wavenumber power-law dependence of generalized and magnetic helicity, $\sim k^{-2}$, in their inverse ranges. This illustrates and confirms the important role of the interplay between large and small scales in the dynamics of turbulent flows. 
\end{abstract}
\maketitle

\section{Introduction}   
\subsection{The interactions of turbulent eddies and waves in atmospheric and oceanic flows}

Turbulence and nonlinear phenomena are characterized by stochastic behavior, nonlinear waves, power-law energy spectra, and by intermittent events with non-Gaussian probability distribution functions 
\cite{newell_01, sagaut_08, buhler_10, mahrt_14, pouquet_17p, gregg_18}. They are present in a multitude of geophysical and astrophysical environments (see {\it e.g.} the recent reviews in the Special Issue of {\sl Earth \& Space Science} (2019) entitled ``{\it Nonlinear Systems in Geophysics: Past Accomplishments and Future Challenges}" ). More specifically, the role of turbulence has  been advocated for example in  the process of rain formation \cite{shaw_01} because of strong local accelerations, in the properties of atmospheric aerosols \cite{lopez_16}, or more recently in the multi-fractality of temperature distributions \cite{lovejoy_10b, kalamaras_19, schertzer_20}.
Similarly, huge variations of the energy dissipation take place locally in the ocean \cite{vanharen_16j} in the vicinity of ridges, as well as in space plasmas such as the solar wind and beyond \cite{sorriso_07} (see below, \S \ref{SS:PLASMAS}). Extreme events are in general at small scales, appearing in the gradients of the velocity,  the density, the temperature and the magnetic field, through vorticity, shear layers, filaments or current sheets. They can also be observed at large scales, as for example with the vertical velocity in the nocturnal, very stable, Planetary Boundary Layer \cite{lenschow_12, mahrt_14}. 

Similarly, the influence of gravity waves over turbulent eddies has been studied over Antarctica (see , {\it e.g.} \cite{cava_15}), and intense gradients are identified as well in that region of the globe \cite{walterscheid_16}.
In fact,  strong vertical winds, as well as vertically sheared horizontal winds, can be viewed as  common features of stably stratified turbulence \cite{rorai_14}, in the presence or not of rotation. Even though such a behavior takes place in a narrow range of the control parameter \cite{feraco_18}, it affects measurably the overall dynamics of the flow, with a slow return to isotropy at small scale  \cite{smyth_00b, sagaut_08, pouquet_19p}, together with strong localized mixing, dissipation and intermittency for Richardson numbers close to the threshold of linear or convective  instabilities \cite{smyth_19, sujovolsky_19, pouquet_19p}. Furthermore, the trajectories of Lagrangian particles are also measurably modified in the vicinity of shear layers (see, {\it e.g.} \cite{buaria_19}). Such a marginal state close to a threshold almost everywhere can be modeled through simplified dynamical systems following field gradients \cite{rorai_14, feraco_18, sujovolsky_19, sujovolsky_20}, in line with classical approaches in turbulence, as reviewed {\it e.g.} in \cite{meneveau_11}. 

Finally, in the presence of rotation in a stably stratified fluid, 
{
several other phenomena can take place. The dynamical exchanges between waves and nonlinear eddies lead to a modified distribution of energy between the kinetic and potential modes, with the dominance of one over the other shifting at a wavenumber that does not depend on the Reynolds number but rather on the Froude number, that is, the ratio of the wave period to the eddy turn-over time \cite{marino_15w} (see \cite{herbert_16} for the case of the inverse cascade of energy). Furthermore,
}
the existence of bi-directional dual cascades of energy towards  large scales {\sl  and} small scales, both with constant energy fluxes, is a clear mechanism coupling nonlinearly all scales and affecting the resulting dissipation. Thus, the dynamical interactions between small and large scales play an essential role in estimating the efficiency of mixing  in such flows \cite{pouquet_13p, marino_15p, alexakis_18}, and it is found to vary linearly with the control parameter, namely the Froude number \cite{pouquet_18, pouquet_19e}.

\subsection{The case of space plasmas} \label{SS:PLASMAS}
Similar phenomena are observed as well for turbulent flows in the presence of magnetic fields. Such fields, together with charged particles, are abundant in the cosmos. At large scales, the magnetohydrodynamic (MHD) approximation, in which the displacement current is neglected in Maxwell's equations, is adequate, and observations of the Solar Wind, dating back to the Voyager spacecraft, confirmed the physical description of a medium governed by the interactions of turbulent, nonlinear eddies and Alfv\'en waves (see, {\it e.g.},  for recent reviews, \cite{bruno_05, veltri_09, matthaeus_15, galtier_18b} and references therein). Turbulence is also found to play a central role  in shaping these media \cite{matthaeus_11, marino_12, pouquet_15b}.

However, as the direct turbulent cascade of energy approaches smaller scales, plasma effects and dispersive waves come into effect, appearing for example through a generalized Ohm's law whose expression  depends on the degree of ionization of the medium, which itself can differ greatly from the solar wind to the interstellar gas. Current spacecraft technologies allow for the resolution of much smaller temporal and spatial scales than what was available previously, and one can now reach the ion inertial length, $\epsilon_H$, and perhaps the electron inertial length (see for definitions the next section, and {\it e.g.} \cite{toth_17}). Other types of waves, kinetic Alfv\'en waves or whistler waves for example,  come into play between the ion and electron scales, and the distribution of energy among modes is altered from a spectrum close to that of Kolmogorov (1941) to substantially steeper scaling laws \cite{sahraoui_13}, leading to marked anisotropies \cite{lacombe_17}.
Using the MMS (Magnetospheric Multi-Scale) suite of four satellites, recent observations indicate the presence of Kelvin-Helmoltz instabilities at large scales. They can drive small-scale turbulence through secondary instabilities (see, {\it e.g.}   \cite{stawarz_16}), reconnection and dissipative processes in shear layers and current sheets. The signature of Kelvin-Helmoltz instabilities and intermittency may well persist in the statistics of such flows \cite{dimare_19}. At even smaller scales, Hall-MHD, as well as electron dynamics are also observed  \cite{lecontel_16, faganello_17, bandyo_18a, stawarz_19}. 
Two-dimensional two-fluid Hall-MHD simulations have shown recently that there is a sizable proportion of the turbulent transfer, and hence of the dissipation, that is localized in  coherent structures such as current sheets which are thin but have  transverse dimensions of the order of the integral scale \cite{camporeale_18p}. Besides losing energy to dissipative processes,  plasmas also exchange energy with particles through {\it e.g.} ion-cyclotron waves, as observed recently in the magnetosphere \cite{kitamura_18}.

In the presence of forcing acting only in the momentum equation, and for small initial magnetic fields, one is faced with the so-called dynamo problem of generation of magnetic fields, as reviewed extensively, {\it e.g.} in  \cite{branden_rev}. Searching for the effect of plasma waves on the growth of both large-scale and small-scale magnetic fields, one finds that, for Hall-MHD, the magnetic field grows faster for intermediate values of the control parameter $\epsilon_H$, with also a dependence on the magnetic Reynolds number, $R_M=U_0L_0/\eta$ with $U_0,L_0$ characteristic velocity and length scales, and $\eta$ the magnetic diffusivity. Specifically, the growth rate is larger when the ion length scale $\epsilon_H$ is close to (but larger than) the dissipation scale  (see for example \cite{mahajan_05, mininni_05} and references therein). Both magnetic helicity and magnetic energy grow, with a flat energy spectrum at large scales and closer to a Kolmogorov spectrum at small scales. 
Numerous  studies have been devoted to the full dynamics of Hall MHD. For example, it is shown in \cite{galtier_07} using shell models that the energy spectrum changes from a classical Kolmogorov law for large eddies to a steeper scaling after the ion inertial length, the slope of which depends on the amount of excess magnetic energy compared to its kinetic counterpart (see also \cite{galtier_06} for a weak turbulence approach). 

Small-scale dynamics in Hall MHD, and how its evolution differs from the pure MHD case, is of prime importance for laboratory and space plasmas, and has been studied extensively. At early times, like in MHD, vorticity and current sheets form, of thickness the dissipation length scale, called the Kolmogorov scale in fluid turbulence and with a $-3/4$ dependence on the kinetic Reynolds number $R_V=U_0L_0/\nu$  with respect to the characteristic length scale of the flow, with $\nu$ the kinematic viscosity. These sheets can roll-up, with a strong local correlation between the velocity, the magnetic field and the current \cite{mininni_06b}. However, the  dissipative scale for MHD is much smaller, for astrophysical Reynolds numbers which are very large, than the ion and even the electron inertial scales which are reached first in the process of transferring the energy to smaller scales. This leads to a second inertial range in which the nonlinearity associated with the Hall current now prevails giving different scaling laws for energy spectra.
 A detailed analysis of dissipative processes in space plasmas can be found in \cite{borovsky_03a, borovsky_03b}.
For example, Reynolds numbers for the solar wind, the magnetosheath and magnetotail can vary roughly from   $10^{11}$ to $10^{14}$. For length scales between a few to a thousand Earth's radii, 
this leads to a (Kolmogorov) dissipation length scale varying from the $mm$, i.e. comparable to the case of the atmosphere, to the meter. These scales are much smaller than the ion gyroradius, estimated to be between 70$km$ and 400$km$, or even the electron gyroradius. 
This results in a substantial change in the  dynamics of the flow at small scales, compared to MHD, giving rise to  more complex small-scale structures, enhanced reconnection and a steepening of  energy spectra, as observed in the solar wind \cite{sahraoui_13}, in models \cite{galtier_07} and in numerical simulations 
\cite{franci_15, gonzalez_19}. We also note that, in the presence of a strong uniform magnetic field, it is shown in \cite{galtier_15} that the  magnetic energy and helicity spectra are constrained by a relation stemming from their conservation, providing a lack of uniqueness in power-law steady-state solutions
(see \cite{grappin_83} for the case of the cross-correlation between the velocity and magnetic field in MHD). 
Finally, in \cite{stawarz_15H}, it was shown that, for the small-scale behavior of Hall MHD in the decaying case, magnetic energy becomes dominant at sub-ionic scales, with narrow and intense current structures in which one observes a strong alignment between the current and the magnetic field (leading to force-free fields), as well as a narrow electric field auto-correlation function.
On the other hand, the large-scale behavior of Hall-MHD, close to the ion inertial scale, has been much less investigated. Thus, in this paper, we wish to address the specific problem of the possible occurrence  and strength of inverse cascades to large scales in Hall MHD, as we vary the ion inertial length. The next section discusses equations and parameters, and we analyze our results in \S \ref{S:RESULTS-T} for temporal data, and in \S \ref{SECTION_SPEC} for 
{growth rates and}
spectral data. We recover 
{some of the scaling results using simple phenomenological}
arguments in \S \ref{SECTION:EXP}, and in \S \ref{SECTION:KF} we briefly describe the effect of varying the ratio of the forcing scale to the ion inertial length.  Finally, the last section presents a short discussion and conclusion.

\section{Problem set-up}  \subsection{Equations and parameters}

{
For a two-species plasma with ions and electrons, the usual Ohm's law relating electric field ${\bf E}$ and current density ${\bf j}=\nabla \times {\bf b}$ has to be generalized \cite{vasyliunas_75,priest_00}, 
depending on the length scale of the gradients {\it w.r.t.} the ion inertial  scale $\epsilon_H$, and where one could have collisionless dissipation mechanisms that limit the gradients even in the quasi-absence of collisions as in space plasmas (see \cite{song_01} for the three-fluid case including neutrals). 
In the Hall MHD model examined here, with ${\bf v}$ the velocity field  and $\eta$ the magnetic diffusivity,  the generalized Ohm's law is given by:
\be 
{\bf E} = -{\bf v} \times {\bf b} + \epsilon_H {\bf j} \times {\bf b} + \eta{\bf  j} \ .
 \label{eq:OHM} \ee
Small-scale dynamics becomes more complex than in MHD, with the breaking of current sheets beyond the ion inertial length (see for example \cite{cothran_05}).
In the case of Hall MHD, a large number of studies have found that the formation of helical coherent structures is enhanced \cite{mahajan_98}, as well as small-scale filamentation \cite{laveder_02}. The Hall current can also affect the rate of growth of the magnetic field and its saturation level \cite{mininni_03,gomez_10}, as well as the level of back-scatter to large scales \cite{mininni_07}. 
Recent high-resolution, multi-spacecraft measurements from MMS have enabled the direct measurement of generalized Ohm's law near small-scale current sheets in greater detail than previously possible 
\cite{torbert_16,webster_18,shuster_19}.
\\
In this context we write }
the  forced incompressible Hall MHD equations, with $\nabla \cdot {\bf v} = 0 \ , \ \nabla \cdot {\bf b} = 0$, as: 

\begin{eqnarray}
\frac{\partial {\bf v}}{\partial t} &=& -{\bf v} \cdot \nabla {\bf v} - \nabla P + {\bf j} \times {\bf b} + \nu \nabla^2 {\bf v} 
+ {\bf f}_v  ,  \label{eq:hall_mom} \\
\frac{\partial {\bf b}}{\partial t} &=& \nabla \times \left( {\bf v} \times {\bf b} \right) - \epsilon_H \nabla \times \left( {\bf j} \times {\bf b} \right) + \eta\nabla^2 {\bf b} + {\bf f}_b  . \label{eq:hall_ind}
\end{eqnarray}
{The energy input in the system, modeled by ${\bf f}_v$ and ${\bf f}_b$  at small (electron) scales can occur through reconnection processes which have been observed in the Earth's magnetotail at these scales \cite{ergun_18}. We also note that the magnetic field ${\bf b}$ is in fact in units of an Alfv\'en velocity,}
 with ${\bf b}={\bf B}/\sqrt{\mu_0 \rho}$, where ${\bf B}, \rho_0, \mu_0$ are respectively the magnetic induction, the density (assumed constant) and the permeability of vacuum.  The velocity ${\bf v}$ and magnetic field ${\bf b}$ are adimensionalized by a characteristic velocity $U_0$; $P$ is the particle pressure, and we take $\nu=\eta$ (unit magnetic Prandtl number).
Finally, ${\bf f}_{v,b}$ are forcing  functions with random phases constrained so as to set the initial  relative amount of kinetic and magnetic helicity, $\sigma_V $ and $\sigma_M$, as desired (see equation (\ref{E:sigma}) below). The initial conditions are identical to the forcing formulation.
We also define the magnetic potential ${\bf a}$, as usual, through ${\bf b}=\nabla \times {\bf a}$. The Hall term is controlled by the dimensionless parameter $\epsilon_H = d_i$ which is the  ion inertial length, measured in terms of the overall dimension of the flow (see \cite{toth_17} for the role of the ion scale in the overall dynamics in numerical approaches). The MHD equations are recovered for $\epsilon_H = 0$.

The code we use is pseudo-spectral and implements a hybrid methodology for parallelization, using both MPI and Open-MP  \cite{mininni_11, hybrid_11}. The runs analyzed in this paper, computed  in a cubic box and with periodic boundary conditions, are summarized in Tables 1 and 2. 
Forcing spectra are centered in the Fourier shells with $19\le k_F \le 21$ for the runs of Table 1, and $7\le k_f \le 9$ for the runs of Table 2. The box is of length $2\pi$, corresponding to a minimum wavenumber $k_{min}=1$; we use a classical 2/3 de-aliasing rule, and thus the maximum wavenumber is $k_{max}=N_p/3$ with $N_p$ the number of grid points in each direction. The amplitude of the forcing is set so that the {\it rms} velocity and magnetic fields are of order unity.
The time step for all the runs varies between  $5 \times 10^{-4}$ and $5 \times 10^{-3}$.

\subsection{The ideal case}
The ideal invariants in Hall-MHD  \citep{turner_86}, for $\nu=\eta=0$, are the total energy 
$E_T=E_V+E_M= \langle |{\bf v}|^2 + |{\bf b}|^2 \rangle/2$, the magnetic helicity 
$H_M=\langle {\bf a} \cdot {\bf b} \rangle/2$ and the generalized helicity $H_G$ defined as:
\begin{equation*}
H_G = \frac{1}{2} \langle \left({\bf a} + \epsilon_H {\bf v} \right) \cdot \left({\bf b} + \epsilon_H \mbox{\boldmath$\omega$\unboldmath} \right) \rangle 
\label{eqhg} \end{equation*}
and with 
\be H_G= H_M + 2 \epsilon_H H_C + \epsilon_H^2 H_V = H_M + \epsilon_H H_X\ ; \ee
   $\mbox{\boldmath$\omega$}=\nabla \times {\bf v}$ is the vorticity, 
 $H_V = \langle {\bf v} \cdot \mbox{\boldmath$\omega$\unboldmath} \rangle/2$  the kinetic helicity (an invariant for ideal neutral fluids), and $H_C = \frac{1}{2} \langle {\bf v} \cdot {\bf b} \rangle$ is the cross-correlation between the velocity and magnetic fields. Note that, because $H_M$ is itself invariant, the combination 
 $H_X=2H_C+\epsilon_H H_V$ is also invariant. For  $\epsilon_H \rightarrow 0$ corresponding to the MHD case, one thus recovers from the invariance of $H_X$ the cross-helicity invariance which can thus be seen as the equivalent  of $H_X$ but for MHD. 
 This change of invariants from the MHD case  may imply as well a change in the dynamics of the flow (see, {\it e.g.}, \cite{servidio_08}). 
Note that in the expression of $H_{G,M}$ appear polarized waves (right and left, respectively); namely, $H_G$ can be written as 
 $H_G={\bf \Gamma} \cdot {\bf \Omega}/2$,   with ${\bf \Gamma}={\bf a} + \epsilon_H {\bf v}, \ {\bf \Omega}= {\bf b}+ \epsilon_H\vomega = \nabla \times \Gamma$ \cite{banerjee_16};
 {$H_G$ is also called ion helicity in \cite{ohsaki_05}. \\
When MHD flows in the Solar Wind are strongly correlated, accelerated particles are more prominent \cite{sorriso_19}; this is likely due to the role played by $H_C$ in the so-called exact laws for MHD \cite{politano_98g} (see \cite{marino_12} for an observation of such laws, and see below, equation (\ref{banerjee}) for the helical case in Hall MHD). It has also been conjectured that $H_C$ can be measured in the solar convection zone \cite{rudiger_11}. Moreover, the cross-helicity in MHD is known to grow with time \cite{pouquet_86}, and it has been shown to be of different signs in the large and small scales, with the so-called pinning effect at the dissipation scale \cite{grappin_82} (see also \cite{passot_19}). This dichotomy is also present in the spatial structures of the flow \cite{perez_09}, with large one-signed lobes of high relative correlation separated in the current sheets by fast oscillating structures \cite{meneguzzi_96}. Thus, $H_C$ can affect both the large scales and hence be a factor in the dynamo effect of generation of large-scale magnetic fields \cite{yokoi_13}, as well as play a role in the small scales modeled through an enhanced magnetic diffusivity which can be associated with fast reconnection \cite{yokoi_13c, titov_19}. Whether $H_G$ plays corresponding roles for scales smaller than $\epsilon_H$ has only been studied recently \cite{miloshevich_17,miloshevich_20}.  For example, on the basis of statistical equilibria, it is shown in  \cite{miloshevich_17} that the direction of the cascade for $H_G$ is ambiguous, as we also argue below noting its  dependence on the ion inertial length, $\epsilon_H$. 
\\
Furthermore, the presence of cross-helicity in MHD can lead to different energy spectra, depending on $\sigma_C$  (see equation (\ref{E:sigma}) below) \cite{grappin_82,politano_89}. Today,  this remains a disputed issue which may depend on the model that is used. A unifying framework, for a two-dimensional formulation of reduced MHD in the presence of a strong uniform magnetic field, from large (MHD) scales to scales below the ion inertial length, has been proposed in \cite{passot_19}, with, in particular, a detailed analysis of the weak (wave) turbulence regime leading to  integro-differential equations with various steady power-law  solutions.}
Exact scaling  laws in terms of structure functions can be derived for Hall MHD. They  represent, in a different form, the conservation of $E_T, H_M$ and $H_G$ \cite{banerjee_16}. For strong Hall currents, and assuming homogeneity (but not isotropy in this formulation), these exact laws reduce to:
\begin{equation}
\tilde{\epsilon}_m = \epsilon_H[\delta[{\bf b} \times {\bf j}] \cdot \delta{\bf b}] \  , \ee
\be
\tilde{\epsilon}_G  =\epsilon_H [\delta[{\bf v} \times {\bf b}] \cdot \delta \vomega +  
\delta[{\bf v} \times \vomega] \cdot \delta {\bf b}] + \epsilon_H^2 \delta[{\bf v} \times \vomega] \cdot \delta \vomega     \ , 
\label{banerjee} \end{equation}
where, for any vector ${\bf F}$, one defines $\delta {\bf F}={\bf F}({\bf x}+{\bf r})- {\bf F}({\bf x})$,  with ${\bf r}$ in the inertial range(s), and where ${\tilde \epsilon}_{[m,G]}$ are the decay rates of $H_{[M,G]}$.
Such exact laws for incompressible Hall MHD, under the further assumptions of large Reynolds number and stationarity,  represent dynamical constraints on the temporal, spatial and spectral evolution  of the flow, that differ from the MHD case, in particular emphasizing a stronger involvement  than in MHD of the  small scales, through the kinetic helicity.

Finally, we define relative helicities which correspond to the relative alignment or anti-alignment of vectors when maximal ($\pm 1$); they are in fact  cosines  functions, namely: 
\begin{equation}
            \sigma_M = \frac{{\bf a} \cdot {\bf b}}{|{\bf a}||{\bf b}|}, 
\qquad \sigma_C = \frac{{\bf v} \cdot {\bf b}}{|{\bf v}||{\bf b}|}, 	
\qquad \sigma_V =  \frac{{\bf v} \cdot \mbox{\boldmath$\omega$\unboldmath}}{|{\bf v}||\mbox{\boldmath$\omega$\unboldmath}|} 
\ee
and
\be
 \sigma_G = \frac{\left({\bf a}+\epsilon_H {\bf v} \right) \cdot \left({\bf b}+\epsilon_H \mbox{\boldmath$\omega$\unboldmath} \right)}{|{\bf a}+\epsilon_H {\bf v} ||{\bf b}+\epsilon_H \mbox{\boldmath$\omega$\unboldmath}|} = \frac{{\bf \Gamma} \cdot {\bf \Omega}}{|{\bf \Gamma}| |{\bf \Omega}|}.
\label{E:sigma} \end{equation}
\begin{table} \caption{
List of the runs, with ID  their identification, $N_p$ the numerical resolution, $\nu$ the viscosity, $\epsilon_H$ the Hall parameter,
 $\sigma_{M,V,C,G}$ the relative rates, for the forcing, of the  magnetic, kinetic, cross and generalized helicities respectively (see equations (\ref{E:sigma})). Finally, $Re$ is the Reynolds number, and $k_{d_{i}}=1/\epsilon_H$ is the ion inertial wavenumber. For these runs, the forcing scale $k_F$ is in the range $19 \le k_F \le 21$.  
}    \centering \begin{tabular}{cccccccccccc}  \toprule
\textbf{ID}	& \textbf{$N_p$} & \textbf{$\nu$} & \textbf{$\epsilon_H$}  & \textbf{$\sigma_M$}  & \textbf{$\sigma_V$} & \textbf{$\sigma_C$} & \textbf{$\sigma_G$} & \textbf{$Re$} & $k_{d_{i}}$ 
\\ \hline 
AM1&  $128^3$  & 0.016 &  0.0       & 0.65 &  0.131 & -0.027 & -- & 15.1 &  --   \\
AH2 & $128^3$   & 0.016 &  0.0667 & 0.65 &   0.131 & -0.027 & 0.295 & 17.2 &  15  \\
AH3  & $128^3$  & 0.016 &  0.0833 & 0.65 &   0.131 & -0.027 & 0.247  & 17.6 &  12   \\
AH4  & $128^3$  & 0.016 &  0.14     & 0.65 &   0.131 & -0.027 & 0.174  & 18.5 &  7   \\
AH5  & $128^3$  & 0.016 &  0.2       & 0.65 &   0.131 & -0.027 & 0.15  & 18.8 &  5  \\
\hline 
\end{tabular}   \end{table} 

\begin{figure*} 
\includegraphics[width=5.1cm, height=5.52cm]{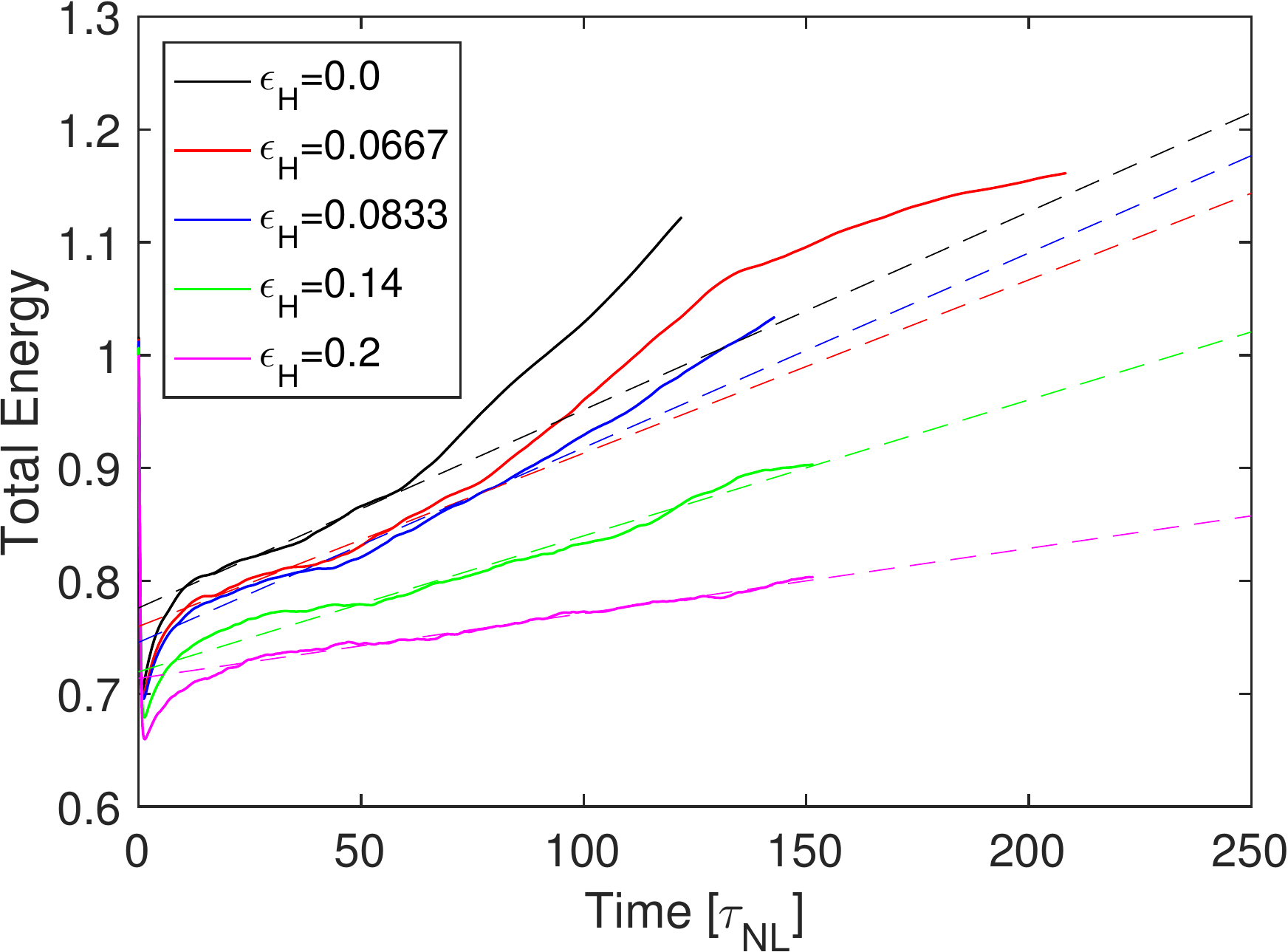}  \hskip0.1truein
\includegraphics[width=5.1cm, height=5.52cm]{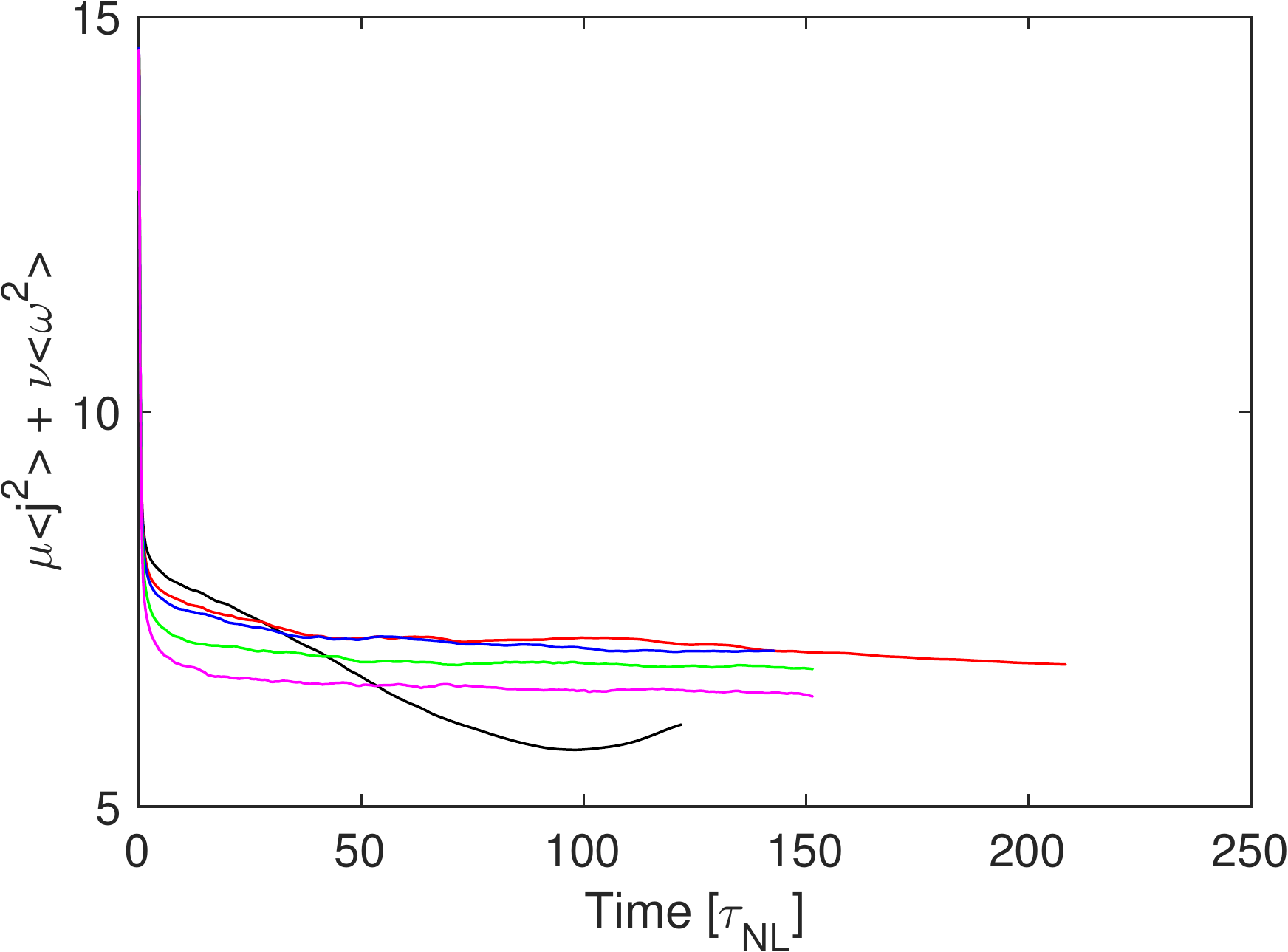}  \hskip0.1truein
\includegraphics[width=5.1cm, height=5.52cm]{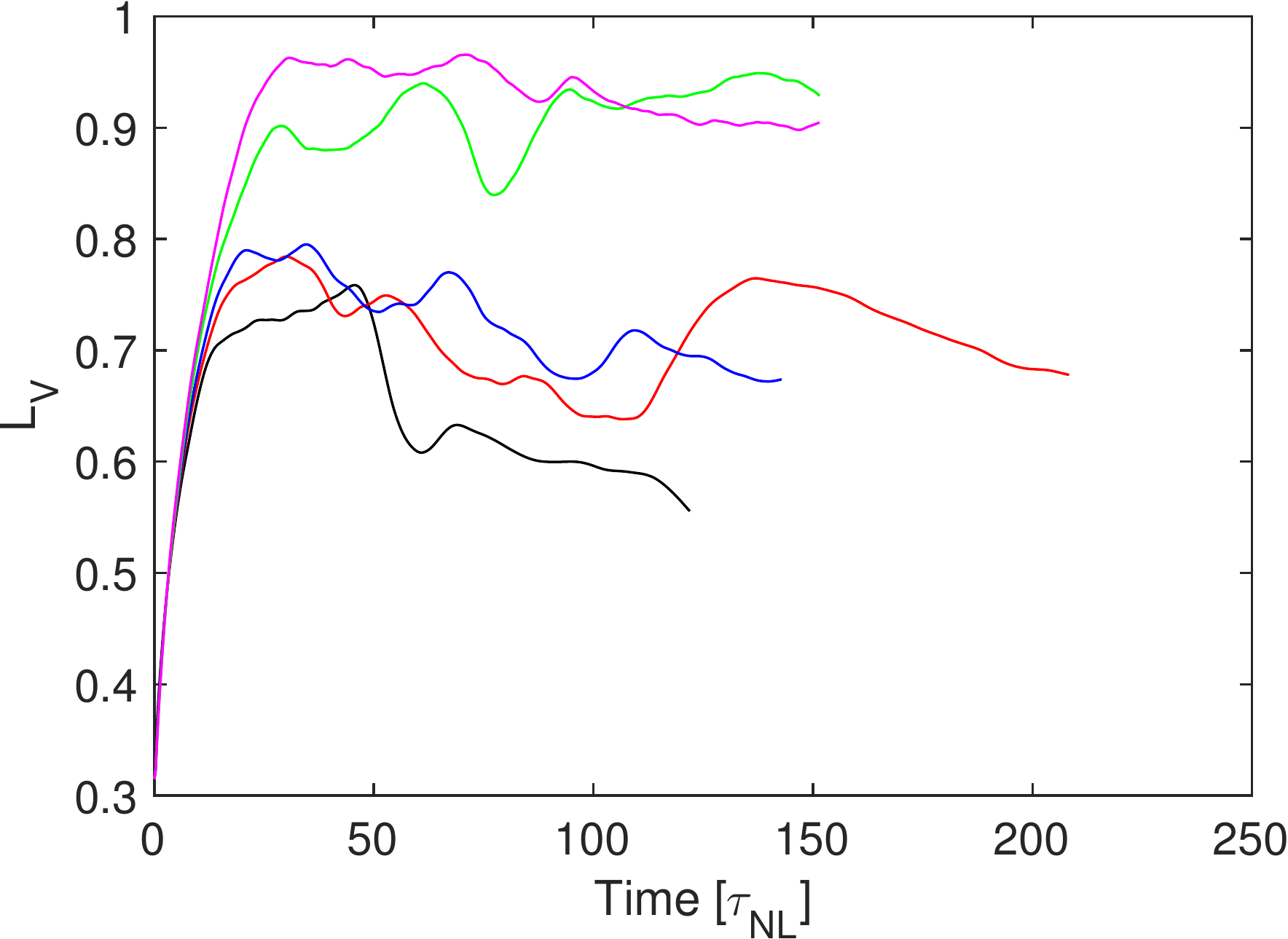}   \vskip0.1truein
\includegraphics[width=5.1cm, height=5.52cm]{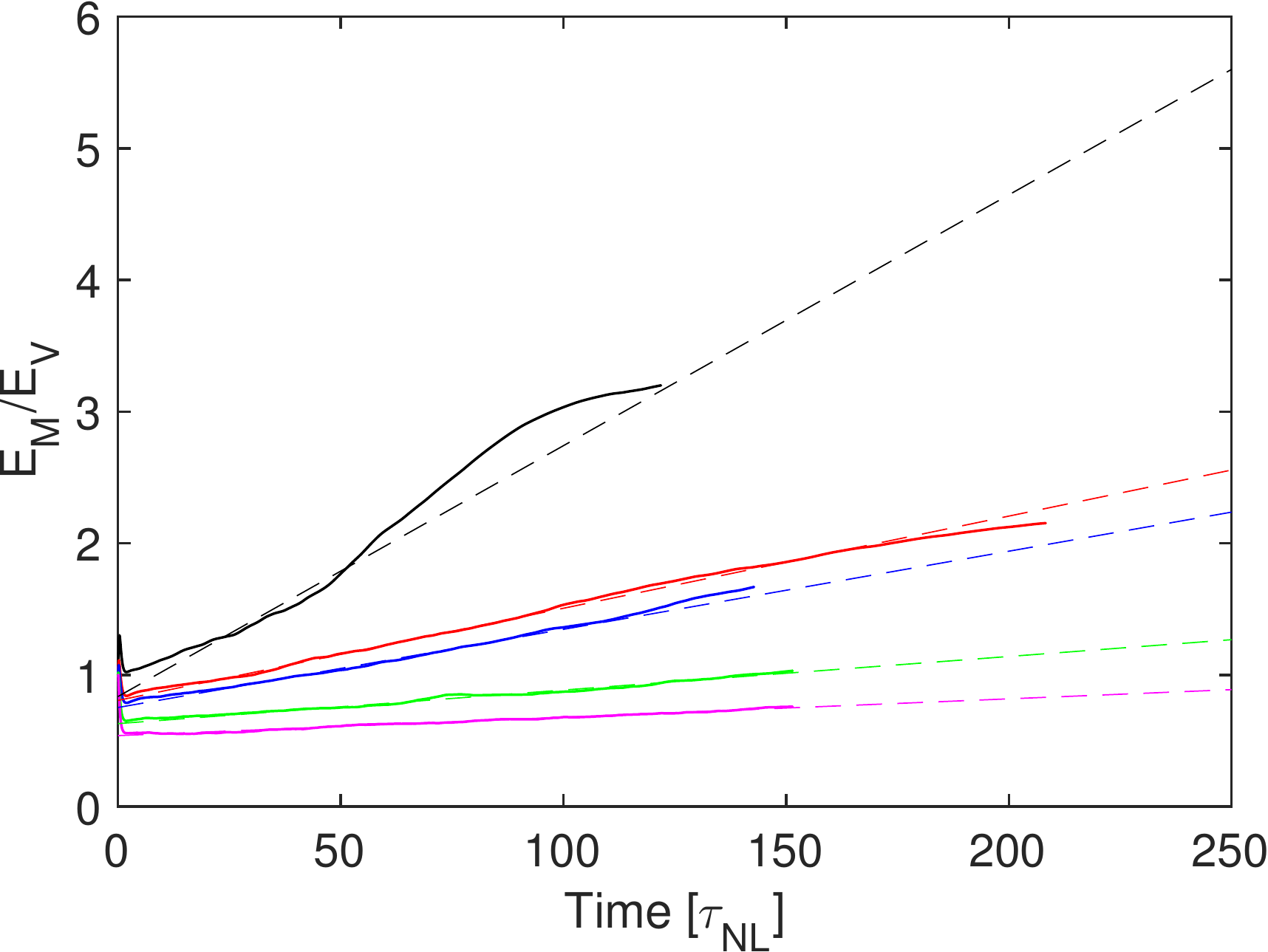}  \hskip0.1truein 
\includegraphics[width=5.1cm, height=5.52cm]{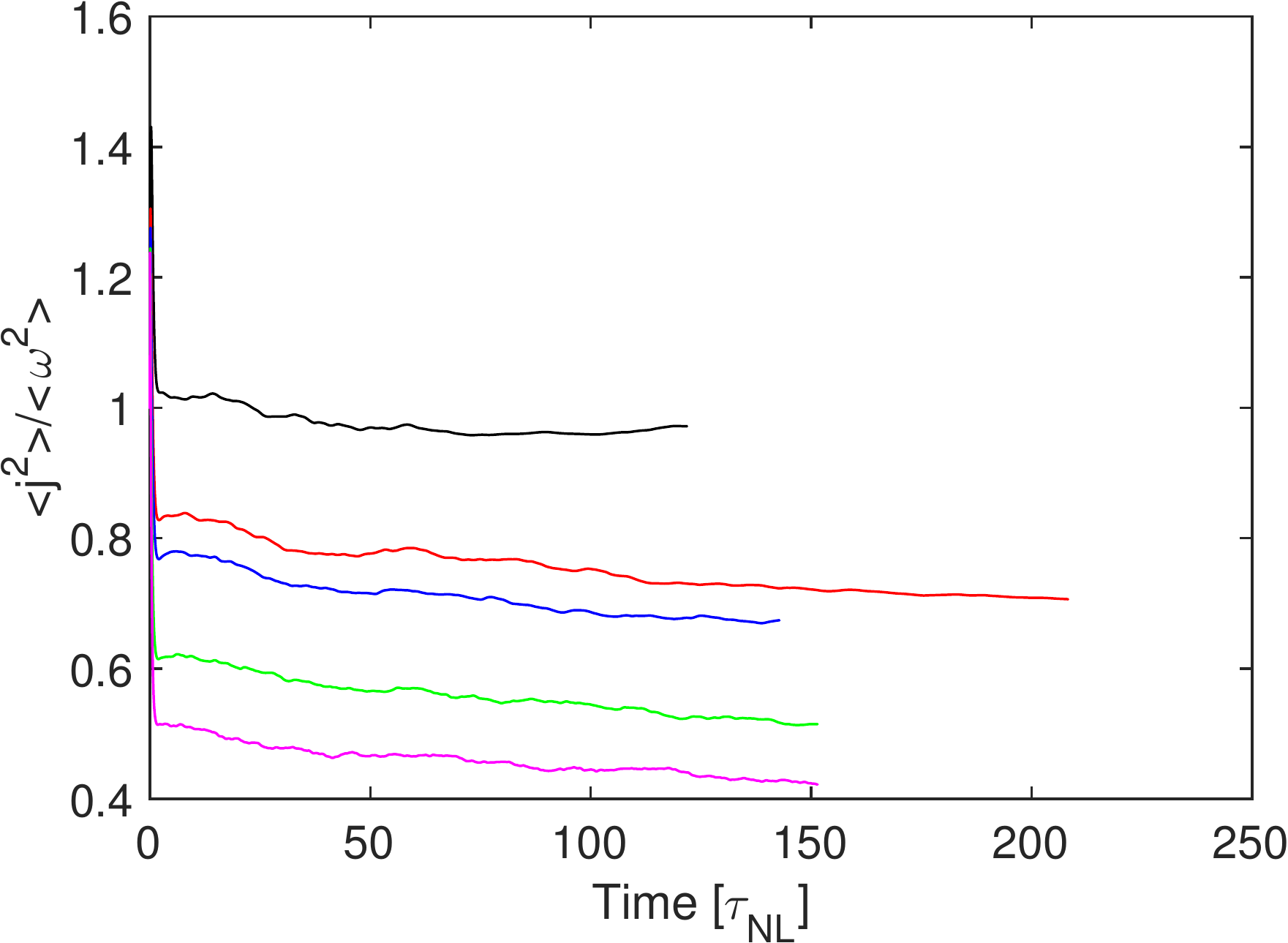}  \hskip0.1truein  
\includegraphics[width=5.1cm, height=5.52cm]{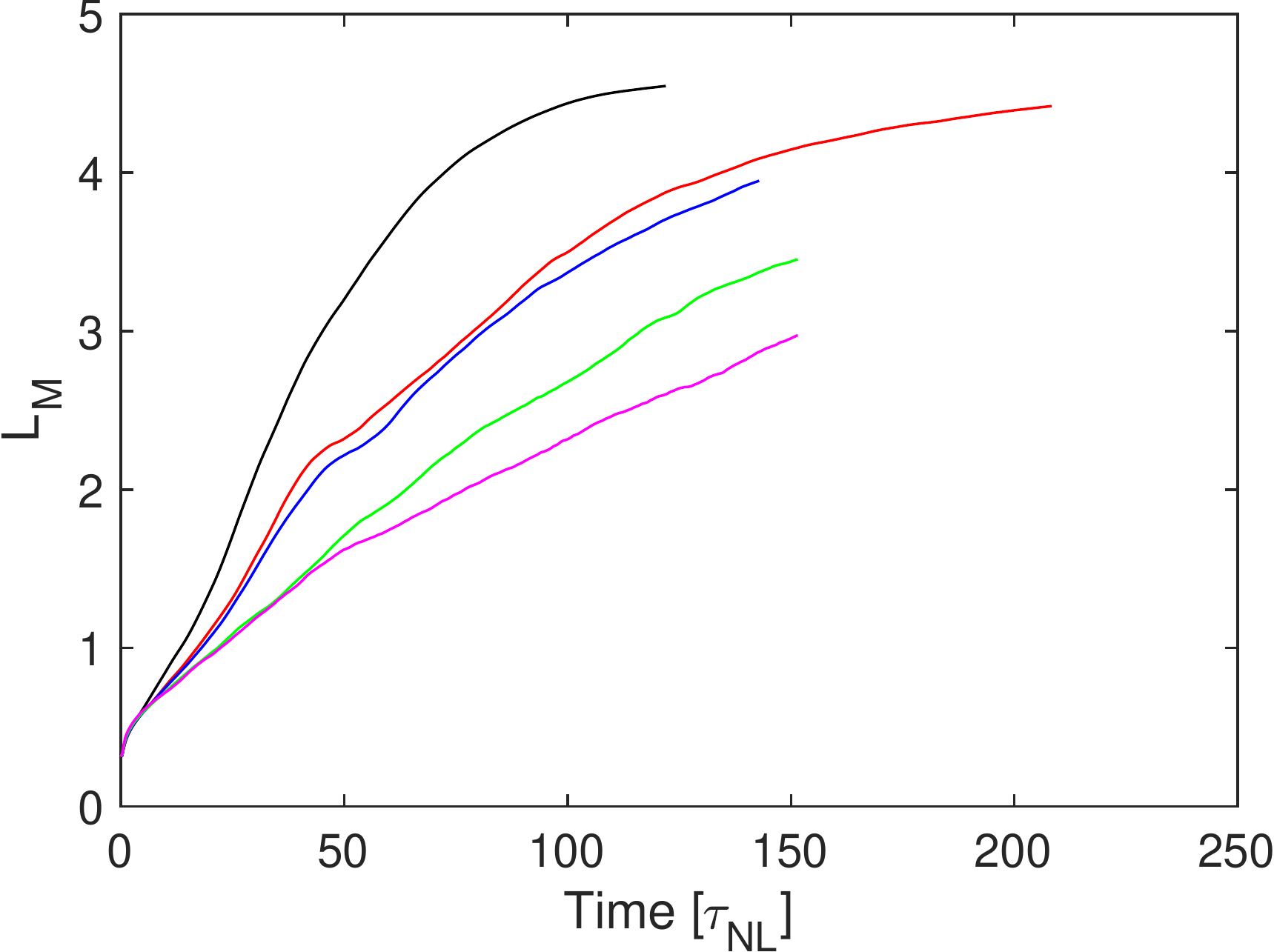}  
\caption{For  runs of Table 1, as a function of time in units of   turn-over time $\tau_{NL}=L_0/U_0$:
Left:  Total energy $E_T$ (top), and ratio of magnetic to kinetic energy, $E_M/E_V$ (bottom).
Middle: Total dissipation (top), and ratio of $L_2$ norms of current and vorticity (bottom).
Right: Integral scales built on the kinetic   energy (top) and  on the magnetic energy  (bottom). Note the different scaling on the vertical axes. Dotted lines indicate linear fits for growth rates.
}\label{f:1}    \end{figure*}

In the linearized case,  two types of waves coexist in Hall MHD \citep{sahraoui_06}. Magnetic polarization is defined as $P_M = \sigma_M\sigma_C$, computed in Fourier space. It measures the direction of circular polarization relative to the magnetic field.  $P_M>0$ ({\it vs.} $P_M<0)$ corresponds  to left ({\it vs.} right) circularly polarized fields  \citep{meyrand_12}. They are called ion-cyclotron and whistler waves, and have different dispersion relations in terms of wavenumbers, which can affect the destabilization of large-scale magnetic fields, as described by the so-called alpha-dynamo in MHD. The  turbulent diffusivity is affected as well by the Hall current and can become negative, unlike the MHD case in three dimensions (see \cite{mininni_07} and references therein). The wavenumber-dependent ratio of magnetic to kinetic energy, at each wavenumber $k$, depends on $\epsilon_H$ and $k$, and the Alfv\'enic state of equipartition typical of MHD is broken by the Hall current, both at large scales and at small scales.

The behavior of dissipation-less ideal systems can be obtained from first principles \cite{lee_52, kraichnan_67, kraichnan_73}, with  the long-time energy  spectrum scaling corresponding to an equipartition between all individual Fourier modes in the simplest case.
However, it has been conjectured, and it has been shown recently numerically, that the  behavior in the ideal case can be in fact a predictor of  their dissipative counterparts, the small-scale thermalized modes acting as an effective viscosity and resistivity on the large scales \cite{cichowlas_05}. Henceforth,  a Kolmogorov spectrum typical of fluid turbulence and as found in  atmospheric flows \cite{nastrom_85}, including for helicity \cite{koprov_05}, is observed in ideal systems at intermediate scales and intermediate times before the system reaches equilibrium. These results have been extended to  other systems, as for example in MHD \cite{krstulovic_11}, and they are believed to be universal \cite{mininni_07b}. 

It is thus of great interest to study such equilibria which can in particular give indications on the directions of turbulent cascades to either small or large scales. 
Statistical equilibria for Hall MHD with a finite number of modes
were derived in \cite{servidio_08} (see also \cite{zhuJZ_14}), revealing several distinguishing features of these idealized systems. In particular, there is, as in MHD, a large-scale condensation, here of  generalized helicity $H_G$, as well as of $H_M$, and, to a lesser extent, also present in the  magnetic energy. Furthermore, the equipartition  between kinetic and magnetic energy, associated with the presence of  Alfv\'en waves, is broken in the presence of non-zero $H_G$, at a wavenumber that depends on $\epsilon_H^{-1/2}$. One can conjecture that, similarly, the helical equipartition (between kinetic and current helicity) is broken as well, when applying a Schwarz inequality. Following up with numerical simulations, these authors also show that large-scale excitation is weaker in Hall MHD with correspondingly more small-scale energy available for dissipative processes \cite{servidio_08}. Note that in the statistical equilibria solutions, the expressions for  $H_{M}$ and $H_G$ are polynomial in $\epsilon_H$. One can thus expect, indeed, that there will be different regimes 
 depending on the generalized temperatures associated with these ideal invariants. 

\begin{figure*} 
\includegraphics[width=5cm, height=5.52cm]{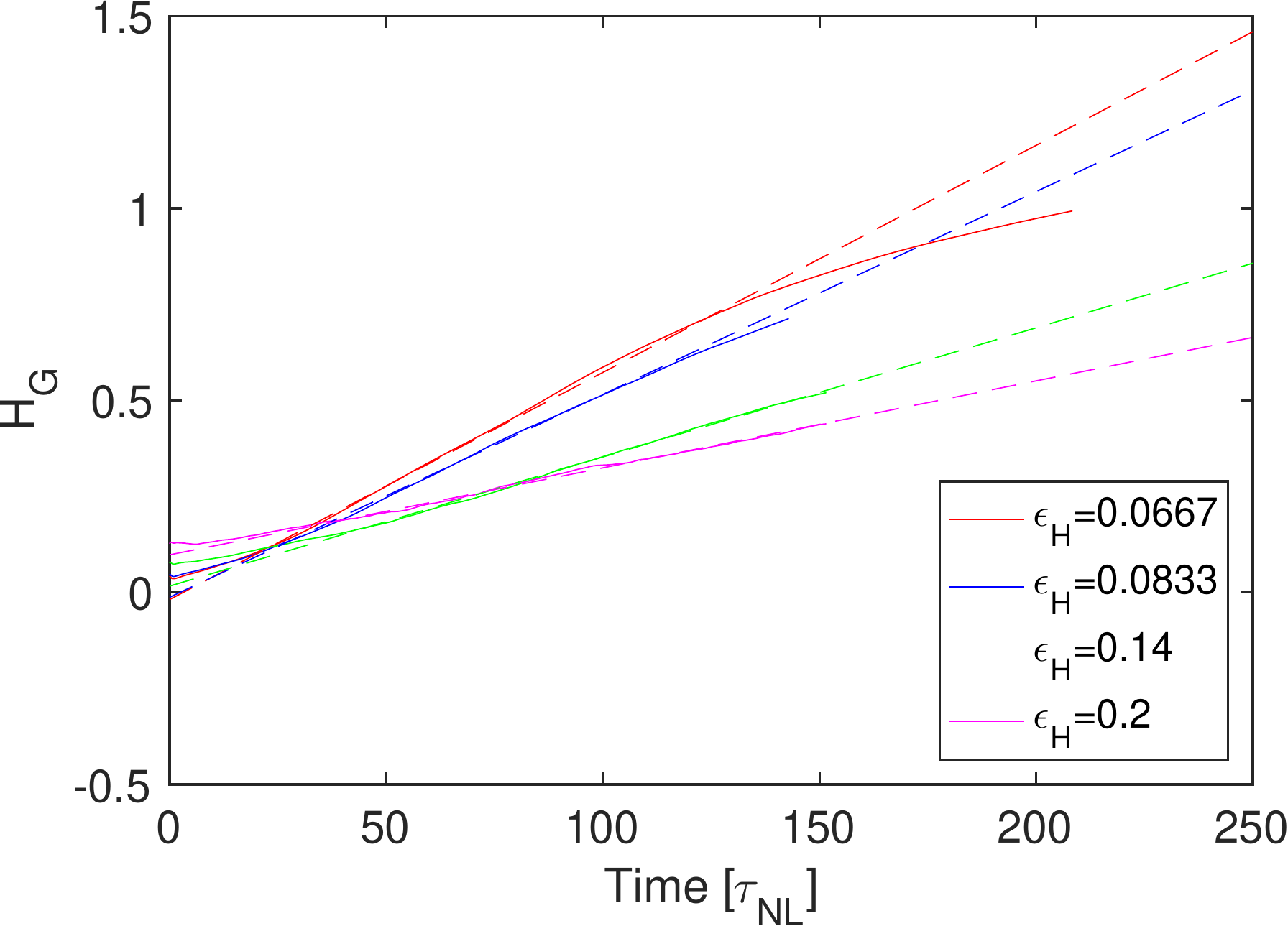}  \hskip0.1truein   
\includegraphics[width=5cm, height=5.52cm]{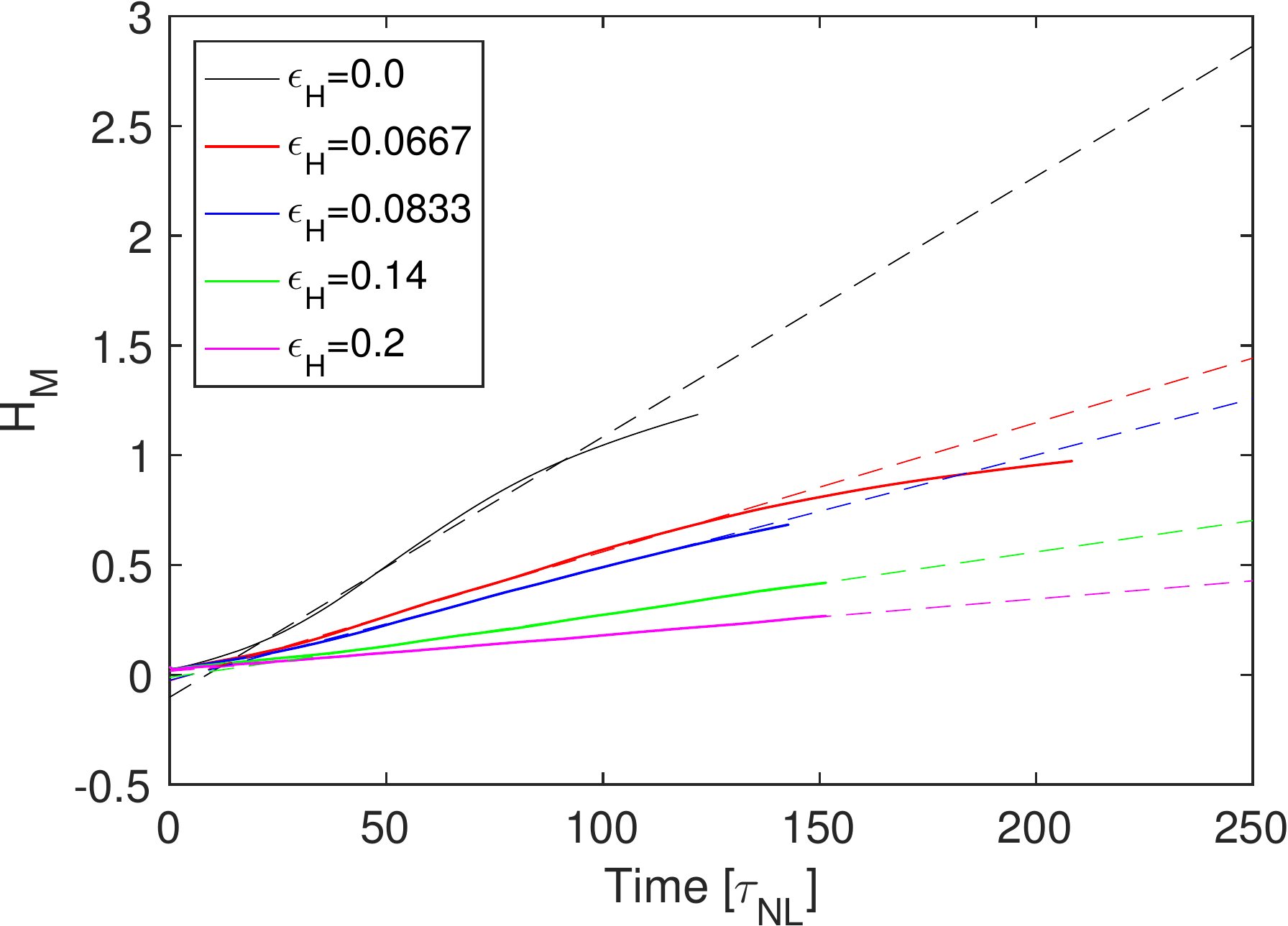}  \hskip0.1truein   
 \includegraphics[width=5cm, height=5.52cm]{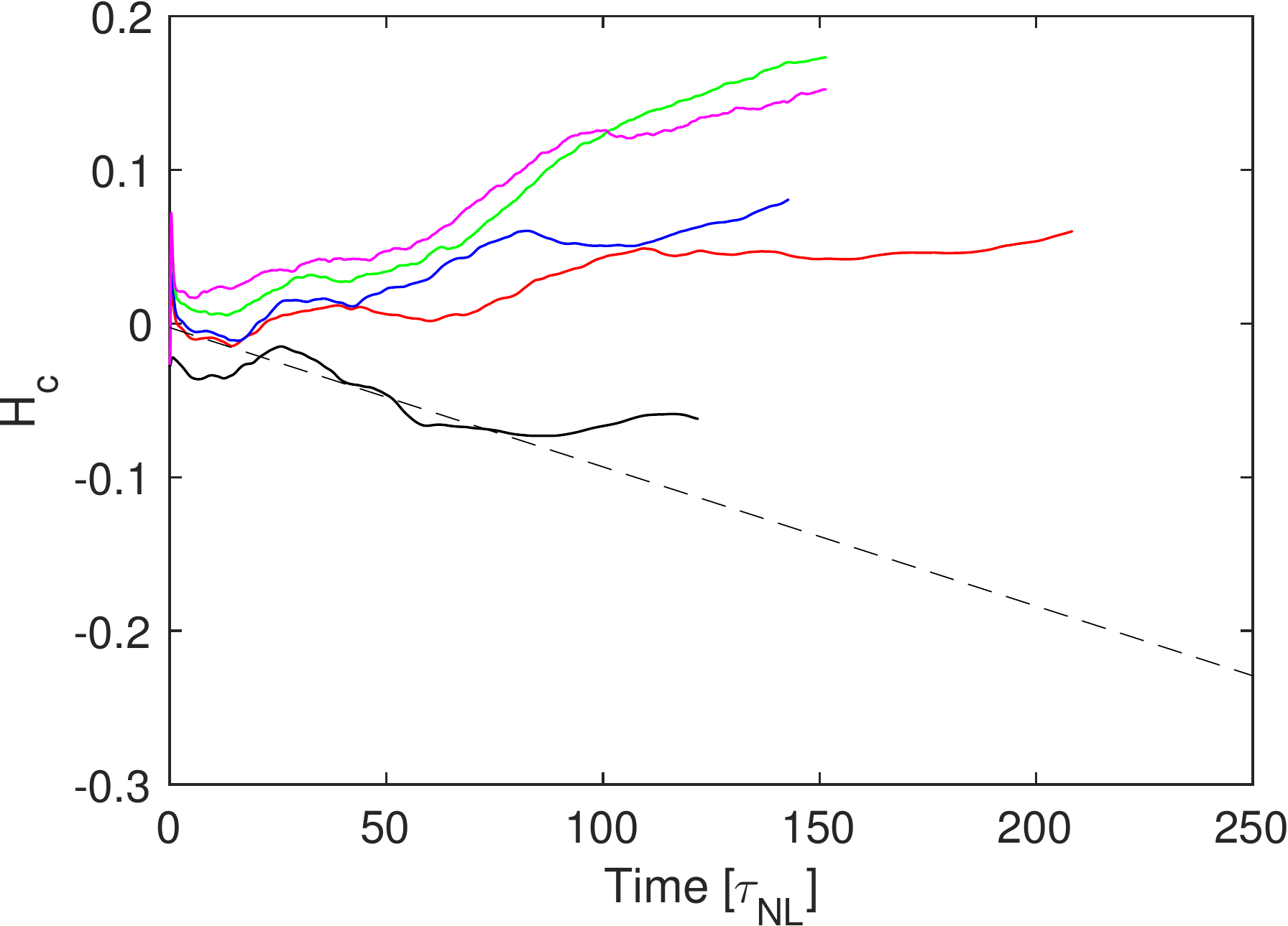}  \vskip0.1truein     
\includegraphics[width=5cm, height=5.52cm]{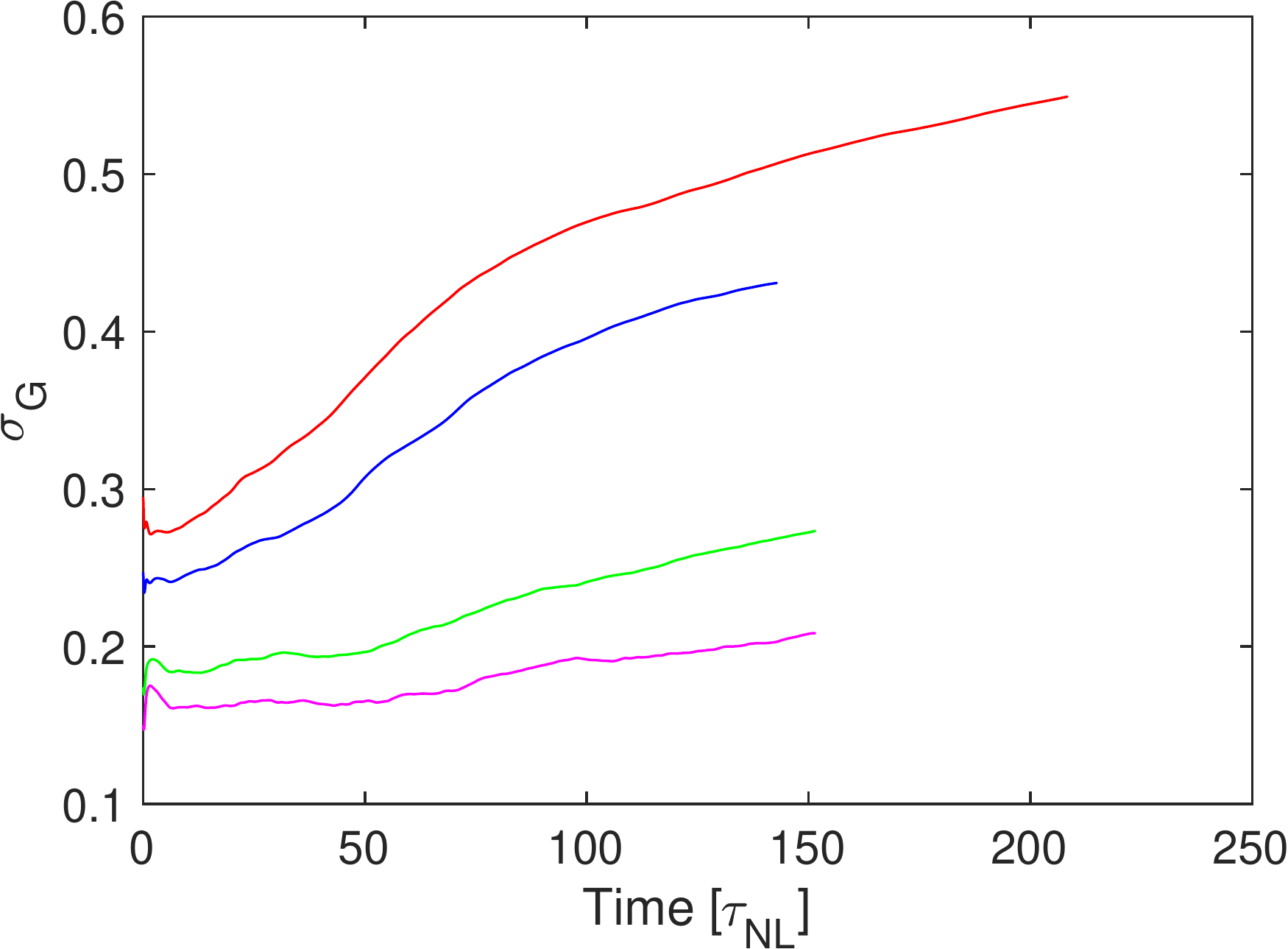}  \hskip0.1truein   
\includegraphics[width=5cm, height=5.52cm]{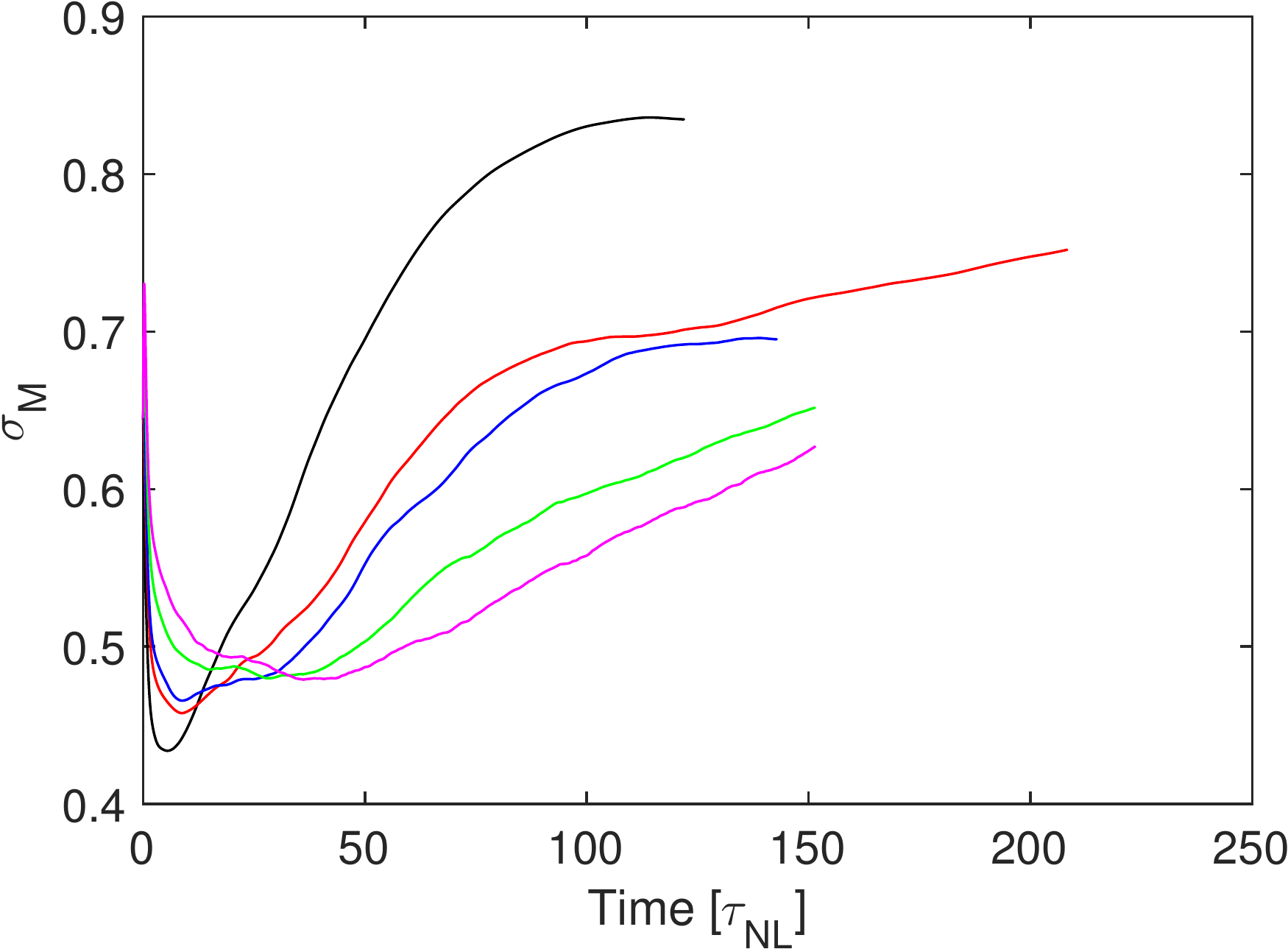}  \hskip0.1truein   
\includegraphics[width=5cm, height=5.52cm]{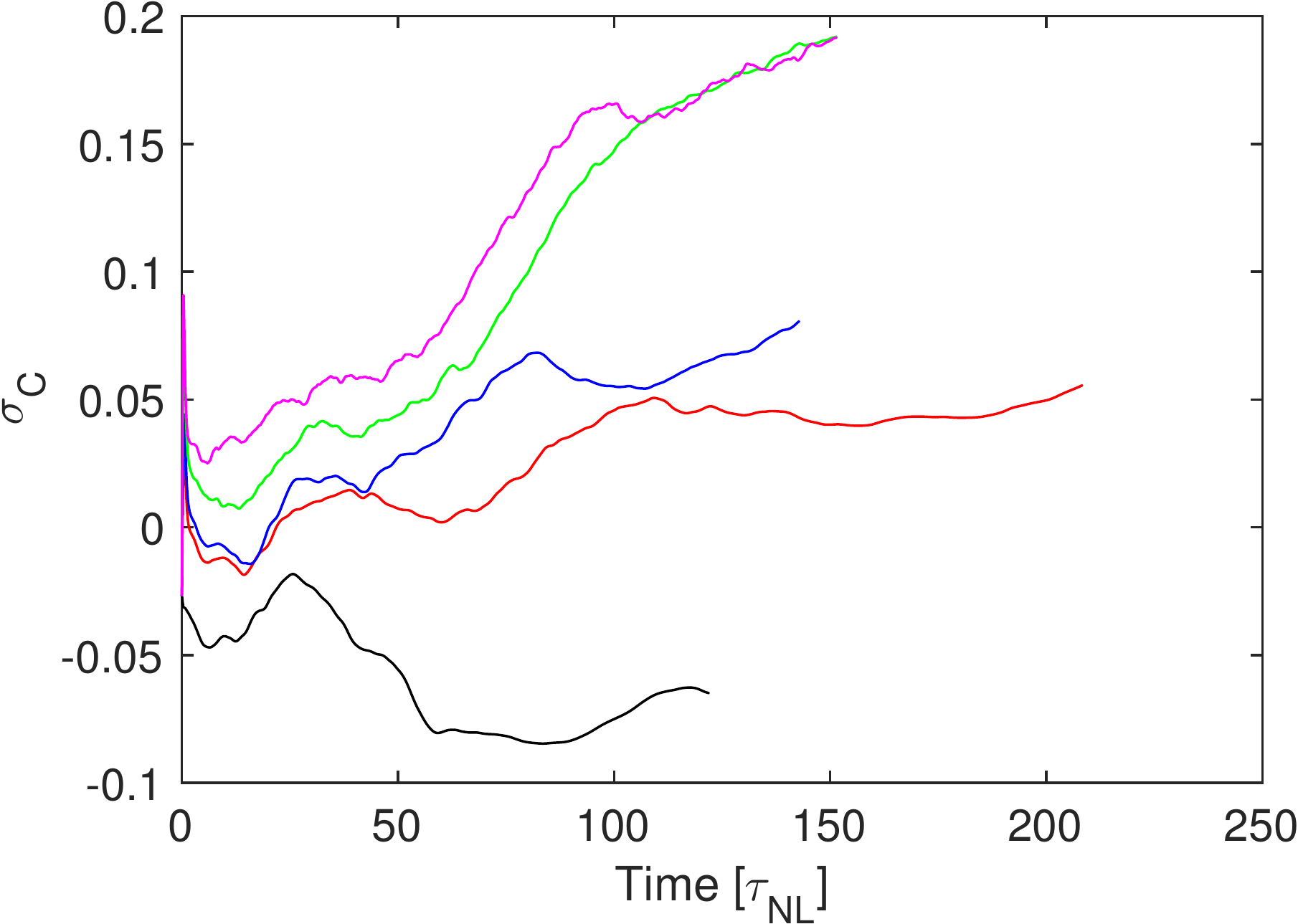}  \hskip0.1truein  
\vskip-0.05truein  
\caption{Temporal data for the runs of Table 1.
 Left column: Generalized helicity $H_G$ (top) and its relative counterpart $\sigma_G$ (bottom).
 Middle:  Total magnetic helicity $H_M$ (top), and its relative counterpart $\sigma_M$  (bottom). 
 Right: cross-helicity $H_C$ (top), and its relative counterpart  $\sigma_C$ (bottom).
}\label{f:2}    \end{figure*}

\section{Large-scale dynamics of Hall MHD: Temporal data} \label{S:RESULTS-T}            
We now examine the behavior of the runs of Table 1 with small-scale forcing.
We first plot in Fig. \ref{f:1} the temporal variations of the total energy (top left) and the total dissipation 
$\epsilon_T=\epsilon_V+\epsilon_M= \nu \left<|\vomega|^2 \right> + \eta  \left<|j|^2 \right>$
(top middle).  The different values of $\epsilon_H$ are given by different colors (see inset), and the dotted lines represent fits to the growth rates of energy (and of $E_M/E_V$). Note that, for all these runs, the ion inertial length is larger than the forcing scale and thus resides in the inverse cascade range. 
Below these plots are given the temporal evolution of the ratio of magnetic to kinetic energy (bottom left), and of $\left<j^2 \right>/\left<\omega^2 \right>$ (bottom middle). 
Because of the growth of $H_M$ and $H_G$ (see below, Fig. \ref{f:2}), and since by Schwarz inequality, $E_M(k) \ge kH_M(k)$, $E_M$ grows as well and thus so does $E_T$, as observed here.
The ratio $E_M/E_V$ also grows (Fig. \ref{f:1}, bottom left), although to a lesser extent for the higher $\epsilon_H$ values, due  to the lesser efficiency of the inverse cascade for strong Hall currents, as well as to the lack of  efficient Alfv\'en waves. In the small  scales, the saturation of dissipation in Hall-MHD is faster than in MHD, occurring at a much earlier time, and at a higher level, at least for low values of $\epsilon_H$. Moreover, the ratio of current to vorticity, close to unity in MHD, is lower in Hall-MHD, again with a sub-dominance of dissipative eddies in current structures the stronger the Hall term  (see Fig. \ref{f:1}, bottom middle). 

For the highest value of $\epsilon_H$, the energy ratio $E_M/E_V$ remains smaller than one at all times. 
This corroborates the important point already noted in \cite{servidio_08} on the basis of statistical equilibria: the Alfv\'en energy equipartition is broken by the Hall term. Indeed, when ${\bf b}= \pm \alpha {\bf v}$ as in an Alfv\'en wave, with $\alpha$ a pseudo-scalar constant in space, the first term in the generalized Ohm's law disappears
{ (see equation (\ref{eq:OHM})),}
but the magnetic induction can still evolve through the Hall current. However, in the momentum equation, the nonlinear terms disappear altogether if as above, 
$\vomega=\alpha {\bf v}, \sigma_V=\pm 1$. 
 This will remain true as long as current and induction do not align (we note however that, in MHD, the alignment between ${\bf b}$ and ${\bf j}$ is very efficient \cite{servidio_08b}). As $\epsilon_H$ grows, the dominance of vorticity over current can be attributed as well to the fact that the kinetic helicity term in $H_G$ gains in importance, controlling the correlations between velocity and vorticity and thus, to some extent, the strength of the vorticity itself. Indeed, it is known that, for neutral fluids, the kinetic helicity follows a $k^{-5/3}$ law and the relative kinetic helicity thus decays slowly, as $1/k$ (for rotating flows, see  \cite{mininni_09c}). 

\begin{figure*} 
 \vskip-0.97truein   \hskip-0.27truein
 \includegraphics[width=9.3cm, height=12.5cm]{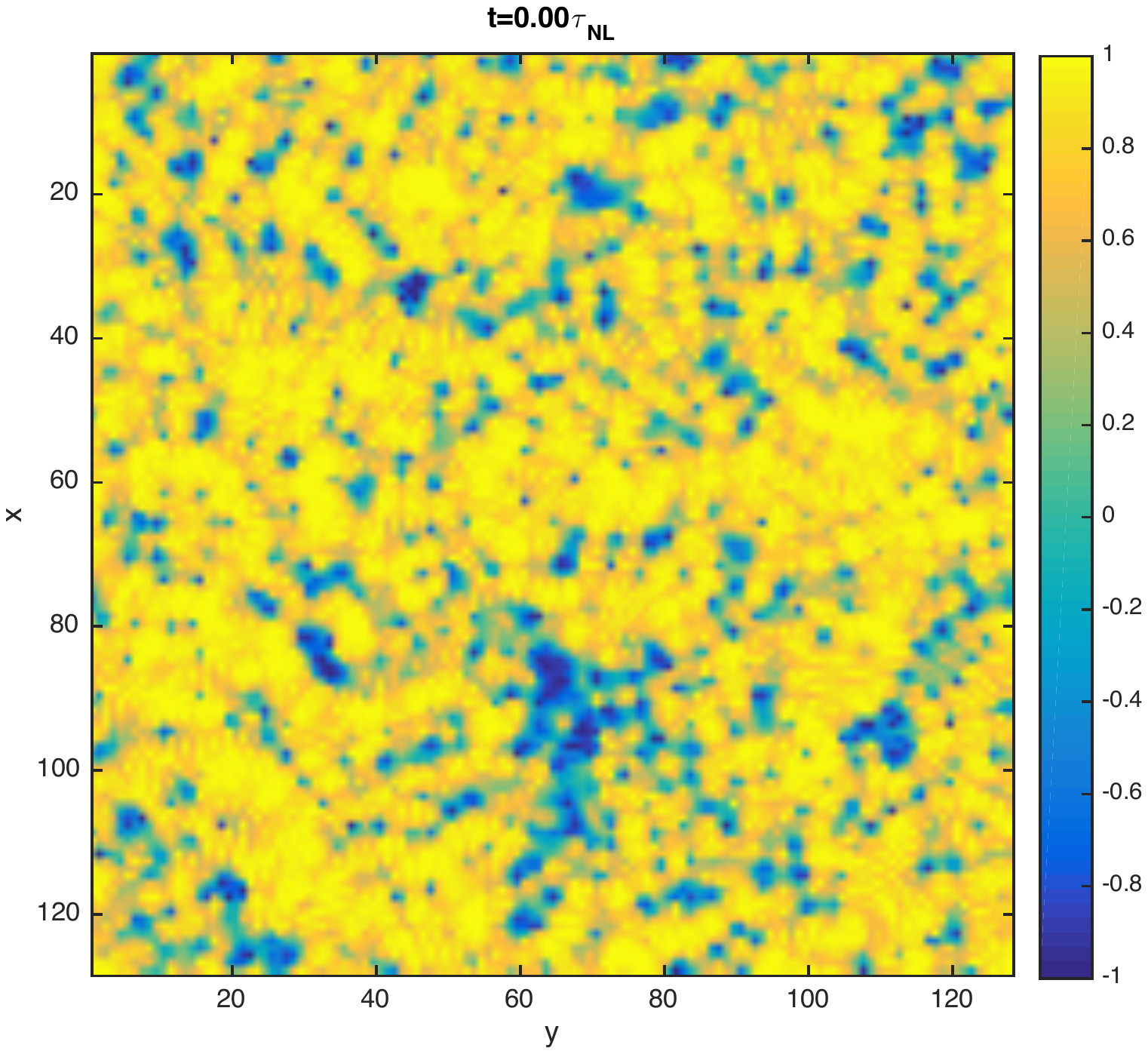}   \hskip-0.7truein  
 \includegraphics[width=9.3cm, height=12.5cm]{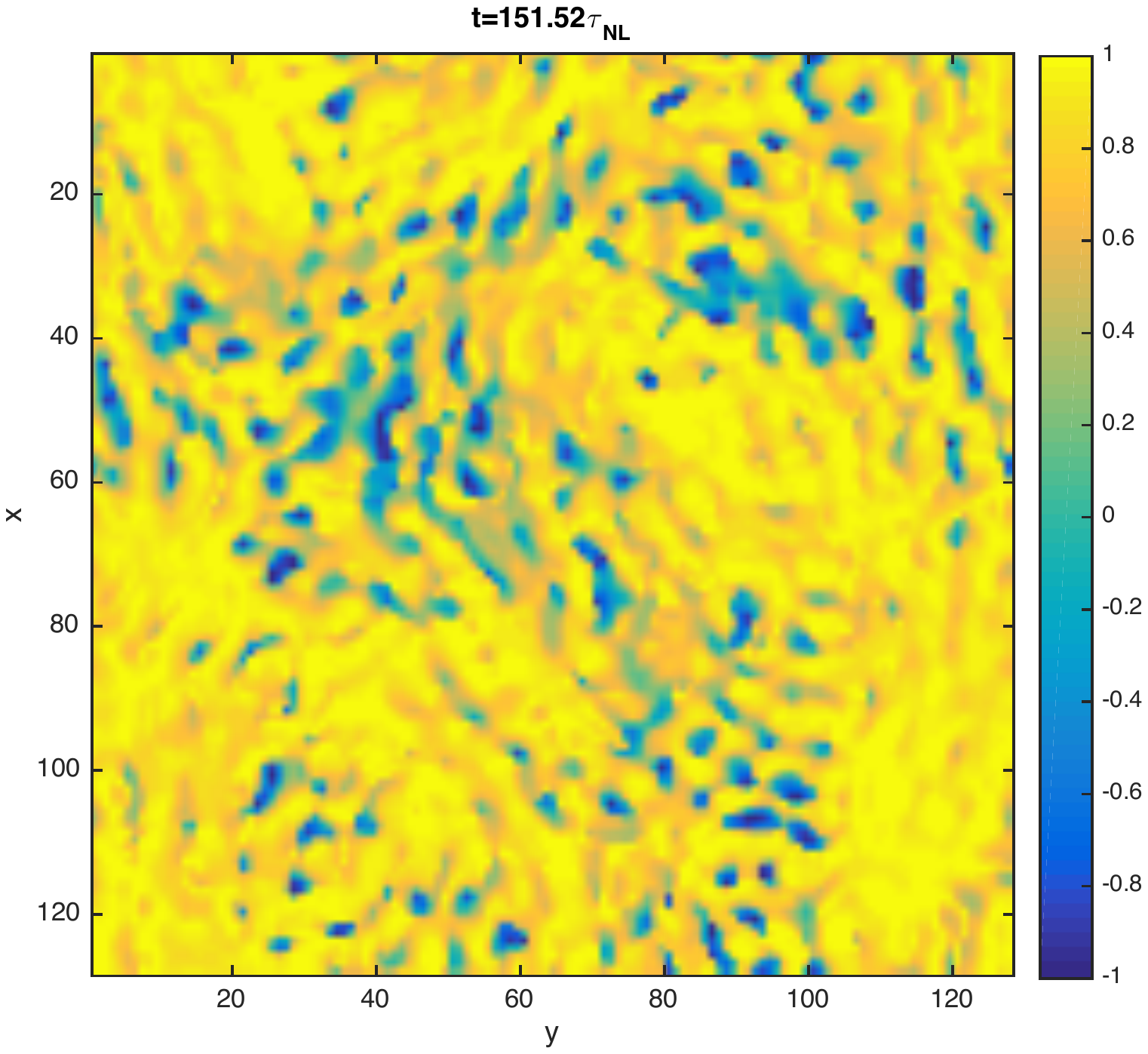}                  
\vskip-1.097truein \caption{Horizontal cut of the point-wise  relative rate of  magnetic helicity $\sigma_M({\bf x})$ at $t=0$ (left) and at $t=150$ (right) for run AH5 of Table 1, with 
$\epsilon_H=0.2, \ \sigma_M=0.65$.  
The signature of the forcing, at $L_F\approx 2\pi/20\approx 0.16$ in units of the size of the box, is visible on both plots, as well as the formation of large-scale structures at long times. 
}\label{f:3}    \end{figure*}

The right-most plots in Fig. \ref{f:1} give the  variations with time of the magnetic (top) and kinetic (bottom) integral scales, defined classically as: 
\be L_{V,M}(t) =\frac{\int [E_{V,M}(k,t)/k]dk}{\int E_{V,M}(k,t)dk} \ . \ee
Note the different magnitudes for $L_V$ and $L_M$ on the vertical axes. As for all other temporal figures, the time is in units of the turn-over time, $\tau_{NL}=L_0/U_0$.  At any given time, the stronger 
$\epsilon_H$, the larger $L_V$ is, and the smaller $L_M$ is, although for all times and all $\epsilon_H$, $L_M$ remains larger than $L_V$. This is again indicative of a lesser efficiency of the inverse cascade of magnetic helicity as the Hall term becomes more preponderant. 
 $L_V$ has a rapid growth, with a rate which is independent of $\epsilon_H$, and it saturates at relatively early times, but at levels (and times) which depend on $\epsilon_H$. On the other hand, $L_M$ grows at rates that differ with $\epsilon_H$ and continues its growth, except for the pure MHD case. It will likely only saturate when $\sigma_M\approx 1$ at $k=k_{min}=1$. Saturation is delayed as $\epsilon_H$ is increased, a signature of the slower growth rate for high $\epsilon_H$. 
 
In Fig. \ref{f:2}, we follow-up with various helical data as a function of time for the runs of Table 1. 
Specifically, we display in the top row the generalized helicity (left), the magnetic helicity which is also an invariant in the ideal case   (middle), and the cross-helicity (right). Their relative rates (see equations (\ref{E:sigma})) are given in the bottom row of Fig. \ref{f:2}.
 All these helical measures  grow, except for $H_C$ in the MHD case.
For $H_M$, the stronger growth is for MHD, and with a saturation that is reached earlier in MHD.
 The cross-correlation $H_C$ grows as well, but with an inversion in the change of rate of growth with $\epsilon_H$: there is no growth in MHD, and the growth rate of $H_C$ increases with $\epsilon_H$, as its role in $H_G$ becomes more important.
 {Another cross-correlation coefficient can be defined, namely $\sigma_C^\prime=H_C/E_T$ \cite{pouquet_86}. Its behavior (not shown) is almost identical to what is displayed here, for both sets of runs in Tables 1 \& 2, and it will thus not be discussed further.
}
As a result, an interesting point may be the following: In MHD, it has never been quite clear whether the cross correlation between velocity and magnetic field cascades to small scales (like the energy), or to large scales, in particular since it is not definite positive; but its physical dimension indicates it should follow the energy itself. In the presence of inverse cascades of helicity, and using Schwarz inequalities, the magnetic energy inevitably follows the magnetic helicity \cite{pouquet_76}, and so does the kinetic energy, entraining now the cross-helicity to large scales, hence its growth. This point deserves further study. We finally note that the resulting polarization $P_m=\sigma_C \sigma_M$ is positive for all the Hall-MHD runs of Table 1, corresponding to left-polarized waves for these flows, with an increase over time from a rather low value $\approx 0.025$  to close to $0.14$. 

The growth of the characteristic scales $L_V$ and $L_M$ is also noticeable when one visualizes the flow, as is done in Fig. \ref{f:3} which displays, at the initial and final time of the AH5 run, the relative rate of magnetic helicity (see also Fig. \ref{f:4} below). The imprint of the forcing scale $\approx 2\pi/20$ is seen in both plots, but at the later time,  larger eddies are also clearly discernible. 

\section{Large-scale dynamics of Hall MHD: Growth rates in inverse cascades and spectral data} \label{SECTION_SPEC}

\begin{figure*} 
\includegraphics[width=7.15cm, height=5.70cm]{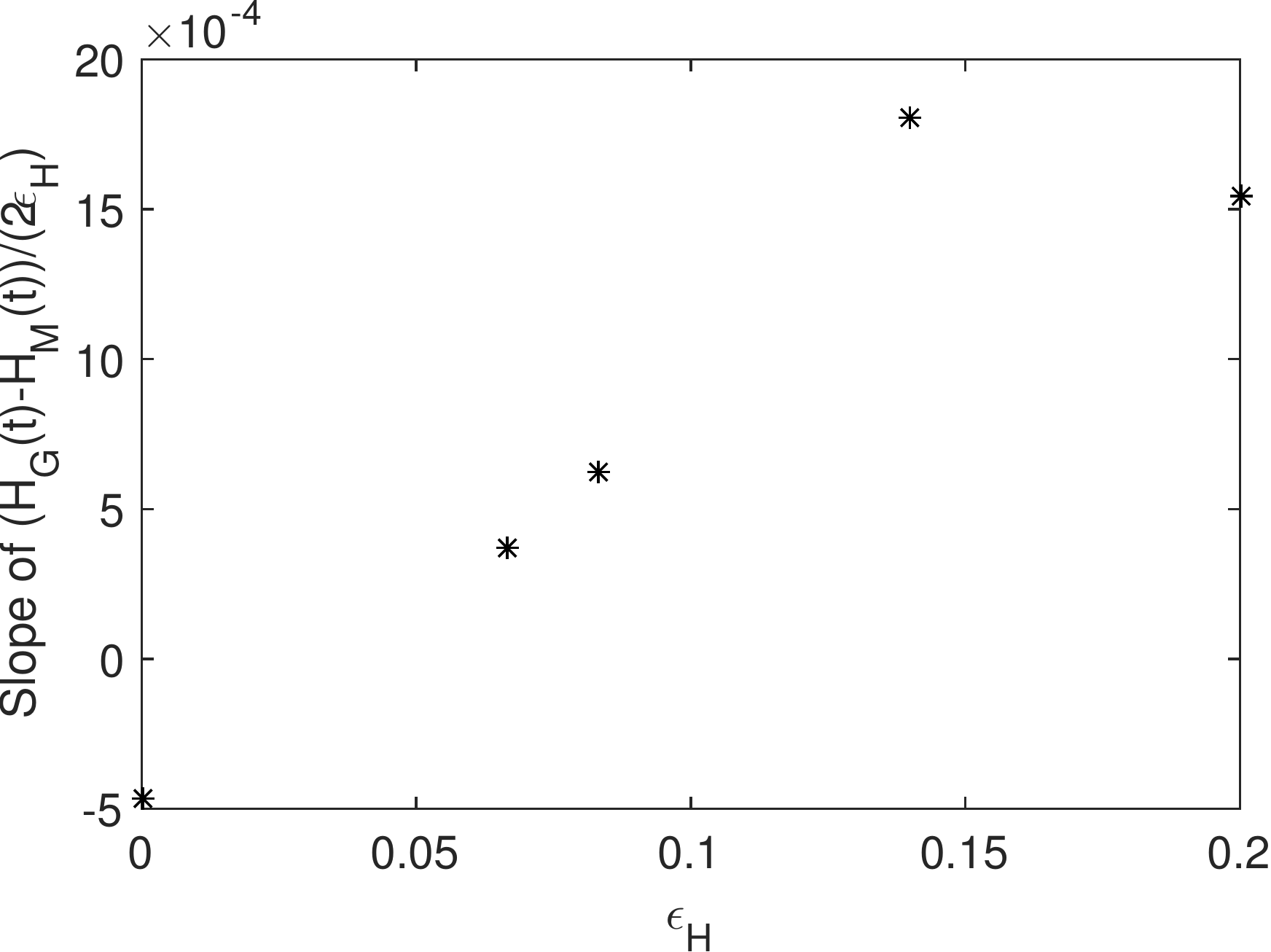}    \hskip0.0750truein      
\includegraphics[width=7.15cm, height=5.70cm]{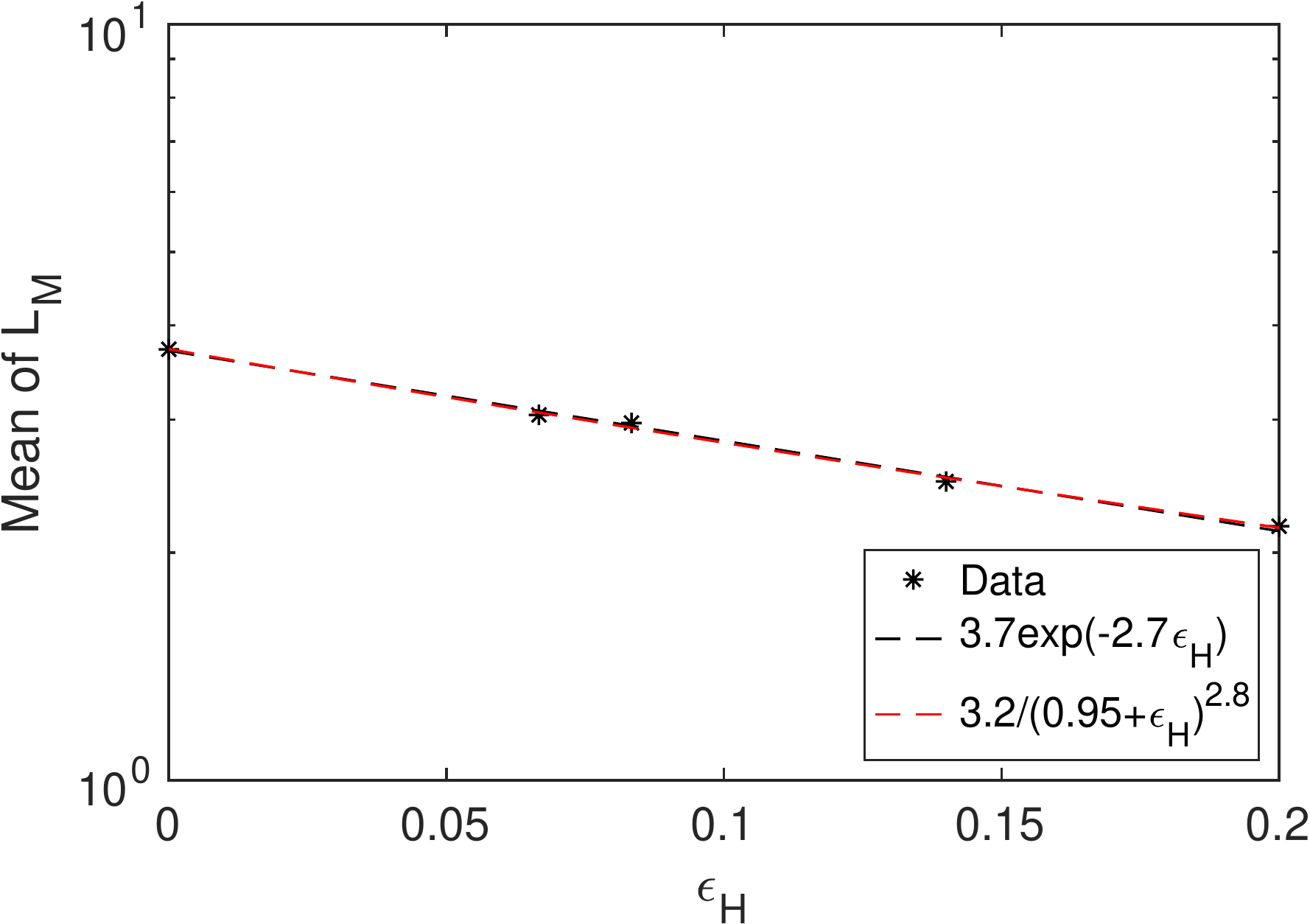}      
 \vskip0.12truein
\includegraphics[width=7.15cm, height=5.70cm]{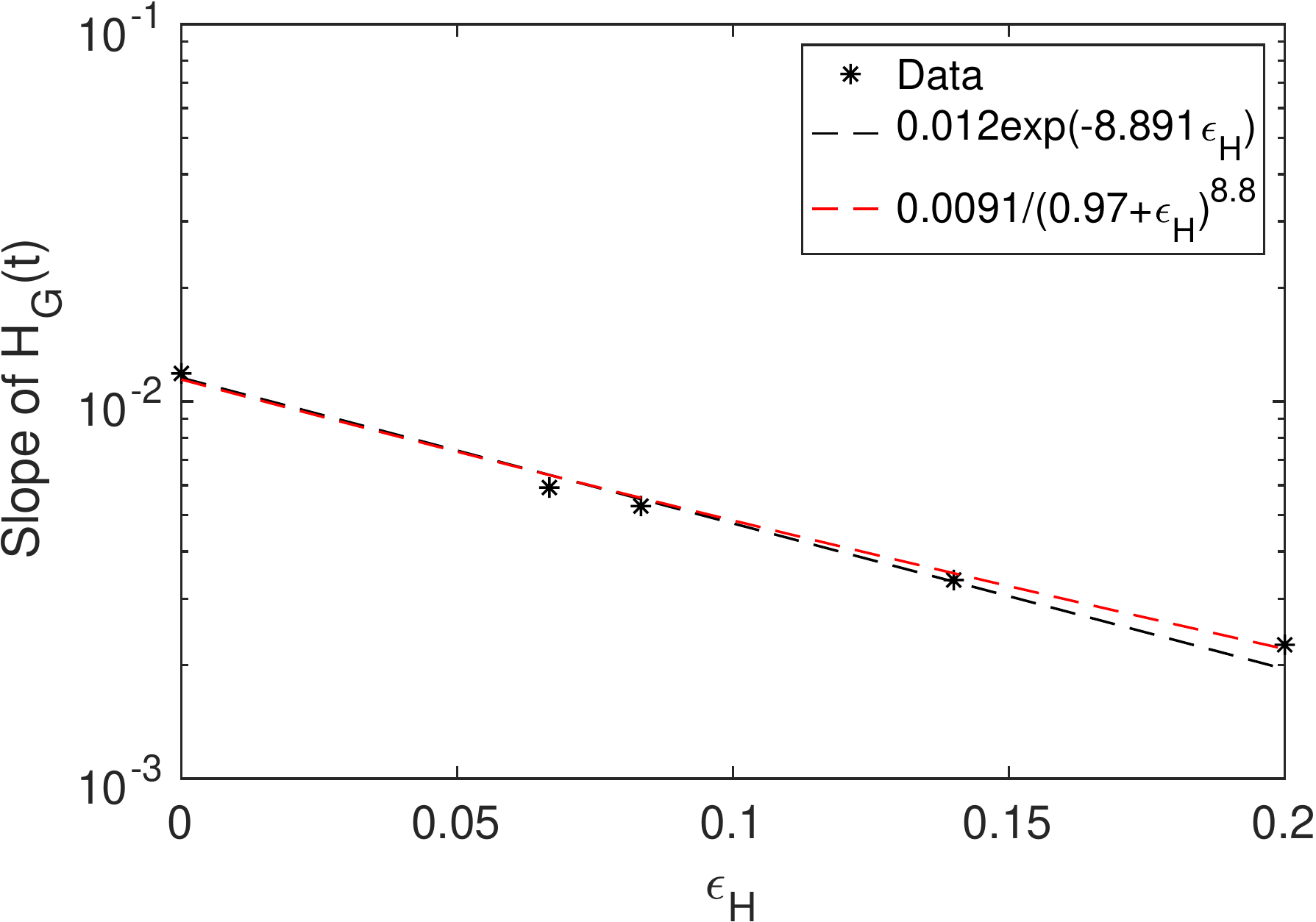}    \hskip0.075truein     
\includegraphics[width=7.15cm, height=5.70cm]{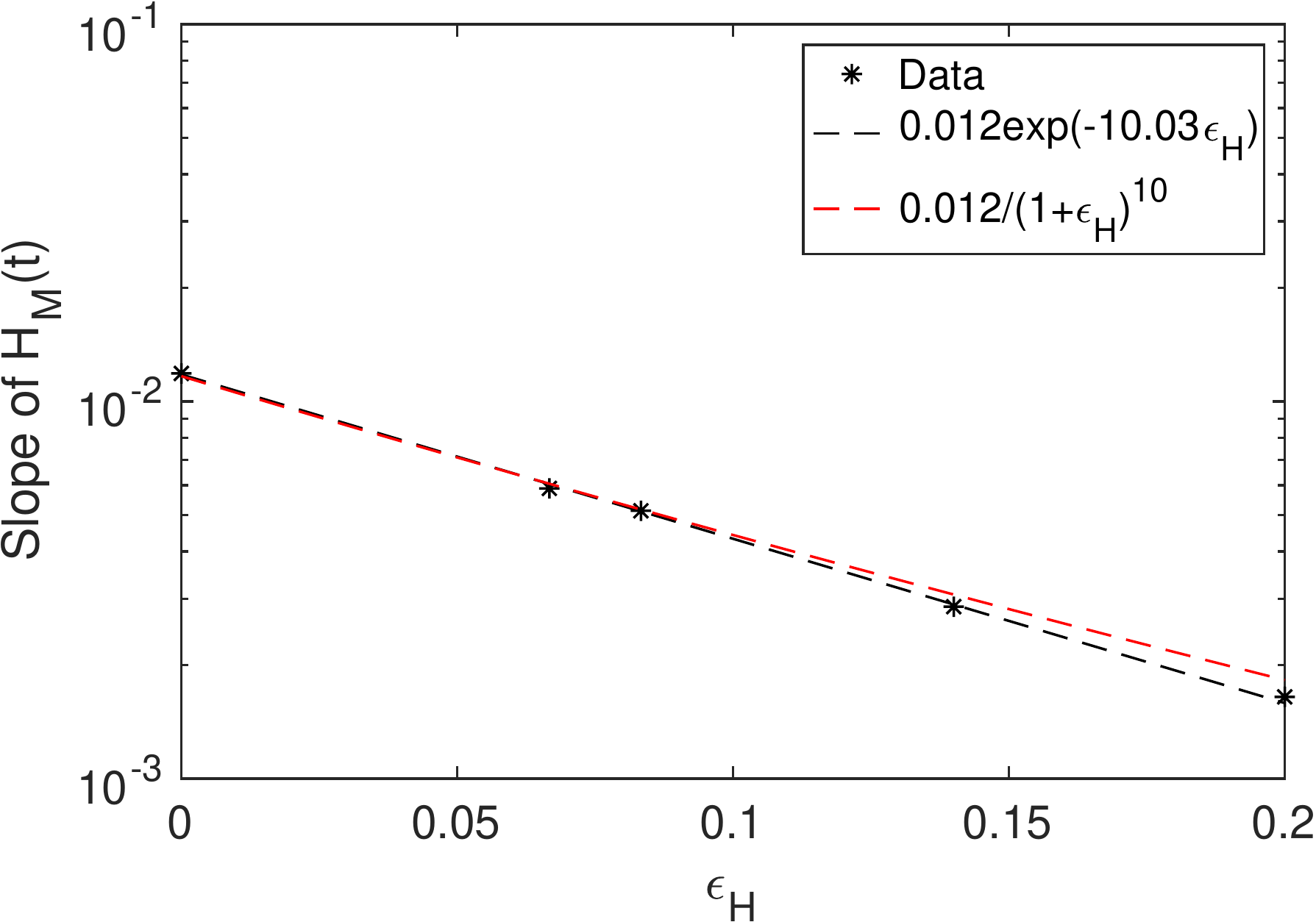}          
\caption{For the runs of Table 1, different scaling laws given as a function of $\epsilon_H$, in lin-log coordinates.
Top left:  temporal growth rate  of $H_C+\epsilon_H H_V/2$ (see equation (\ref{eqhg})).
Top right:  temporal mean of the magnetic integral scale $\left< L_M \right>_t$.
Bottom left:  growth rate of $H_G$.  
Bottom right: : growth rate of  $H_M$. When appropriate, least-square fits are 
{done as indicated with dash lines (see insets). In black is an exponential form $a\ e^{-b\epsilon_H}$, for which a simple argument is given in \S \ref{SECTION:EXP}, and in red, a fit to 
$\alpha/(\beta + \epsilon_H)^\gamma$.}
} \label{f:4}    \end{figure*}

Magnetic helicity is viewed as a large-scale correlation since it involves the magnetic potential; the kinetic helicity, on the other hand, favors the small scales since it involves the vorticity, whereas the cross-correlation is dimensionally comparable to the total  energy. In Hall MHD, as in MHD, $H_M$ controls the dynamics of the large scales, but $H_G$ is hybrid scale-wise since it depends on the ion inertial length. For small $\epsilon_H$, $H_G\approx H_M+ 2\epsilon_H H_C$ and since $H_M$ is invariant separately, so is $H_C$, approximately at least; thus, the inverse cascade of generalized helicity has to be less efficient since the flow dynamics also has to conserve $H_C$, increasingly so as $\epsilon_H$ increases. 
In fact, when $\epsilon_H$ becomes larger than unity, the dominant term in $H_G$ is now the kinetic helicity which, dimensionally, is bound to have a direct cascade, as found in numerous studies of fluid turbulence. Thus, we can expect a complex dynamics of inverse cascades when $\epsilon_H$ is varied.
This leads to a non-monotonic variation of the efficiency of inverse cascades in Hall MHD, as already argued by several authors, and as shown in Fig. \ref{f:4} (top left)  in the variation of the rate of growth of generalized helicity with $\epsilon_H$. All plots here are in lin-log coordinates. The intermediate scales embodied in $H_C$ and the small scales embodied in $H_V$ come into play as a constraint on the small-scale and large-scale dynamics as they become progressively relevant in this generalized helicity invariant. 

As the inverse cascade proceeds, characteristic length scales increase as well, at various rates depending on the strength of the Hall term, as we saw before and as illustrated by the next plot in Fig. \ref{f:4} (top right) giving the variation with $\epsilon_H$ of the temporal mean of the magnetic integral scale. 
We also give in Fig. \ref{f:4} the scaling with the ion inertial length of the growth rate of the generalized helicity (bottom left) and of its magnetic counter part  (bottom right). For this range of $\epsilon_H$ values, these growth rates both have a monotonic variation with comparable factors in the exponential decrease.

{Two fits -- one exponential, using $a\ e^{-b\epsilon_H}$, and one of the rational form 
$\alpha/(\beta + \epsilon_H)^\gamma$ -- are indicated in the plots with respectively black and red dashed lines;
$1/b$ and $\beta$, like $\epsilon_H$, have the physical dimensions of a length scale. The coefficients 
$(a,b),\  (\alpha,\beta,\gamma)$ are given in the insets for each fit. Note that (i) power-law indices $\gamma$ are high for the two rates (between 8.8 and 10.);  (ii) the fits are comparable, and in fact very close for $L_M$; and (iii) $b\approx \gamma, \ \alpha \approx 1$. This latter result, using a Taylor expansion, is not unexpected as long as $\epsilon_H$ remains small. 
However, we  note that the range of values for which such fits are available is not large, preventing a better estimate of these functional forms.
The  expression $\alpha^\prime/(\beta^\prime - \epsilon_H)^{\gamma^\prime}$ was also tried on the data. It does not fit quite as well for the helical rates of growth, but gives an equivalently good fit for $L_M$, but note that this expression is singular (here, for $\epsilon_H\approx 0.79$, not shown). Finally, note 
 that we  give in the next section a phenomenological  argument for the exponential form of the fits, using a simple  model based on the scaling of the helicity spectra. 
 }

\begin{figure*} 
\includegraphics[width=7.0cm, height=5.72cm]{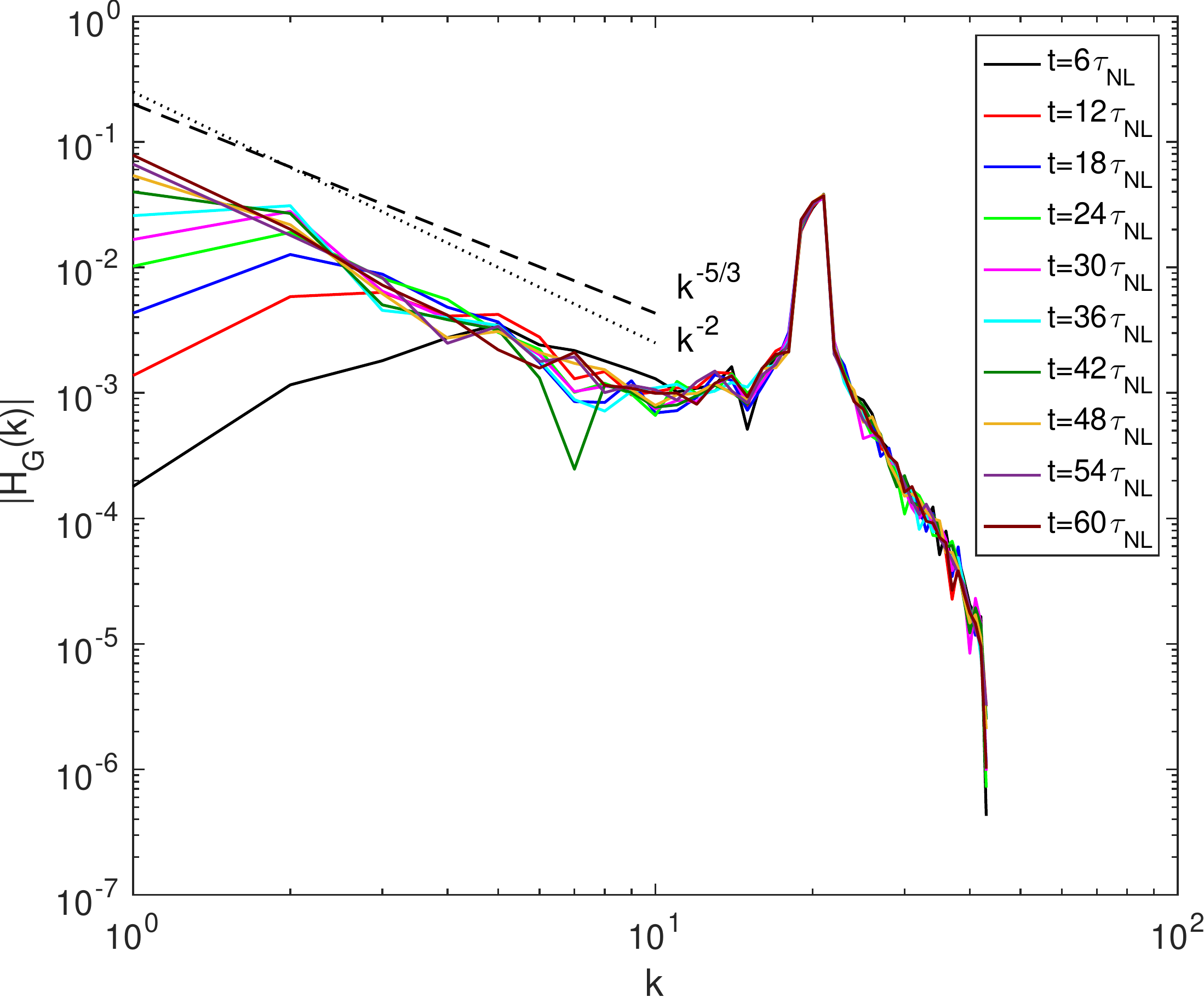}    \hskip0.05truein   
\includegraphics[width=7.0cm, height=5.72cm]{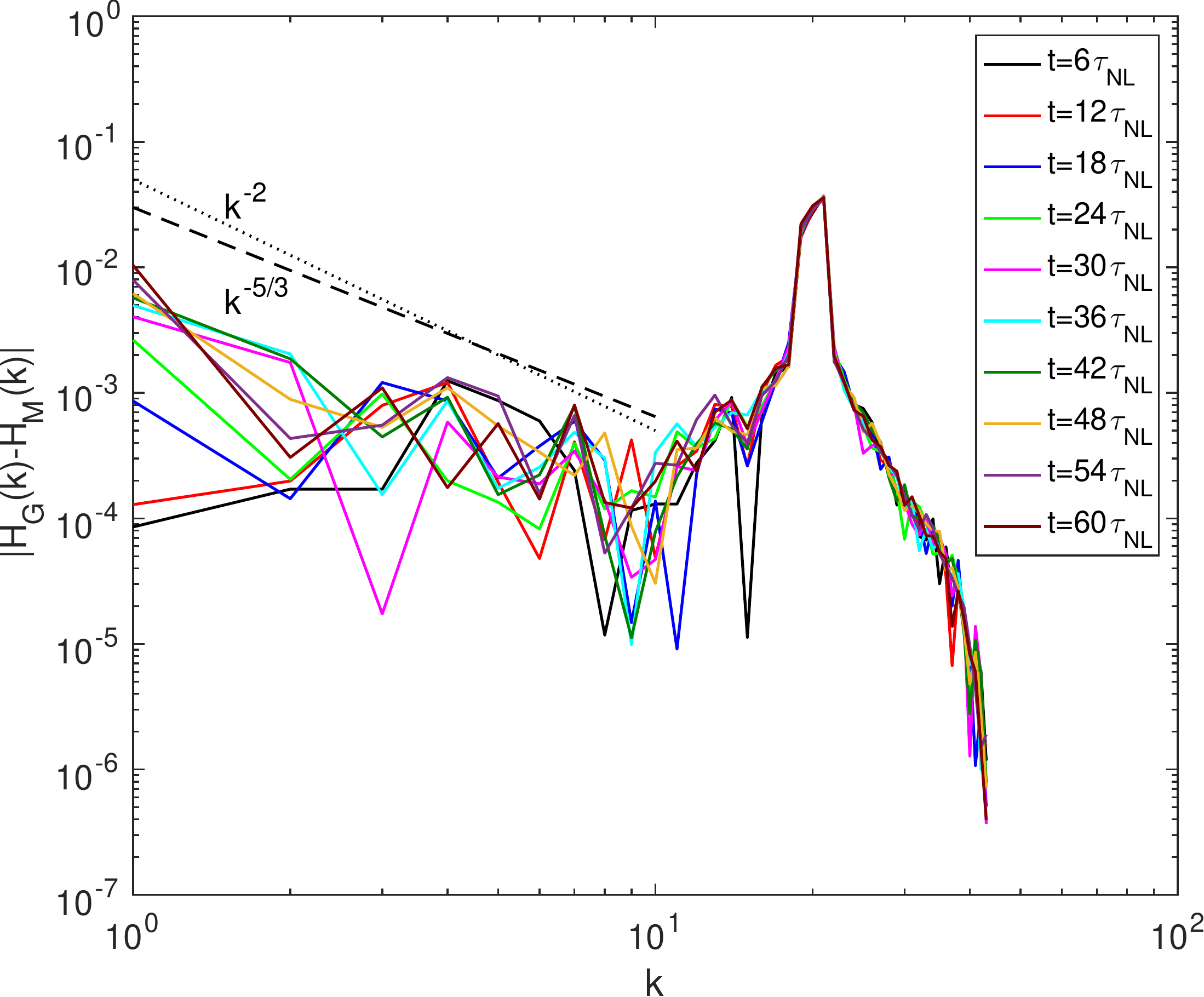}   \vskip-0.025truein  
\includegraphics[width=7.0cm, height=5.72cm]{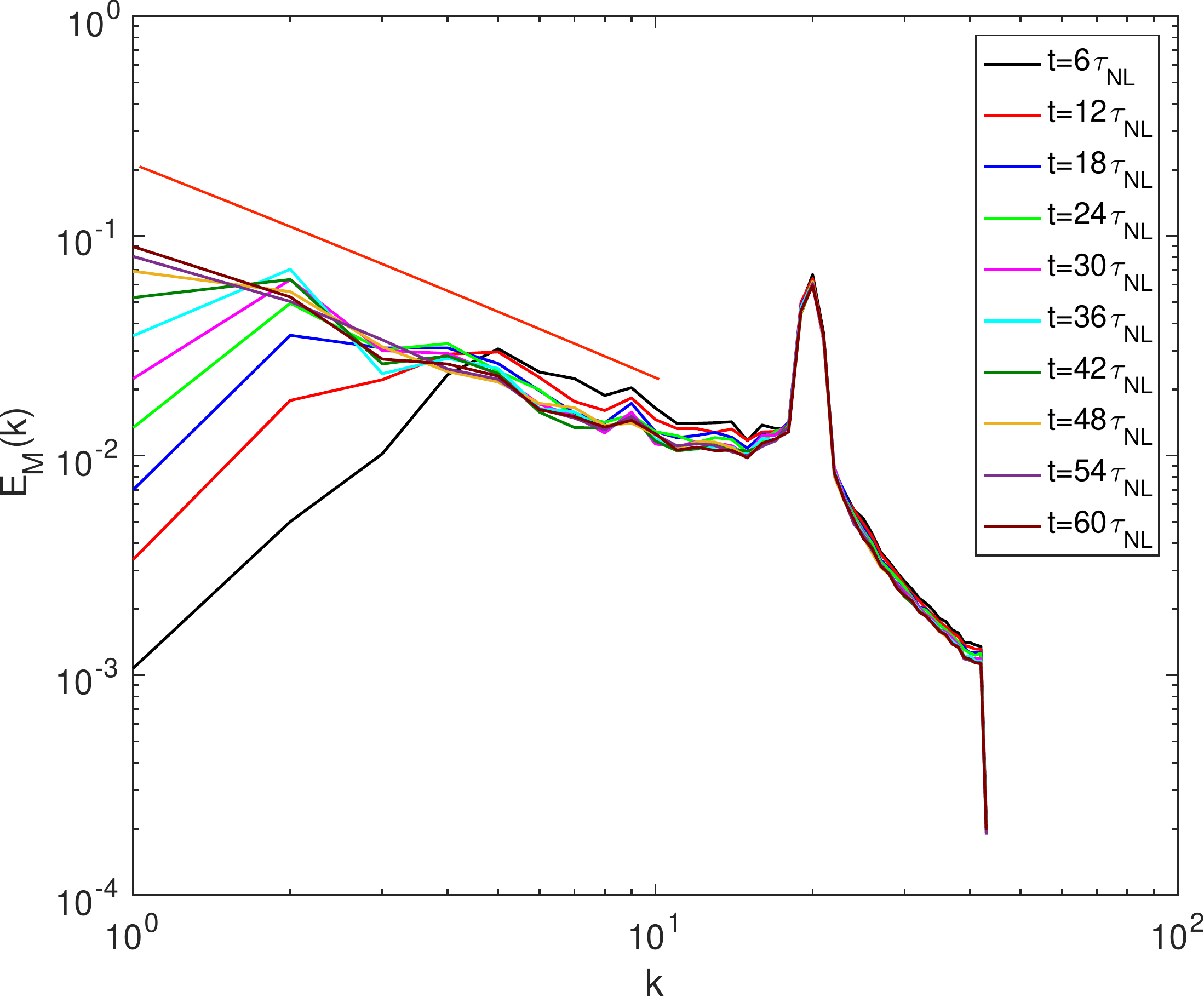}    \hskip0.05truein   
\includegraphics[width=7.0cm, height=5.72cm]{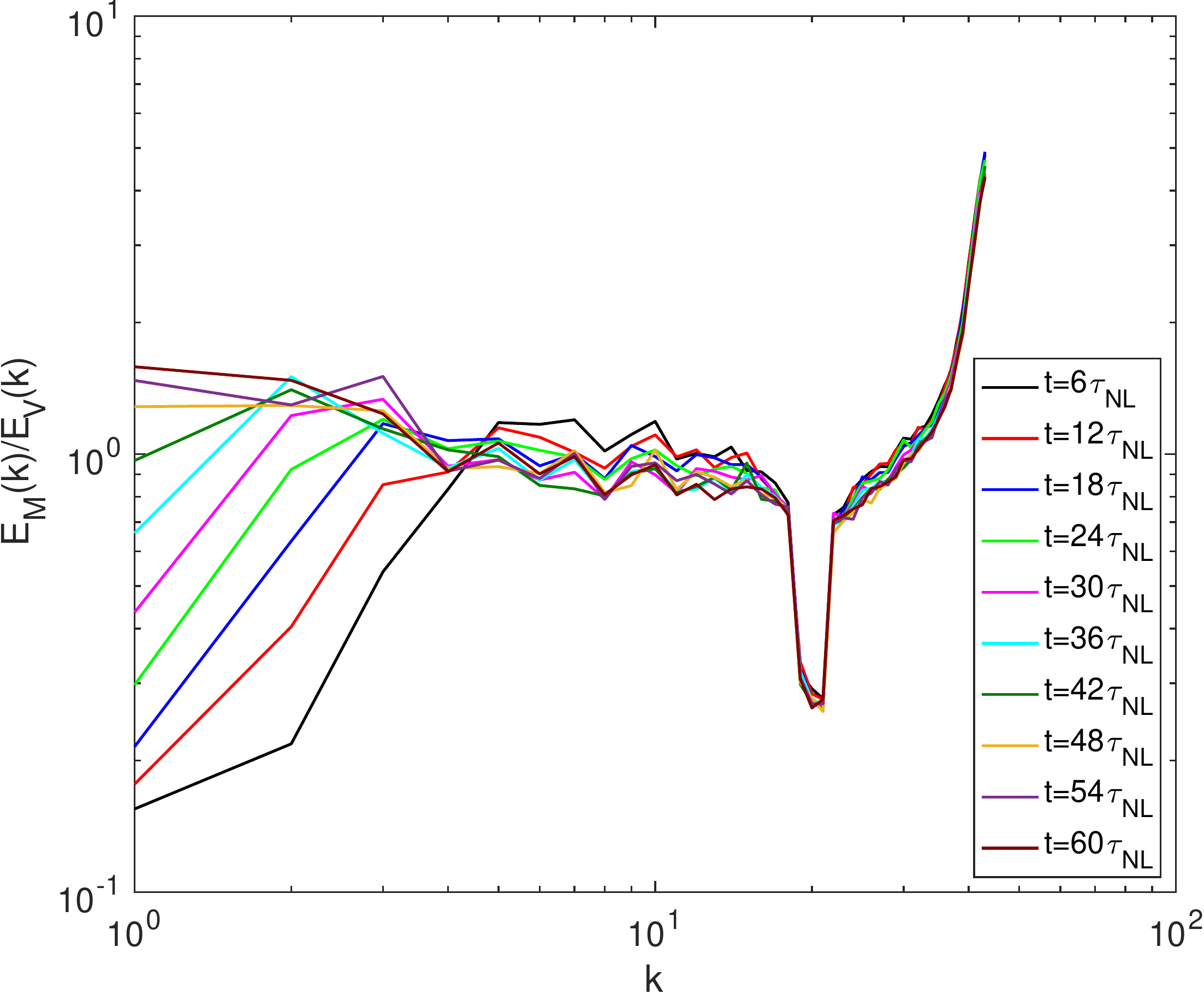}                                 
\caption{Top: Spectra of $H_G(k)$ (left) and  
{$H_G-H_M= \epsilon_H H_X= 2 \epsilon_H H_C + \epsilon_H^2 H_V$ (right).} 
Bottom:  Spectra of $E_M(k)$ (left) and $E_M(k)/E_V(k)$ (right). 
All data is for run AH5 of Table 1 at different times, in units of turn-over times.
Reference power laws are also provided
}\label{f:5}    \end{figure*}

Examining now spectral information, we observe that the build-up with time of the inverse cascades towards larger scales is progressive, with quasi-stationarity at intermediate scales once the inertial-range scaling is reached, as shown in Fig. \ref{f:5} (top left) for the generalized helicity Fourier spectrum for various times for Run AH5 of Table 1 (see insets). The $k^{-2}$ scaling is that predicted  on dimensional grounds for magnetic helicity \cite{pouquet_76} (see also next section); a Kolmogorov $-5/3$ spectrum is  indicated as well, for comparison. The magnetic helicity spectra behave in similar ways (not shown). We also give in Fig. \ref{f:5} (top right), and for the same times,  the spectra for 
{$H_G-H_M= \epsilon_H H_X = \epsilon_H [ 2 H_C +\epsilon_H H_V]$,}
 {\it i.e.} the other formulation of an helical invariant in Hall MHD. A build-up in $H_X$ is visible as well, but with a rather flat spectrum at scales larger than but close to the forcing scale, and with a possible $k^{-2}$ scaling at the largest scales at the latest times.

Because of a Schwarz inequality, namely $E_M(k)\ge kH_M(k)$, the magnetic energy has to follow the magnetic helicity to large scales, as shown in Fig. \ref{f:5} (bottom left). Moreover, we find that $E_M\sim k^{-1}$, a scaling corresponding to a fully helical state ($|\sigma_M| \approx 1$), with stationarity at intermediate scales as the inverse cascade builds up.  
Finally, the magnetic to kinetic  energy ratio shown in Fig. \ref{f:5} (bottom right) is close to an equipartition value in the large scales, as in the case of MHD \cite{pouquet_76}; this large-scale equipartition builds up with time as the inverse cascades of both $H_G$ and $H_M$ proceed.
On the other hand, in the small scales, magnetic energy dominates; however, no inertial range is discernible due to the lack of scale separation between $k_F\approx 20$ and the  wavenumber corresponding to the grid size, $k_{max}\approx 43$. Small-scale dynamics and its possible influence on the large-scale dynamics for a sufficiently large Reynolds number will require a separate study.

\section{
Exponential decrease with Hall parameter of the  growth rate of $H_G$ and $H_M$ in inverse cascades} \label{SECTION:EXP}

As shown in the preceding sections, there is a clear growth of various physical quantities in these runs, and their growth rates vary with the magnitude  of the Hall term.
One striking result of  Fig. \ref{f:4} is that we observe  an exponential decay with $\epsilon_H$ of the growth rates of $H_G$ and $H_M$.

These exponential scaling laws can in fact be recovered through a simple dimensional argument which we now derive. Let us first write the equation for the temporal evolution of the magnetic helicity $H_M$. 
Point-wise, starting from equation (\ref{eq:hall_ind}) in the absence of dissipation and forcing, we have:
\be
\partial_t [{\bf a} \cdot {\bf b}] ({\bf x}) = \partial_t  H_M({\bf x}) = 
{\bf a} \cdot \nabla  \times  [({\bf v} - \epsilon_H {\bf j}) \times {\bf b}]
+ {\bf b} \cdot [\partial_t {\bf a} ({\bf x})] \ .
\ee
First we remark  that both terms in the time derivative of $H_M$ contribute equally upon integration over space, and performing an integration by part; indeed, with the curl operator, there is no change of sign, namely $\int {\bf m} \cdot \nabla \times {\bf n}\ d^3 {\bf x} = + \int {\bf n} \cdot \nabla \times {\bf m}\ d^3 {\bf x}$. So, taking for example the Coulomb gauge, we have $D_t H_M  \equiv {\tilde{\epsilon}_m} =0$, and the temporal evolution of the total magnetic helicity will stem from a competition, and an eventual balance, between dissipation and forcing.

The second step is to recall the scaling of the inverse magnetic helicity cascade, namely \cite{pouquet_76}:
\be H_M(k)\sim  {\tilde{\epsilon}_m}^{2/3} k^{-2} , \label{strong} \ee
with ${\tilde{\epsilon}_m}$ of physical dimension$[L^3][T^{-3}]$, ${\bf b}$ having the dimensions of a velocity. This stems from an analysis under the assumption that the cascade is governed by ${\tilde{\epsilon}}_m$ and the wavenumber $k$, under the assumption of isotropy. This is not an entirely trivial statement, and in fact it has been proven to be irrelevant in at least two instances. On the one hand, in the neutral fluid case, the equivalent scaling based on the injection (and dissipation) rate of kinetic helicity, ${\tilde{\epsilon}_v}\equiv DH_v/Dt$, is 
$H_V(k)\sim {\tilde{\epsilon}_v}^{2/3} k^{-4/3}$ with $E_V(k)\sim   {\tilde{\epsilon}_v}^{2/3} k^{-7/3}$ 
\cite{brissaud_73}. This scaling has never been observed, except possibly in the framework of rotating stratified turbulence as occurs in the atmosphere \cite{baerenzung_11}. The generic turbulence case for fluids leads rather to a passively advected kinetic helicity with $H_V(k) \sim  {\tilde{\epsilon}_v} \epsilon_v^{-1/3} k^{-5/3}$,  where now $\epsilon_v$ is the injection rate of kinetic energy. This scaling results in a relative helicity 
$\sigma_V\sim 1/k$, corresponding to a relatively slow return to full isotropy with scale.

The second instance where the straightforward dimensional argument for the inverse cascade of helicity may be failing in some cases takes place for MHD in three dimensions: it has been shown that other spectra can be observed, differing from the $k^{-2}$ scaling mentioned above, both at small scales and at large scales, namely $H_M(k)\sim k^{-3}$ or steeper \cite{mininni_09, mueller_12}. This  change in the pure inverse cascade scaling may stem from non-local interactions between widely separated scales, 
{which are strong for spectra steeper than $k^{-3}$.}
The reason for the existence of such different solutions from what is advocated in equation (\ref{strong}) remains unknown at this time, although a general but somewhat {\it ad hoc} argument can be given to justify it on the basis of what the prevailing time-scales could be in the dynamical evolution of these systems  \cite{mininni_09, mueller_12}. This point will need further investigations.
 
 The generalized helicity $H_G$ has the same physical dimensions as $H_M$ and thus the same analysis leads  straightforwardly to, with ${\tilde \epsilon}_G = dH_G/Dt$:
\be H_G(k)\sim {\tilde \epsilon}_G^{2/3} k^{-2} \ .  \label{hfk-2} \ee
Note that $ {\tilde \epsilon}_G$ and $ {\tilde \epsilon}_m$ are not independent, since 
$\dot H_G=\dot H_M+ 2\epsilon_H \dot H_C + \epsilon_H^2 \dot H_V$.

The third step in the argument to arrive at an exponential scaling is to write dimensionally, in symbolic terms,  that ${\tilde{\epsilon}_m} \sim (a,v,b)/L_H - \epsilon_H (a,j,b)/L_H$ where $L_H$ is a (constant) characteristic length, and where $(A,B,C)={\bf A} \cdot [{\bf B} \times {\bf C}]$ (together with circular permutations) denotes a vector triple product. Note that this expression is, of course, compatible with the exact law given in equation (\ref{banerjee}). In this simple formulation, taking the derivative with respect to $\epsilon_H$ and using the scaling of the magnetic helicity spectrum given in equation (\ref{strong}),  leads to:   
\be
\frac{D{\tilde{\epsilon}_m}}{D\epsilon_H} \sim - \frac{(a,j,b)}{L_H} \sim - \frac{b^3}{L_H}
 \sim - \frac{ {\tilde{\epsilon}_m}}{L_H},
\label{eureka} \ee
 with the assumption that  the inverse cascade of magnetic helicity is (eventually) fully helical, or $E_M(k)\sim {\tilde{\epsilon}_m}^{2/3}k^{-1}$, thus $E_M\sim kE_M(k) \sim b^2 \sim {\tilde{\epsilon}_m}^{2/3}$, neglecting logarithmic corrections. The data of Fig. \ref{f:2} seems indeed to indicate that $|\sigma_M|$  approaches unity for long times.
From equation (\ref{eureka}), one then immediately obtains,  with ${\tilde{\epsilon}}_{m,0} = {\tilde{\epsilon}}_m (\epsilon_H=0)$ the rate of growth for MHD:  
\be 
{\tilde{\epsilon}_m} / {\tilde{\epsilon}}_{m,0} =  e^{-\epsilon_H/L_H}\ ,
\label{exp} \ee
in agreement with Fig. \ref{f:4}. Similarly, one can write 
\be {\tilde{\epsilon}_G} / {\tilde{\epsilon}}_{G,0} =  e^{-\epsilon_H/L_H} \ee
 for intermediate values of $\epsilon_H$ when the kinetic helicity component of $H_G$ is still negligible.
Note that these exponential behaviors all depend crucially on the scaling relationships of the magnetic and generalized helicity spectra, and on the fact that such spectra converge and thus one can express these fields locally in scale. Specifically,  magnetic helicity spectra steeper than $k^{-3}$, as sometimes observed in MHD
 \cite{mininni_09, mueller_12, mueller_13}  and as mentioned above, would not allow for this exponential behavior.
 
 What is $L_H$ in the above expressions? It is likely proportional to $L_F$, the scale at which kinetic and magnetic energy and magnetic helicity are being injected, and the only fixed large-scale of the flow, except for $k_{min}=1$; $L_F$ is also the smallest scale in the inertial ranges of the inverse cascades. The empirical fit to the data (see Fig. \ref{f:4})  indicates $L_H\approx 0.1$, whereas $L_F =2\pi/ k_F \approx 0.3$. 
We note that the numerical simulations analyzed herein are performed at a constant and rather low Reynolds number, since it is well-known that the inverse cascade can develop for Reynolds number of order unity, providing the necessary nonlinearity at least at the forcing scale, and at larger scales of course. However, this supposes locality of nonlinear interactions, and this may not hold in Hall MHD since, in that case,  there are interactions between small scales and large scales \cite{mininni_07}. It also supposes that the invariant cascading to larger scales does not include smaller-scale features, which is not a correct assumption for $H_G$ as we noted before, since it involves, for higher value of $\epsilon_H$, the kinetic helicity.
These points will thus need further studies. Another remark is that the assumption of maximal helicity may be too strong for the present case (see Fig. \ref{f:2}).

The temporal mean of the integral scale based on the magnetic energy spectrum, $L_M$, on the other hand, displays a different, but still exponential, scaling. Taken over a long time after the initial growth phase, it decreases with $\epsilon_H$ (top right plot in Fig. \ref{f:4}).
It can be seen as a consequence of the lesser efficiency of the inverse cascade of magnetic helicity as 
$\epsilon_H$ increases.
A simple argument for this scaling goes as follows. One can show that, in the inverse cascade of magnetic helicity, the wavenumber $k(t_i)$ reached at a given time $t_i$ is found to be proportional to  \cite{pouquet_76}:
\be  k(t_i) \sim [1/\tilde \epsilon_M^{1/3}]\ t_i^{-1} \ . \label{eqT1} \ee
Replacing ${\tilde \epsilon}_m$ by its expression in terms of the Hall parameter $\epsilon_H$, one can conclude that the largest scale in the system (for $k_{min}=1$) in the Hall-MHD inverse cascade of magnetic helicity $H_M$, is reached at a time varying with $\epsilon_H$ as
\be T_{k_{min}=1}\sim e^{\ +\epsilon_H/[3L_H]} \ . \label{eqT2} \ee
Thus,  the stronger the Hall term, the longer it takes to reach the size of the box, or any scale in the inverse cascade for that matter. It follows that a temporal average of the magnetic integral scale will also decay with 
$\epsilon_H$, but with a third the rate of the decrease of magnetic helicity (with possibly a logarithmic correction coming from the magnetic energy). This is consistent with what is observed in Fig. \ref{f:4} 
{for both functional fits.}
Of course, as the excitation reaches the size of the box, the formation of large-scale coherent structures takes place. Their presence and further temporal dynamics may alter the scaling just derived, as shown recently  for example in the case of two-dimensional fluids \cite{frishman_18}. This could interfere as well with the inverse cascade scaling at late times.

Finally, we also note that we observe such an exponential variation with $\epsilon_H$ for the growth rate of 
$\left< a^2 \right>$, with an exponent of $\approx -9.015$ (not shown), 
and of the temporal rate of growth of the kinetic integral scale $L_V$ (not shown).

\section{Variation of the forcing wavenumber} \label{SECTION:KF}

\begin{table} \caption{
Same as  Table 1  with  forcing for $7\le k_f\approx \le 9$. In  runs AH2f--AH4f, the ion inertial scale is smaller than the forcing scale, contrary to runs of Table 1. The fit presented in Fig. \ref{f:6} (right) is done for runs AH5f to AH9f.
}    \centering \begin{tabular}{cccccccccccc} \toprule
\textbf{ID}	& \textbf{$N_p$} & \textbf{$\nu$} & \textbf{$\epsilon_H$}  & \textbf{$\sigma_M$}  & \textbf{$\sigma_V$} & \textbf{$\sigma_C$} & \textbf{$\sigma_G$} & \textbf{$Re$} & $k_{d_{i}}$ 
\\ \hline 
AM1f&  $48^3$  & 0.016 &  0.0      & 0.11 &  0.20 & 0.15 & 0.64 & 34.8 &  --   \\
AH2f & $48^3$   & 0.016 &  0.0667    & 0.11 &   0.24 & 0.10 & 0.52 & 34.7 &  15.   \\
AH3f  & $48^3$  & 0.016 &  0.0833    & 0.11 &   0.24 & 0.10 & 0.48  & 34.7 &  12.   \\
AH4f  & $48^3$  & 0.016 &  0.14    & 0.11 &  0.24 & 0.10 & 0.36  & 34.7 &  7.2   \\
\hline
AH5f  & $48^3$  & 0.016 &  0.20    & 0.11 &   0.24 & 0.10 & 0.27  & 34.7 &  5. \\
AH6f  & $48^3$  & 0.016 &  0.25    & 0.11 &   0.20 & 0.15 & 0.23  & 34.8 &  4.   \\
AH7f  & $48^3$  & 0.016 &  0.30    & 0.11 &   0.22 & 0.19 & 0.21  & 34.9 & 3.3   \\
AH8f  & $48^3$  & 0.016 &  0.45    & 0.11 &   0.20 & 0.15 & 0.15  & 34.8 &  2.2   \\
AH9f  & $48^3$  & 0.016 &  0.60 & 0.11 &   0.22 & 0.19 & 0.15  & 34.9 &  1.7 \\
\hline
AH10f  & $48^3$& 0.016 &  0.90  & 0.11 &   0.22 & 0.19 & 0.13  & 34.9 & 1.1  \\
AH11f  & $48^3$& 0.016 &  1.2  & 0.11 &   0.22 & 0.19 & 0.12  & 34.9 & 0.8  \\
\hline  \bottomrule \end{tabular}   \end{table} 

\begin{figure*} 
 \vskip-0.92truein   \hskip-0.2truein
\includegraphics[width=4.95cm, height=11.9cm]{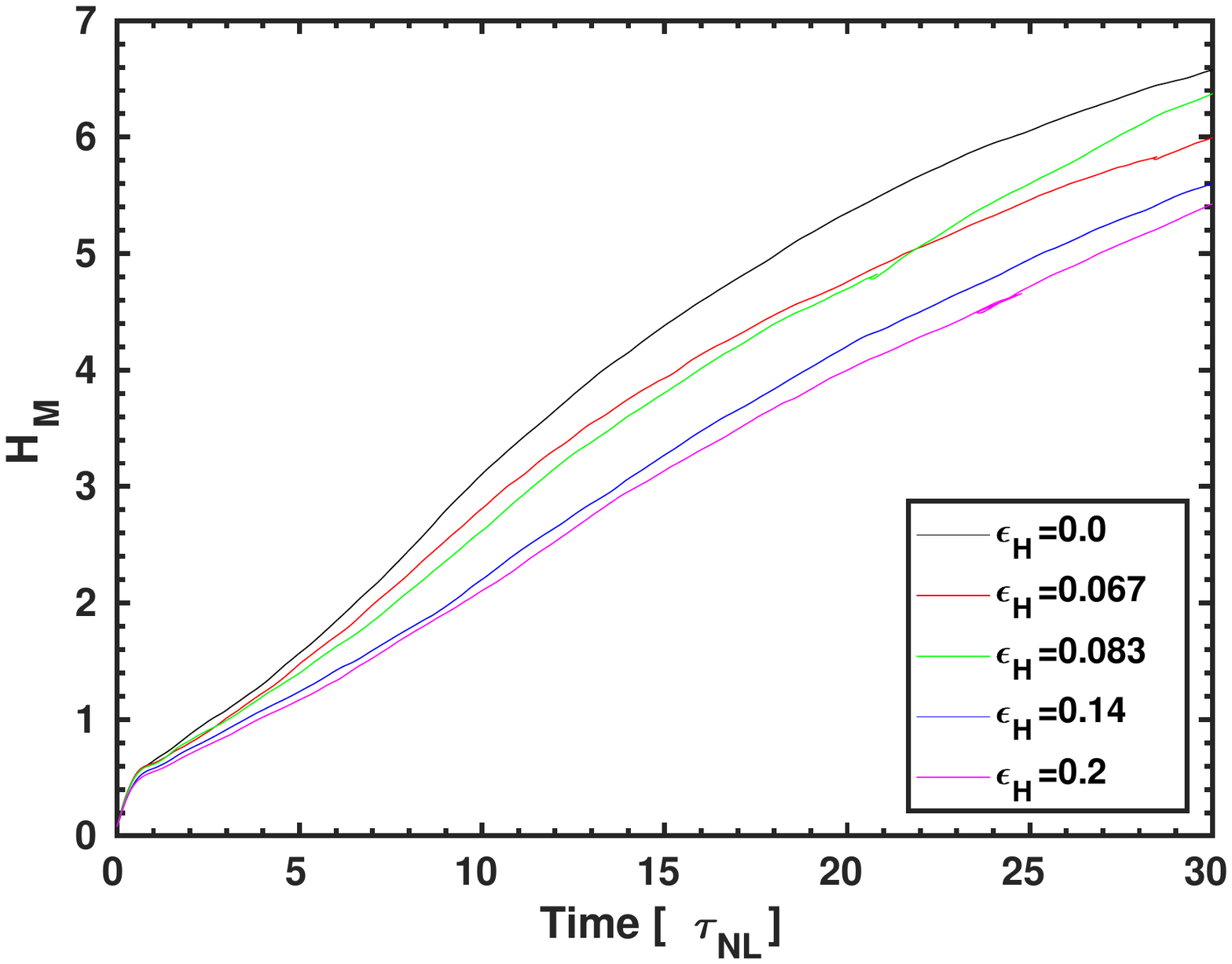}  \hskip-0.1truein    
\includegraphics[width=4.95cm, height=11.9cm]{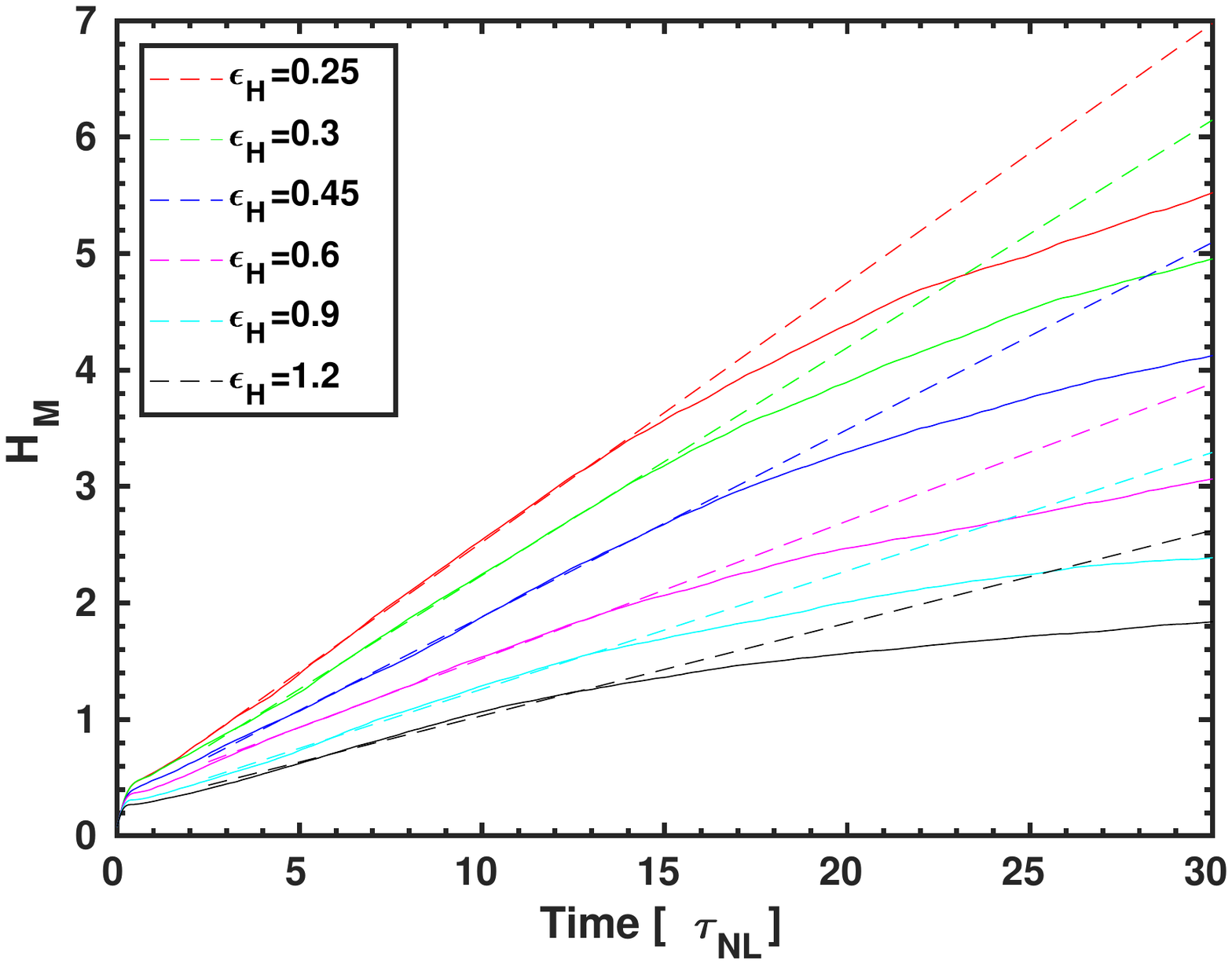}  \hskip-0.1truein.    
\includegraphics[width=5.95cm, height=11.9cm]{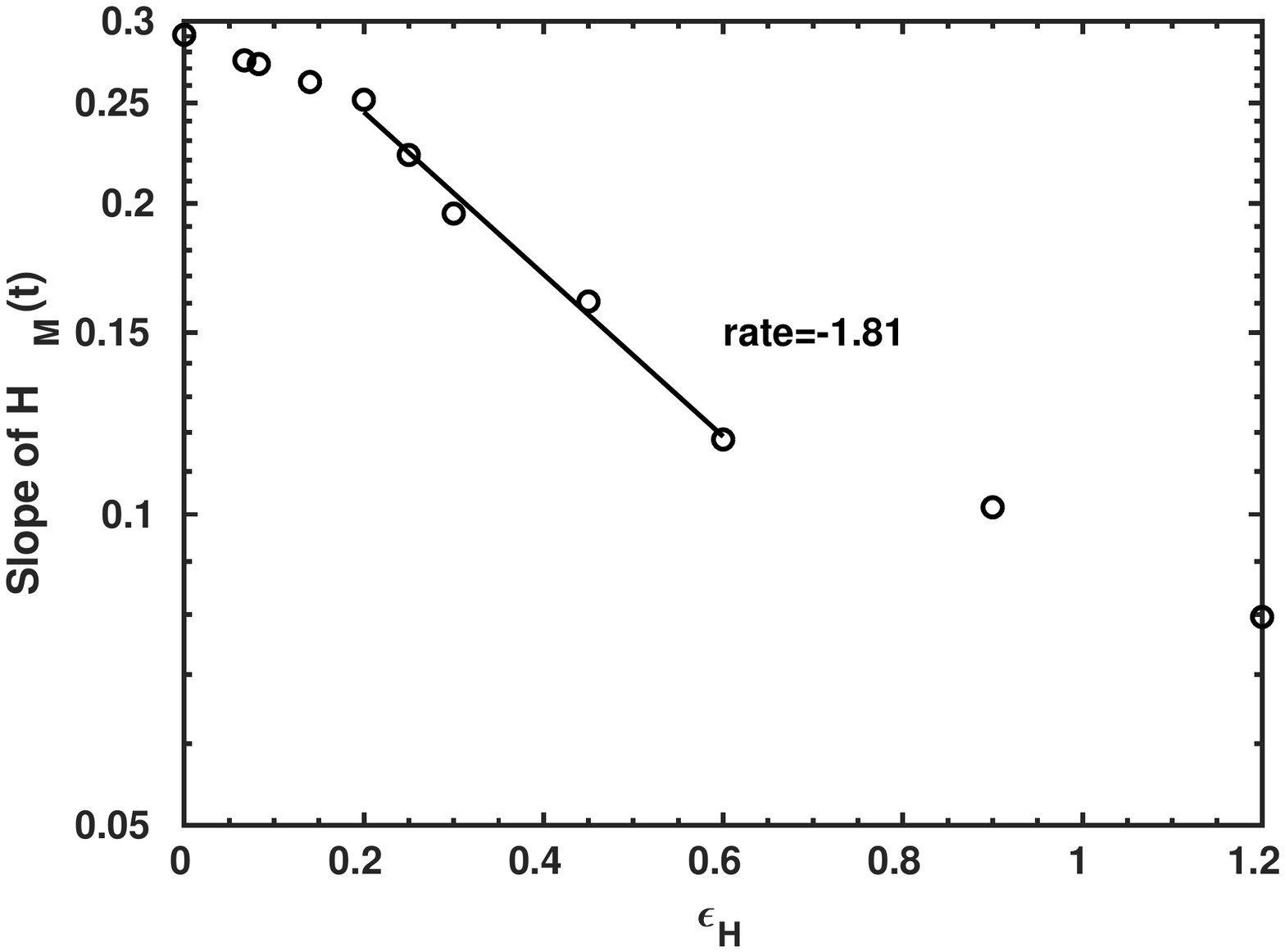}   
  \vskip-0.912truein     
\caption{Left: Total magnetic helicity  as a function of time for a subset of the runs of Table 2 with $k_F\approx 8$ and $\epsilon_H \le 0.2$.
Middle: The same with values of $\epsilon_H$ extended to ${\cal O}(1)$ (see insets); dotted lines indicate temporal fits.
Right: Variation of the rate of growth of $H_M$ for the same runs, in lin-log coordinates, with a fit in the intermediate range of values of $\epsilon_H$. 
Note the three regimes, with an exponential decay at intermediate values of $\epsilon_H$. 
}\label{f:6}    \end{figure*}

We performed a second series of runs but  now with $ 7\le k_F\le 9$ (see Table 2). The runs are computed on grids of $48^3$ points so as to preserve, comparing with the runs of Table 1, the same resolution of the small-scale dynamics. In that case, for  runs with 
$\epsilon_H < 0.2$, the ion inertial length scale is smaller than the forcing scale and, as expected because of the locality of nonlinear interactions in the inverse cascade, all runs see a similar growth rate, independent of $\epsilon_H$ and corresponding roughly to that of MHD (see Fig. \ref{f:6}, left). We also note that, for longer times,  the saturation level of $H_M$ does depend on 
$\epsilon_H$, and is lower the larger $\epsilon_H$, as expected from the arguments developed in the preceding section (see also \cite{turner_86} where it is argued that the relaxed state for long times need not be force-free in Hall MHD). 

When  extending these runs to higher values of $\epsilon_H$, the ion inertial length is now again in the inverse cascade range and the growth rate of magnetic helicity is clearly smaller for higher 
$\epsilon_H$ (Fig. \ref{f:6}, middle). The variation of the growth rate of  magnetic helicity with $\epsilon_H$ for all runs of Table 2 is given in Fig. \ref{f:6} (right). 
The resulting scaling is again an exponential decrease which, when taking intermediate values, has a 
$-1.81$ exponent, with a saturation at both ends of the spectrum of $\epsilon_H$ values (when including all values of $\epsilon_H$, the exponent is $-1.17$, not shown).

We do  observe qualitatively that for a larger forcing scale, the decay has a smaller exponent, as argued in \S \ref{SECTION:EXP},  but a quantitative agreement is clearly lacking: the scaling  for the runs of Table 1 is almost five times larger than for the runs of Table 2, although the ratio in forcing scales is only a factor of 3 between the two sets of runs. Several elements could explain this discrepancy, given the fact that we argue in the preceding section that the length appearing in the scaling exponent is that of the forcing. At high values of $\epsilon_H$, the difference  is probably due to the fact that for $\epsilon_H \ge 1$, the Hall-MHD range is not fully resolved since, in that case, $L_0=2\pi < \epsilon_H$. Moreover,  the effect of small scales in the ideal conservation laws, for $H_G$ in particular, is felt through the contribution to its evaluation of both $H_C$ and $H_V$, but nonlinear interactions at small scales are barely present in the runs of Tables 1 and 2. Indeed, another intervening factor may well be the lack of resolution of the direct inertial range in a problem in which, as $\epsilon_H$ increases, the small scales  play a more prominent role in the inverse cascade through the invariance of $H_G$, a problem not present in pure MHD flows. 
{Yet another factor may be the amount of cross helicity present in the flow: completely negligible for the runs of Table 1 (with $\sigma_C\approx 0.03$), it is more significant for the runs of Table 2 (with $0.1 \le \sigma_C \le 0.2$). As analyzed in \cite{miloshevich_17} on the basis of statistical equilibria for extended MHD, the amount of cross-correlation between the velocity and the magnetic field  may have a measurable effect on the strength of the inverse cascades.} These issues are left for future work.

\section{Discussion and conclusion}

In the solar wind, the regime of Hall MHD arises at small scales, starting at the ion inertial length. It has been studied thoroughly in the context of the change to small-scale dynamics, reconnection and dissipative processes due to the presence of dispersive plasma waves. It leads to a steepening of the energy spectra in the direct cascade, and to strong small-scale structures, all phenomena observed in the solar wind, and more recently in the magnetosheath \cite{alexandrova_08, huang_17, chasapis_18, phan_18, stawarz_19, bandyopadhyay_19}. 
In this paper, we are concerned with the occurrence within such a system of large-scale phenomena due to inverse cascades which are known to exist thanks to pioneering studies of idealized Hall-MHD  \cite{servidio_08}. Such inverse cascades can also affect small-scale dynamics because of the strong non-locality of global nonlinear transfer \cite{alexakis_06}, even if the nonlinear interactions within the inverse cascades are local. 

We show that, as a function of the ion inertial length, there is an exponential decrease of the rate of growth of magnetic and generalized helicity, $H_M$ and $H_G$, as the controlling parameter for Hall MHD is increased. Moreover, this phenomenon is explained through a simple dimensional argument that relies on the scaling of the magnetic and generalized helicity spectra. Exponential scaling can also be found, in simulations of reduced MHD  turbulence, for the fraction of (global) energy dissipation, in terms of the vorticity and current (or equivalently in terms of the curl of the Elsass\"er variables $\vomega_\pm=\vomega \pm {\bf j}$), when expressed as a function of the fraction of volume occupied by dissipative structures \cite{zhdankin_16}. 

Indeed, the subsequent energy and helicity input towards large scales can in turn affect the complex small-scale dynamics and the ensuing energy dissipation. In particular, it was stated in \cite{servidio_08} that the inverse cascade in Hall MHD is weaker than in the MHD case, a result confirmed by the present analysis at least for positive polarity, $P_M>0$. This can be related to the fact that, in Hall MHD, the magnetic field is not so efficient at creating a large-scale force-free structure, with a resulting 
$\sigma_M\approx 1$. 
Furthermore, it was shown in \cite{mininni_13} for the problem of two-dimensional Navier-Stokes turbulence, that inverse transfer is effective even when no forcing is acting on the flow.  This is due to the fact that, since invariants are quadratic, one has detailed balance, {\it i.e.} conservation of the invariants for each individual set of triadic interactions; as such, this represents a huge constraint on the resulting nonlinear dynamics. Hence the magnitude of inverse transfer in Hall-MHD, which depends on $\epsilon_H$, is bound to affect the dissipative structures at small scales.

The  correlation between the velocity and the magnetic field grows as well, in both absolute and relative terms.
It is not an invariant except  in the limit $\epsilon_H\rightarrow 0$, when $H_X$ reduces to $H_C$ (see equation (3)). In MHD, it has been known for a long time that $H_C$ affects the amount of dissipation present in the fluid \cite{pouquet_88}, so it may be the case as well here.
Furthermore, an intriguing possibility is whether or not one obtains, for some values of the controlling parameter at a given Reynolds number, a dual, bi-directional cross-helicity cascade, as already observed for the total energy in the atmosphere in the presence of both rotation and stratification \cite{pouquet_13p, marino_15p}. 
Such two-signed constant fluxes have been found as well in oceanic data \cite{scott_05} and 
in numerical models of the atmosphere \cite{skamarock_14}. Similarly,  bi-directional cascades were analyzed in the case of  MHD turbulence both in two dimensions and in three dimensions (see the reviews in \cite{alexakis_18, pouquet_19e} and references therein). Further study of the role of $H_C$ and of the Reynolds number in the dynamics of Hall MHD is reserved for future work.
{
Theories of wave turbulence (or closures in the strongly nonlinear case) will be useful to achieve higher Reynolds numbers with substantial scale separation in order to unravel the different phenomena at play. These could also give access to formulations of transport coefficients, such as eddy viscosity and eddy noise for these complex problems, and see how they depend on the control parameters such as $\epsilon_H$ and the relative helicities. We note that, recently, a model for low ratios of magnetic to electron pressure has also detected the possibility of an inverse cascade of (generalized) cross-helicity  in the context of kinetic Alfv\'en wave interactions \cite{passot_18,miloshevich_20} (see also \cite{schekochihin_09,cho_11}). 
}

It would also be of interest to investigate the dynamics of inverse cascades for left-circular polarized waves, with $P_M>0$, in which case the magnetic energy may become more prominent. It is known that the whistler waves have a stronger effect than the ion-cyclotron waves on transport coefficients and in particular on the effective diffusivity, which can in fact become negative \cite{mininni_07}. 
Similarly, it was shown in \cite{ji_99b} that the plasma $\beta$ ({\it i.e.}, the ratio of thermal to  magnetic pressure) can affect the interactions between large and small scales and thus the inverse cascades in magneto-fluids and space plasmas. In particular, it can make them less efficient in the presence of a strong Hall current, as found here for $P_M<0$. 
One could also look at these questions from the slightly less-demanding problem, from a numerical stand-point, of electron MHD (or EMHD \cite{galtier_03b, cho_11, zhuJZ_14, pouquet_19e}), in which one only deals with the evolution of the magnetic induction. EMHD is the limit of Hall MHD that obtains for small velocities and large ion inertial scales, and is known to have an inverse cascade of magnetic helicity  \cite{cho_11, kim_15}.
For example, is the cascade in fact bi-directional? Is there more reconnection as well, due to non-local effects between large and small scales?  These points are left for future work.

\begin{acknowledgments}   {\tiny }  
The runs analyzed in this paper have used an open allocation on the Janus super-computer at LASP/CU, which is gratefully acknowledged, together with time on a local cluster. 
{We thank reviewers for useful remarks.}
NCAR  is supported by the National Science Foundation. Support for AP, from LASP and in particular from Bob Ergun, is gratefully acknowledged as well. 
JES is supported by STFC(UK) grant  ST/S000364/1.       
 \end{acknowledgments}  
\bibliography{ap_20_jan20}  

\begin{thebibliography}{133}
\expandafter\ifx\csname natexlab\endcsname\relax\def\natexlab#1{#1}\fi
\expandafter\ifx\csname bibnamefont\endcsname\relax
  \def\bibnamefont#1{#1}\fi
\expandafter\ifx\csname bibfnamefont\endcsname\relax
  \def\bibfnamefont#1{#1}\fi
\expandafter\ifx\csname citenamefont\endcsname\relax
  \def\citenamefont#1{#1}\fi
\expandafter\ifx\csname url\endcsname\relax
  \def\url#1{\texttt{#1}}\fi
\expandafter\ifx\csname urlprefix\endcsname\relax\def\urlprefix{URL }\fi
\providecommand{\bibinfo}[2]{#2}
\providecommand{\eprint}[2][]{\url{#2}}

\bibitem[{\citenamefont{Newell et~al.}(2001)\citenamefont{Newell, Nazarenko,
  and Biven}}]{newell_01}
\bibinfo{author}{\bibfnamefont{A.}~\bibnamefont{Newell}},
  \bibinfo{author}{\bibfnamefont{S.}~\bibnamefont{Nazarenko}},
  \bibnamefont{and} \bibinfo{author}{\bibfnamefont{L.}~\bibnamefont{Biven}},
  \bibinfo{journal}{Physica D} \textbf{\bibinfo{volume}{152-153}},
  \bibinfo{pages}{520} (\bibinfo{year}{2001}).

\bibitem[{\citenamefont{Sagaut and Cambon}(2008)}]{sagaut_08}
\bibinfo{author}{\bibfnamefont{P.}~\bibnamefont{Sagaut}} \bibnamefont{and}
  \bibinfo{author}{\bibfnamefont{C.}~\bibnamefont{Cambon}},
  \emph{\bibinfo{title}{Homogeneous Turbulence Dynamics}}
  (\bibinfo{publisher}{Cambridge University Press, Cambridge},
  \bibinfo{year}{2008}).

\bibitem[{\citenamefont{B\"uhler}(2010)}]{buhler_10}
\bibinfo{author}{\bibfnamefont{O.}~\bibnamefont{B\"uhler}},
  \bibinfo{journal}{Ann. Rev. FLuid Mech.} \textbf{\bibinfo{volume}{42}},
  \bibinfo{pages}{205} (\bibinfo{year}{2010}).

\bibitem[{\citenamefont{Mahrt}(2014)}]{mahrt_14}
\bibinfo{author}{\bibfnamefont{L.}~\bibnamefont{Mahrt}}, \bibinfo{journal}{Ann.
  Rev. Fluid Mech.} \textbf{\bibinfo{volume}{46}}, \bibinfo{pages}{23}
  (\bibinfo{year}{2014}).

\bibitem[{\citenamefont{Pouquet et~al.}(2017)\citenamefont{Pouquet, Marino,
  Mininni, and Rosenberg}}]{pouquet_17p}
\bibinfo{author}{\bibfnamefont{A.}~\bibnamefont{Pouquet}},
  \bibinfo{author}{\bibfnamefont{R.}~\bibnamefont{Marino}},
  \bibinfo{author}{\bibfnamefont{P.~D.} \bibnamefont{Mininni}},
  \bibnamefont{and}
  \bibinfo{author}{\bibfnamefont{D.}~\bibnamefont{Rosenberg}},
  \bibinfo{journal}{Phys. Fluids} \textbf{\bibinfo{volume}{29}}
  (\bibinfo{year}{2017}).

\bibitem[{\citenamefont{Gregg et~al.}(2018)\citenamefont{Gregg, D'{A}saro,
  Riley, and Kunze}}]{gregg_18}
\bibinfo{author}{\bibfnamefont{M.}~\bibnamefont{Gregg}},
  \bibinfo{author}{\bibfnamefont{E.}~\bibnamefont{D'{A}saro}},
  \bibinfo{author}{\bibfnamefont{J.}~\bibnamefont{Riley}}, \bibnamefont{and}
  \bibinfo{author}{\bibfnamefont{E.}~\bibnamefont{Kunze}},
  \bibinfo{journal}{Ann. Rev. Marine Sci.} \textbf{\bibinfo{volume}{10}},
  \bibinfo{pages}{9} (\bibinfo{year}{2018}).

\bibitem[{\citenamefont{Shaw and Oncley}(2001)}]{shaw_01}
\bibinfo{author}{\bibfnamefont{R.}~\bibnamefont{Shaw}} \bibnamefont{and}
  \bibinfo{author}{\bibfnamefont{S.~P.} \bibnamefont{Oncley}},
  \bibinfo{journal}{Atmos. Res.} \textbf{\bibinfo{volume}{59-60}},
  \bibinfo{pages}{77} (\bibinfo{year}{2001}).

\bibitem[{\citenamefont{Lopez et~al.}(2016)\citenamefont{Lopez, Rabbani,
  Crosbie, Raman, Jr., and Sorooshian}}]{lopez_16}
\bibinfo{author}{\bibfnamefont{D.~H.} \bibnamefont{Lopez}},
  \bibinfo{author}{\bibfnamefont{M.~R.} \bibnamefont{Rabbani}},
  \bibinfo{author}{\bibfnamefont{E.}~\bibnamefont{Crosbie}},
  \bibinfo{author}{\bibfnamefont{A.}~\bibnamefont{Raman}},
  \bibinfo{author}{\bibfnamefont{A.~F.~A.} \bibnamefont{Jr.}},
  \bibnamefont{and}
  \bibinfo{author}{\bibfnamefont{A.}~\bibnamefont{Sorooshian}},
  \bibinfo{journal}{Atmosphere} \textbf{\bibinfo{volume}{7}},
  \bibinfo{pages}{1} (\bibinfo{year}{2016}).

\bibitem[{\citenamefont{Lovejoy and Schertzer}(2010)}]{lovejoy_10b}
\bibinfo{author}{\bibfnamefont{S.}~\bibnamefont{Lovejoy}} \bibnamefont{and}
  \bibinfo{author}{\bibfnamefont{D.}~\bibnamefont{Schertzer}},
  \bibinfo{journal}{J. Atmos.} \textbf{\bibinfo{volume}{96}},
  \bibinfo{pages}{1} (\bibinfo{year}{2010}).

\bibitem[{\citenamefont{Kalamaras et~al.}(2019)\citenamefont{Kalamaras, Tzanis,
  Deligiorgi, Philippopoulos, and Koutsogiannis}}]{kalamaras_19}
\bibinfo{author}{\bibfnamefont{N.}~\bibnamefont{Kalamaras}},
  \bibinfo{author}{\bibfnamefont{C.~G.} \bibnamefont{Tzanis}},
  \bibinfo{author}{\bibfnamefont{D.}~\bibnamefont{Deligiorgi}},
  \bibinfo{author}{\bibfnamefont{K.}~\bibnamefont{Philippopoulos}},
  \bibnamefont{and}
  \bibinfo{author}{\bibfnamefont{I.}~\bibnamefont{Koutsogiannis}},
  \bibinfo{journal}{Atmosphere} \textbf{\bibinfo{volume}{10}},
  \bibinfo{pages}{1} (\bibinfo{year}{2019}).

\bibitem[{\citenamefont{Schertzer and Tchiguirinskaia}(2020)}]{schertzer_20}
\bibinfo{author}{\bibfnamefont{D.}~\bibnamefont{Schertzer}} \bibnamefont{and}
  \bibinfo{author}{\bibfnamefont{I.}~\bibnamefont{Tchiguirinskaia}},
  \bibinfo{journal}{Earth Space Science, Preprint, to appear}
  (\bibinfo{year}{2020}).

\bibitem[{\citenamefont{van Haren and Gostiaux}(2016)}]{vanharen_16j}
\bibinfo{author}{\bibfnamefont{H.}~\bibnamefont{van Haren}} \bibnamefont{and}
  \bibinfo{author}{\bibfnamefont{L.}~\bibnamefont{Gostiaux}},
  \bibinfo{journal}{J. Mar. Res.} \textbf{\bibinfo{volume}{74}},
  \bibinfo{pages}{161} (\bibinfo{year}{2016}).

\bibitem[{\citenamefont{Sorriso-Valvo et~al.}(2007)\citenamefont{Sorriso-Valvo,
  Marino, Carbone, Noullez, Lepreti, Veltri, Bruno, Bavassano, and
  Pietropaolo}}]{sorriso_07}
\bibinfo{author}{\bibfnamefont{L.}~\bibnamefont{Sorriso-Valvo}},
  \bibinfo{author}{\bibfnamefont{R.}~\bibnamefont{Marino}},
  \bibinfo{author}{\bibfnamefont{V.}~\bibnamefont{Carbone}},
  \bibinfo{author}{\bibfnamefont{A.}~\bibnamefont{Noullez}},
  \bibinfo{author}{\bibfnamefont{F.}~\bibnamefont{Lepreti}},
  \bibinfo{author}{\bibfnamefont{P.}~\bibnamefont{Veltri}},
  \bibinfo{author}{\bibfnamefont{R.}~\bibnamefont{Bruno}},
  \bibinfo{author}{\bibfnamefont{B.}~\bibnamefont{Bavassano}},
  \bibnamefont{and}
  \bibinfo{author}{\bibfnamefont{E.}~\bibnamefont{Pietropaolo}},
  \bibinfo{journal}{Phys. Rev. Lett.} \textbf{\bibinfo{volume}{99}}
  (\bibinfo{year}{2007}).

\bibitem[{\citenamefont{Lenschow et~al.}(2012)\citenamefont{Lenschow, Lothon,
  Mayor, Sullivan, and Canut}}]{lenschow_12}
\bibinfo{author}{\bibfnamefont{D.~H.} \bibnamefont{Lenschow}},
  \bibinfo{author}{\bibfnamefont{M.}~\bibnamefont{Lothon}},
  \bibinfo{author}{\bibfnamefont{S.~D.} \bibnamefont{Mayor}},
  \bibinfo{author}{\bibfnamefont{P.~P.} \bibnamefont{Sullivan}},
  \bibnamefont{and} \bibinfo{author}{\bibfnamefont{G.}~\bibnamefont{Canut}},
  \bibinfo{journal}{Bound. Lay. Met.} \textbf{\bibinfo{volume}{143}},
  \bibinfo{pages}{107} (\bibinfo{year}{2012}).

\bibitem[{\citenamefont{Cava et~al.}(2015)\citenamefont{Cava, Giostra, and
  Katul}}]{cava_15}
\bibinfo{author}{\bibfnamefont{D.}~\bibnamefont{Cava}},
  \bibinfo{author}{\bibfnamefont{U.}~\bibnamefont{Giostra}}, \bibnamefont{and}
  \bibinfo{author}{\bibfnamefont{G.}~\bibnamefont{Katul}},
  \bibinfo{journal}{Atmosphere} \textbf{\bibinfo{volume}{6}},
  \bibinfo{pages}{1271} (\bibinfo{year}{2015}).

\bibitem[{\citenamefont{Walterscheid et~al.}(2016)\citenamefont{Walterscheid,
  Gelinas, Mechoso, and Schubert}}]{walterscheid_16}
\bibinfo{author}{\bibfnamefont{R.~L.} \bibnamefont{Walterscheid}},
  \bibinfo{author}{\bibfnamefont{L.~J.} \bibnamefont{Gelinas}},
  \bibinfo{author}{\bibfnamefont{C.~R.} \bibnamefont{Mechoso}},
  \bibnamefont{and} \bibinfo{author}{\bibfnamefont{G.}~\bibnamefont{Schubert}},
  \bibinfo{journal}{J. Geophys. Res.} \textbf{\bibinfo{volume}{121}},
  \bibinfo{pages}{1} (\bibinfo{year}{2016}).

\bibitem[{\citenamefont{Rorai et~al.}(2014)\citenamefont{Rorai, Mininni, and
  Pouquet}}]{rorai_14}
\bibinfo{author}{\bibfnamefont{C.}~\bibnamefont{Rorai}},
  \bibinfo{author}{\bibfnamefont{P.}~\bibnamefont{Mininni}}, \bibnamefont{and}
  \bibinfo{author}{\bibfnamefont{A.}~\bibnamefont{Pouquet}},
  \bibinfo{journal}{Phys. Rev. E} \textbf{\bibinfo{volume}{89}},
  \bibinfo{pages}{043002} (\bibinfo{year}{2014}).

\bibitem[{\citenamefont{Feraco et~al.}(2018)\citenamefont{Feraco, Marino,
  Pumir, Primavera, Mininni, Pouquet, and Rosenberg}}]{feraco_18}
\bibinfo{author}{\bibfnamefont{F.}~\bibnamefont{Feraco}},
  \bibinfo{author}{\bibfnamefont{R.}~\bibnamefont{Marino}},
  \bibinfo{author}{\bibfnamefont{A.}~\bibnamefont{Pumir}},
  \bibinfo{author}{\bibfnamefont{L.}~\bibnamefont{Primavera}},
  \bibinfo{author}{\bibfnamefont{P.}~\bibnamefont{Mininni}},
  \bibinfo{author}{\bibfnamefont{A.}~\bibnamefont{Pouquet}}, \bibnamefont{and}
  \bibinfo{author}{\bibfnamefont{D.}~\bibnamefont{Rosenberg}},
  \bibinfo{journal}{Eur. Phys. Lett.} \textbf{\bibinfo{volume}{123}},
  \bibinfo{pages}{44002} (\bibinfo{year}{2018}).

\bibitem[{\citenamefont{Smyth and Moum}(2000)}]{smyth_00b}
\bibinfo{author}{\bibfnamefont{W.}~\bibnamefont{Smyth}} \bibnamefont{and}
  \bibinfo{author}{\bibfnamefont{J.}~\bibnamefont{Moum}},
  \bibinfo{journal}{Phys. Fluids} \textbf{\bibinfo{volume}{12}},
  \bibinfo{pages}{1343} (\bibinfo{year}{2000}).

\bibitem[{\citenamefont{Pouquet
  et~al.}(2019{\natexlab{a}})\citenamefont{Pouquet, Rosenberg, and
  Marino}}]{pouquet_19p}
\bibinfo{author}{\bibfnamefont{A.}~\bibnamefont{Pouquet}},
  \bibinfo{author}{\bibfnamefont{D.}~\bibnamefont{Rosenberg}},
  \bibnamefont{and} \bibinfo{author}{\bibfnamefont{R.}~\bibnamefont{Marino}},
  \bibinfo{journal}{Phys. Fluids} \textbf{\bibinfo{volume}{31}},
  \bibinfo{pages}{105116} (\bibinfo{year}{2019}{\natexlab{a}}).

\bibitem[{\citenamefont{Smyth et~al.}(2019)\citenamefont{Smyth, Nash, and
  Moum}}]{smyth_19}
\bibinfo{author}{\bibfnamefont{W.}~\bibnamefont{Smyth}},
  \bibinfo{author}{\bibfnamefont{J.}~\bibnamefont{Nash}}, \bibnamefont{and}
  \bibinfo{author}{\bibfnamefont{J.}~\bibnamefont{Moum}},
  \bibinfo{journal}{Sci. Rep.} \textbf{\bibinfo{volume}{9}},
  \bibinfo{pages}{3747} (\bibinfo{year}{2019}).

\bibitem[{\citenamefont{Sujovolsky and Mininni}(2019)}]{sujovolsky_19}
\bibinfo{author}{\bibfnamefont{N.}~\bibnamefont{Sujovolsky}} \bibnamefont{and}
  \bibinfo{author}{\bibfnamefont{P.}~\bibnamefont{Mininni}},
  \bibinfo{journal}{Phys. Rev. Fluids} \textbf{\bibinfo{volume}{4}},
  \bibinfo{pages}{052402} (\bibinfo{year}{2019}).

\bibitem[{\citenamefont{Buaria et~al.}(2019)\citenamefont{Buaria, Pumir,
  Feraco, Marino, Pouquet, Rosenberg, and Primavera}}]{buaria_19}
\bibinfo{author}{\bibfnamefont{D.}~\bibnamefont{Buaria}},
  \bibinfo{author}{\bibfnamefont{A.}~\bibnamefont{Pumir}},
  \bibinfo{author}{\bibfnamefont{F.}~\bibnamefont{Feraco}},
  \bibinfo{author}{\bibfnamefont{R.}~\bibnamefont{Marino}},
  \bibinfo{author}{\bibfnamefont{A.}~\bibnamefont{Pouquet}},
  \bibinfo{author}{\bibfnamefont{D.}~\bibnamefont{Rosenberg}},
  \bibnamefont{and}
  \bibinfo{author}{\bibfnamefont{L.}~\bibnamefont{Primavera}},
  \bibinfo{journal}{Preprint, see ArXiv:1909.12433}  (\bibinfo{year}{2019}).

\bibitem[{\citenamefont{Sujovolsky and Mininni}(2020)}]{sujovolsky_20}
\bibinfo{author}{\bibfnamefont{N.}~\bibnamefont{Sujovolsky}} \bibnamefont{and}
  \bibinfo{author}{\bibfnamefont{P.}~\bibnamefont{Mininni}},
  \bibinfo{journal}{Preprint, ArXiv:1912.03160v1}  (\bibinfo{year}{2020}).

\bibitem[{\citenamefont{Meneveau}(2011)}]{meneveau_11}
\bibinfo{author}{\bibfnamefont{C.}~\bibnamefont{Meneveau}},
  \bibinfo{journal}{Ann. Rev. Fluid Mech.} \textbf{\bibinfo{volume}{43}},
  \bibinfo{pages}{219} (\bibinfo{year}{2011}).

\bibitem[{\citenamefont{Marino et~al.}(2015{\natexlab{a}})\citenamefont{Marino,
  Rosenberg, Herbert, and Pouquet}}]{marino_15w}
\bibinfo{author}{\bibfnamefont{R.}~\bibnamefont{Marino}},
  \bibinfo{author}{\bibfnamefont{D.}~\bibnamefont{Rosenberg}},
  \bibinfo{author}{\bibfnamefont{C.}~\bibnamefont{Herbert}}, \bibnamefont{and}
  \bibinfo{author}{\bibfnamefont{A.}~\bibnamefont{Pouquet}},
  \bibinfo{journal}{EuroPhys. Lett.} \textbf{\bibinfo{volume}{112}},
  \bibinfo{pages}{49001} (\bibinfo{year}{2015}{\natexlab{a}}).

\bibitem[{\citenamefont{Herbert et~al.}(2016)\citenamefont{Herbert, Marino,
  Pouquet, and Rosenberg}}]{herbert_16}
\bibinfo{author}{\bibfnamefont{C.}~\bibnamefont{Herbert}},
  \bibinfo{author}{\bibfnamefont{R.}~\bibnamefont{Marino}},
  \bibinfo{author}{\bibfnamefont{A.}~\bibnamefont{Pouquet}}, \bibnamefont{and}
  \bibinfo{author}{\bibfnamefont{D.}~\bibnamefont{Rosenberg}},
  \bibinfo{journal}{J. Fluid Mech.} \textbf{\bibinfo{volume}{806}},
  \bibinfo{pages}{165} (\bibinfo{year}{2016}).

\bibitem[{\citenamefont{Pouquet and Marino}(2013)}]{pouquet_13p}
\bibinfo{author}{\bibfnamefont{A.}~\bibnamefont{Pouquet}} \bibnamefont{and}
  \bibinfo{author}{\bibfnamefont{R.}~\bibnamefont{Marino}},
  \bibinfo{journal}{Phys. Rev. Lett.} \textbf{\bibinfo{volume}{111}},
  \bibinfo{pages}{234501} (\bibinfo{year}{2013}).

\bibitem[{\citenamefont{Marino et~al.}(2015{\natexlab{b}})\citenamefont{Marino,
  Pouquet, and Rosenberg}}]{marino_15p}
\bibinfo{author}{\bibfnamefont{R.}~\bibnamefont{Marino}},
  \bibinfo{author}{\bibfnamefont{A.}~\bibnamefont{Pouquet}}, \bibnamefont{and}
  \bibinfo{author}{\bibfnamefont{D.}~\bibnamefont{Rosenberg}},
  \bibinfo{journal}{Phys. Rev. Lett.} \textbf{\bibinfo{volume}{114}},
  \bibinfo{pages}{114504} (\bibinfo{year}{2015}{\natexlab{b}}).

\bibitem[{\citenamefont{Alexakis and Biferale}(2018)}]{alexakis_18}
\bibinfo{author}{\bibfnamefont{A.}~\bibnamefont{Alexakis}} \bibnamefont{and}
  \bibinfo{author}{\bibfnamefont{L.}~\bibnamefont{Biferale}},
  \bibinfo{journal}{Physics Reports} \textbf{\bibinfo{volume}{762}},
  \bibinfo{pages}{1} (\bibinfo{year}{2018}).

\bibitem[{\citenamefont{Pouquet et~al.}(2018)\citenamefont{Pouquet, Rosenberg,
  Marino, and Herbert}}]{pouquet_18}
\bibinfo{author}{\bibfnamefont{A.}~\bibnamefont{Pouquet}},
  \bibinfo{author}{\bibfnamefont{D.}~\bibnamefont{Rosenberg}},
  \bibinfo{author}{\bibfnamefont{R.}~\bibnamefont{Marino}}, \bibnamefont{and}
  \bibinfo{author}{\bibfnamefont{C.}~\bibnamefont{Herbert}},
  \bibinfo{journal}{J. Fluid Mech.} \textbf{\bibinfo{volume}{844}},
  \bibinfo{pages}{519} (\bibinfo{year}{2018}).

\bibitem[{\citenamefont{Pouquet
  et~al.}(2019{\natexlab{b}})\citenamefont{Pouquet, Rosenberg, Stawarz, and
  Marino}}]{pouquet_19e}
\bibinfo{author}{\bibfnamefont{A.}~\bibnamefont{Pouquet}},
  \bibinfo{author}{\bibfnamefont{D.}~\bibnamefont{Rosenberg}},
  \bibinfo{author}{\bibfnamefont{J.}~\bibnamefont{Stawarz}}, \bibnamefont{and}
  \bibinfo{author}{\bibfnamefont{R.}~\bibnamefont{Marino}},
  \bibinfo{journal}{Earth Space Sci.} \textbf{\bibinfo{volume}{6}},
  \bibinfo{pages}{1} (\bibinfo{year}{2019}{\natexlab{b}}).

\bibitem[{\citenamefont{Bruno and Carbone}(2005)}]{bruno_05}
\bibinfo{author}{\bibfnamefont{R.}~\bibnamefont{Bruno}} \bibnamefont{and}
  \bibinfo{author}{\bibfnamefont{V.}~\bibnamefont{Carbone}},
  \bibinfo{journal}{Living Rev. Solar Phys.} \textbf{\bibinfo{volume}{2}},
  \bibinfo{pages}{4} (\bibinfo{year}{2005}).

\bibitem[{\citenamefont{Veltri et~al.}(2009)\citenamefont{Veltri, Carbone,
  Lepreti, and Nigro}}]{veltri_09}
\bibinfo{author}{\bibfnamefont{P.}~\bibnamefont{Veltri}},
  \bibinfo{author}{\bibfnamefont{V.}~\bibnamefont{Carbone}},
  \bibinfo{author}{\bibfnamefont{F.}~\bibnamefont{Lepreti}}, \bibnamefont{and}
  \bibinfo{author}{\bibfnamefont{G.}~\bibnamefont{Nigro}},
  \bibinfo{journal}{Encyclopedia of Complexity and System Science}
  \textbf{\bibinfo{volume}{R.A. Meyers Ed., Springer}} (\bibinfo{year}{2009}).

\bibitem[{\citenamefont{Matthaeus et~al.}(2015)\citenamefont{Matthaeus, Wan,
  Servidio, Greco, Osman, Oughton, and Dmitruk}}]{matthaeus_15}
\bibinfo{author}{\bibfnamefont{W.~H.} \bibnamefont{Matthaeus}},
  \bibinfo{author}{\bibfnamefont{M.}~\bibnamefont{Wan}},
  \bibinfo{author}{\bibfnamefont{S.}~\bibnamefont{Servidio}},
  \bibinfo{author}{\bibfnamefont{A.}~\bibnamefont{Greco}},
  \bibinfo{author}{\bibfnamefont{K.~T.} \bibnamefont{Osman}},
  \bibinfo{author}{\bibfnamefont{S.}~\bibnamefont{Oughton}}, \bibnamefont{and}
  \bibinfo{author}{\bibfnamefont{P.}~\bibnamefont{Dmitruk}},
  \bibinfo{journal}{Phil. Trans. R. Soc. A} \textbf{\bibinfo{volume}{373}}
  (\bibinfo{year}{2015}).

\bibitem[{\citenamefont{Galtier}(2018)}]{galtier_18b}
\bibinfo{author}{\bibfnamefont{S.}~\bibnamefont{Galtier}}, \bibinfo{journal}{J.
  Phys. A.: Math. Theor.} \textbf{\bibinfo{volume}{51}} (\bibinfo{year}{2018}).

\bibitem[{\citenamefont{Matthaeus and Velli}(2011)}]{matthaeus_11}
\bibinfo{author}{\bibfnamefont{W.}~\bibnamefont{Matthaeus}} \bibnamefont{and}
  \bibinfo{author}{\bibfnamefont{M.}~\bibnamefont{Velli}},
  \bibinfo{journal}{Space Sci. Rev.} \textbf{\bibinfo{volume}{160}},
  \bibinfo{pages}{145} (\bibinfo{year}{2011}).

\bibitem[{\citenamefont{Marino et~al.}(2012)\citenamefont{Marino,
  Sorriso-Valvo, D{'}{A}micis, Carbone, Bruno, and Veltri}}]{marino_12}
\bibinfo{author}{\bibfnamefont{R.}~\bibnamefont{Marino}},
  \bibinfo{author}{\bibfnamefont{L.}~\bibnamefont{Sorriso-Valvo}},
  \bibinfo{author}{\bibfnamefont{R.}~\bibnamefont{D{'}{A}micis}},
  \bibinfo{author}{\bibfnamefont{V.}~\bibnamefont{Carbone}},
  \bibinfo{author}{\bibfnamefont{R.}~\bibnamefont{Bruno}}, \bibnamefont{and}
  \bibinfo{author}{\bibfnamefont{P.}~\bibnamefont{Veltri}},
  \bibinfo{journal}{Astrophys. J.} \textbf{\bibinfo{volume}{750}},
  \bibinfo{pages}{41} (\bibinfo{year}{2012}).

\bibitem[{\citenamefont{Pouquet}(2015)}]{pouquet_15b}
\bibinfo{author}{\bibfnamefont{A.}~\bibnamefont{Pouquet}}, in
  \emph{\bibinfo{booktitle}{Lecture Notes, Festival de Th\'eorie,
  Aix-en-Provence}}, edited by
  \bibinfo{editor}{\bibfnamefont{P.}~\bibnamefont{Ghendrih}} \bibnamefont{and}
  \bibinfo{editor}{\bibfnamefont{P.}~\bibnamefont{Diamond}}
  (\bibinfo{publisher}{World Scientific}, \bibinfo{year}{2015}), pp.
  \bibinfo{pages}{45--79}.

\bibitem[{\citenamefont{T\'oth et~al.}(2017)\citenamefont{T\'oth, Chen,
  Gombosi, Cassak, Markidis, and Peng}}]{toth_17}
\bibinfo{author}{\bibfnamefont{G.}~\bibnamefont{T\'oth}},
  \bibinfo{author}{\bibfnamefont{Y.}~\bibnamefont{Chen}},
  \bibinfo{author}{\bibfnamefont{T.~I.} \bibnamefont{Gombosi}},
  \bibinfo{author}{\bibfnamefont{P.}~\bibnamefont{Cassak}},
  \bibinfo{author}{\bibfnamefont{S.}~\bibnamefont{Markidis}}, \bibnamefont{and}
  \bibinfo{author}{\bibfnamefont{I.~B.} \bibnamefont{Peng}},
  \bibinfo{journal}{J. Geophys. Res.} \textbf{\bibinfo{volume}{122}},
  \bibinfo{pages}{10,336} (\bibinfo{year}{2017}).

\bibitem[{\citenamefont{Sahraoui et~al.}(2013)\citenamefont{Sahraoui, Huang,
  Belmont, Goldstein, R\'etino, Robert, and de~{P}atoul}}]{sahraoui_13}
\bibinfo{author}{\bibfnamefont{F.}~\bibnamefont{Sahraoui}},
  \bibinfo{author}{\bibfnamefont{S.}~\bibnamefont{Huang}},
  \bibinfo{author}{\bibfnamefont{G.}~\bibnamefont{Belmont}},
  \bibinfo{author}{\bibfnamefont{M.~L.} \bibnamefont{Goldstein}},
  \bibinfo{author}{\bibfnamefont{A.}~\bibnamefont{R\'etino}},
  \bibinfo{author}{\bibfnamefont{P.}~\bibnamefont{Robert}}, \bibnamefont{and}
  \bibinfo{author}{\bibfnamefont{J.}~\bibnamefont{de~{P}atoul}},
  \bibinfo{journal}{Astrophys. J.} \textbf{\bibinfo{volume}{777}},
  \bibinfo{pages}{15} (\bibinfo{year}{2013}).

\bibitem[{\citenamefont{Lacombe et~al.}(2017)\citenamefont{Lacombe,
  Alexandrova, and Matteini}}]{lacombe_17}
\bibinfo{author}{\bibfnamefont{C.}~\bibnamefont{Lacombe}},
  \bibinfo{author}{\bibfnamefont{O.}~\bibnamefont{Alexandrova}},
  \bibnamefont{and} \bibinfo{author}{\bibfnamefont{L.}~\bibnamefont{Matteini}},
  \bibinfo{journal}{The Astrophysical Journal} \textbf{\bibinfo{volume}{848}},
  \bibinfo{pages}{45} (\bibinfo{year}{2017}),
  \urlprefix\url{http://stacks.iop.org/0004-637X/848/i=1/a=45}.

\bibitem[{\citenamefont{Stawarz et~al.}(2016)\citenamefont{Stawarz, Eriksson,
  Wilder, Ergun, Schwartz, Pouquet, Burch, Giles, Khotyaintsev, {Le\ }Contel
  et~al.}}]{stawarz_16}
\bibinfo{author}{\bibfnamefont{J.~E.} \bibnamefont{Stawarz}},
  \bibinfo{author}{\bibfnamefont{S.}~\bibnamefont{Eriksson}},
  \bibinfo{author}{\bibfnamefont{F.~D.} \bibnamefont{Wilder}},
  \bibinfo{author}{\bibfnamefont{R.~E.} \bibnamefont{Ergun}},
  \bibinfo{author}{\bibfnamefont{S.~J.} \bibnamefont{Schwartz}},
  \bibinfo{author}{\bibfnamefont{A.}~\bibnamefont{Pouquet}},
  \bibinfo{author}{\bibfnamefont{J.~L.} \bibnamefont{Burch}},
  \bibinfo{author}{\bibfnamefont{B.~L.} \bibnamefont{Giles}},
  \bibinfo{author}{\bibfnamefont{Y.}~\bibnamefont{Khotyaintsev}},
  \bibinfo{author}{\bibfnamefont{O.}~\bibnamefont{{Le\ }Contel}},
  \bibnamefont{et~al.}, \bibinfo{journal}{J. Geophys. Res. Space Phys.}
  \textbf{\bibinfo{volume}{121}}, \bibinfo{pages}{11,021}
  (\bibinfo{year}{2016}).

\bibitem[{\citenamefont{Mare et~al.}(2019)\citenamefont{Mare, Sorriso-Valvo,
  Retin\`o, Malara, and Hasegawa}}]{dimare_19}
\bibinfo{author}{\bibfnamefont{F.~D.} \bibnamefont{Mare}},
  \bibinfo{author}{\bibfnamefont{L.}~\bibnamefont{Sorriso-Valvo}},
  \bibinfo{author}{\bibfnamefont{A.}~\bibnamefont{Retin\`o}},
  \bibinfo{author}{\bibfnamefont{F.}~\bibnamefont{Malara}}, \bibnamefont{and}
  \bibinfo{author}{\bibfnamefont{H.}~\bibnamefont{Hasegawa}},
  \bibinfo{journal}{Atmosphere} \textbf{\bibinfo{volume}{10}},
  \bibinfo{pages}{561} (\bibinfo{year}{2019}).

\bibitem[{\citenamefont{{{L}}e Contel et~al.}(2016)\citenamefont{{{L}}e Contel,
  Retin\'o, Breuillard, Mirioni, Robert, Chasapis, Lavraud, Chust, Rezeau,
  Wilder et~al.}}]{lecontel_16}
\bibinfo{author}{\bibfnamefont{O.}~\bibnamefont{{{L}}e Contel}},
  \bibinfo{author}{\bibfnamefont{A.}~\bibnamefont{Retin\'o}},
  \bibinfo{author}{\bibfnamefont{H.}~\bibnamefont{Breuillard}},
  \bibinfo{author}{\bibfnamefont{L.}~\bibnamefont{Mirioni}},
  \bibinfo{author}{\bibfnamefont{P.}~\bibnamefont{Robert}},
  \bibinfo{author}{\bibfnamefont{A.}~\bibnamefont{Chasapis}},
  \bibinfo{author}{\bibfnamefont{B.}~\bibnamefont{Lavraud}},
  \bibinfo{author}{\bibfnamefont{T.}~\bibnamefont{Chust}},
  \bibinfo{author}{\bibfnamefont{L.}~\bibnamefont{Rezeau}},
  \bibinfo{author}{\bibfnamefont{F.~D.} \bibnamefont{Wilder}},
  \bibnamefont{et~al.}, \bibinfo{journal}{Geophys. Res. Lett.}
  \textbf{\bibinfo{volume}{43}}, \bibinfo{pages}{5943} (\bibinfo{year}{2016}).

\bibitem[{\citenamefont{Faganello and Califano}(2017)}]{faganello_17}
\bibinfo{author}{\bibfnamefont{M.}~\bibnamefont{Faganello}} \bibnamefont{and}
  \bibinfo{author}{\bibfnamefont{F.}~\bibnamefont{Califano}},
  \bibinfo{journal}{J. Plasma Phys.} \textbf{\bibinfo{volume}{83}}
  (\bibinfo{year}{2017}).

\bibitem[{\citenamefont{Bandyopadhyay et~al.}(2018)\citenamefont{Bandyopadhyay,
  Chasapis, Chhiber, Parashar, Matthaeus, Shay, Maruca, Burch, Moore, Pollock
  et~al.}}]{bandyo_18a}
\bibinfo{author}{\bibfnamefont{R.}~\bibnamefont{Bandyopadhyay}},
  \bibinfo{author}{\bibfnamefont{A.}~\bibnamefont{Chasapis}},
  \bibinfo{author}{\bibfnamefont{R.}~\bibnamefont{Chhiber}},
  \bibinfo{author}{\bibfnamefont{T.~N.} \bibnamefont{Parashar}},
  \bibinfo{author}{\bibfnamefont{W.~H.} \bibnamefont{Matthaeus}},
  \bibinfo{author}{\bibfnamefont{M.~A.} \bibnamefont{Shay}},
  \bibinfo{author}{\bibfnamefont{B.~A.} \bibnamefont{Maruca}},
  \bibinfo{author}{\bibfnamefont{J.~L.} \bibnamefont{Burch}},
  \bibinfo{author}{\bibfnamefont{T.~E.} \bibnamefont{Moore}},
  \bibinfo{author}{\bibfnamefont{C.~J.} \bibnamefont{Pollock}},
  \bibnamefont{et~al.}, \bibinfo{journal}{Astrophys. J.}
  \textbf{\bibinfo{volume}{866}} (\bibinfo{year}{2018}).

\bibitem[{\citenamefont{Stawarz et~al.}(2019)\citenamefont{Stawarz, Gershman,
  Eastwood, Phan, Gingell, Shay, Burch, Ergun, Giles, Contel
  et~al.}}]{stawarz_19}
\bibinfo{author}{\bibfnamefont{J.~E.} \bibnamefont{Stawarz}},
  \bibinfo{author}{\bibfnamefont{D.~J.} \bibnamefont{Gershman}},
  \bibinfo{author}{\bibfnamefont{J.~P.} \bibnamefont{Eastwood}},
  \bibinfo{author}{\bibfnamefont{T.~D.} \bibnamefont{Phan}},
  \bibinfo{author}{\bibfnamefont{I.~L.} \bibnamefont{Gingell}},
  \bibinfo{author}{\bibfnamefont{M.~A.} \bibnamefont{Shay}},
  \bibinfo{author}{\bibfnamefont{J.~L.} \bibnamefont{Burch}},
  \bibinfo{author}{\bibfnamefont{R.~E.} \bibnamefont{Ergun}},
  \bibinfo{author}{\bibfnamefont{B.~L.} \bibnamefont{Giles}},
  \bibinfo{author}{\bibfnamefont{O.~L.} \bibnamefont{Contel}},
  \bibnamefont{et~al.}, \bibinfo{journal}{Astrophys. J. Lett.}
  \textbf{\bibinfo{volume}{877}}, \bibinfo{pages}{L37(7 pp)}
  (\bibinfo{year}{2019}).

\bibitem[{\citenamefont{Camporeale et~al.}(2018)\citenamefont{Camporeale,
  Sorriso-Valvo, Califano, and Retin\'o}}]{camporeale_18p}
\bibinfo{author}{\bibfnamefont{E.}~\bibnamefont{Camporeale}},
  \bibinfo{author}{\bibfnamefont{L.}~\bibnamefont{Sorriso-Valvo}},
  \bibinfo{author}{\bibfnamefont{F.}~\bibnamefont{Califano}}, \bibnamefont{and}
  \bibinfo{author}{\bibfnamefont{A.}~\bibnamefont{Retin\'o}},
  \bibinfo{journal}{Phys. Rev. Lett.} \textbf{\bibinfo{volume}{120}}
  (\bibinfo{year}{2018}).

\bibitem[{\citenamefont{Kitamura et~al.}(2018)\citenamefont{Kitamura, Kitahara,
  Shoji, Miyoshi, Hasegawa, Nakamura, Katoh, Saito, Yokota, Gershman
  et~al.}}]{kitamura_18}
\bibinfo{author}{\bibfnamefont{N.}~\bibnamefont{Kitamura}},
  \bibinfo{author}{\bibfnamefont{M.}~\bibnamefont{Kitahara}},
  \bibinfo{author}{\bibfnamefont{M.}~\bibnamefont{Shoji}},
  \bibinfo{author}{\bibfnamefont{Y.}~\bibnamefont{Miyoshi}},
  \bibinfo{author}{\bibfnamefont{H.}~\bibnamefont{Hasegawa}},
  \bibinfo{author}{\bibfnamefont{S.}~\bibnamefont{Nakamura}},
  \bibinfo{author}{\bibfnamefont{Y.}~\bibnamefont{Katoh}},
  \bibinfo{author}{\bibfnamefont{Y.}~\bibnamefont{Saito}},
  \bibinfo{author}{\bibfnamefont{S.}~\bibnamefont{Yokota}},
  \bibinfo{author}{\bibfnamefont{D.~J.} \bibnamefont{Gershman}},
  \bibnamefont{et~al.}, \bibinfo{journal}{Science}
  \textbf{\bibinfo{volume}{361}}, \bibinfo{pages}{1000} (\bibinfo{year}{2018}).

\bibitem[{\citenamefont{Brandenburg and Subramanian}(2005)}]{branden_rev}
\bibinfo{author}{\bibfnamefont{A.}~\bibnamefont{Brandenburg}} \bibnamefont{and}
  \bibinfo{author}{\bibfnamefont{K.}~\bibnamefont{Subramanian}},
  \bibinfo{journal}{Phys. Rep.} \textbf{\bibinfo{volume}{{\bf 417}}}
  (\bibinfo{year}{2005}).

\bibitem[{\citenamefont{Mahajan et~al.}(2005)\citenamefont{Mahajan, Mininni,
  and G\'omez}}]{mahajan_05}
\bibinfo{author}{\bibfnamefont{S.~M.} \bibnamefont{Mahajan}},
  \bibinfo{author}{\bibfnamefont{P.~D.} \bibnamefont{Mininni}},
  \bibnamefont{and} \bibinfo{author}{\bibfnamefont{D.~O.}
  \bibnamefont{G\'omez}}, \bibinfo{journal}{Astrophys. J.}
  \textbf{\bibinfo{volume}{619}}, \bibinfo{pages}{1014} (\bibinfo{year}{2005}).

\bibitem[{\citenamefont{Mininni et~al.}(2005)\citenamefont{Mininni, G\'omez,
  and Mahajan}}]{mininni_05}
\bibinfo{author}{\bibfnamefont{P.~D.} \bibnamefont{Mininni}},
  \bibinfo{author}{\bibfnamefont{D.}~\bibnamefont{G\'omez}}, \bibnamefont{and}
  \bibinfo{author}{\bibfnamefont{S.}~\bibnamefont{Mahajan}},
  \bibinfo{journal}{Astrophys. J.} \textbf{\bibinfo{volume}{619}},
  \bibinfo{pages}{1019} (\bibinfo{year}{2005}).

\bibitem[{\citenamefont{Galtier and Buchlin}(2007)}]{galtier_07}
\bibinfo{author}{\bibfnamefont{S.}~\bibnamefont{Galtier}} \bibnamefont{and}
  \bibinfo{author}{\bibfnamefont{E.}~\bibnamefont{Buchlin}},
  \bibinfo{journal}{Astrophys. J.} \textbf{\bibinfo{volume}{656}},
  \bibinfo{pages}{560} (\bibinfo{year}{2007}).

\bibitem[{\citenamefont{Galtier}(2006)}]{galtier_06}
\bibinfo{author}{\bibfnamefont{S.}~\bibnamefont{Galtier}}, \bibinfo{journal}{J.
  Plasma Phys.} \textbf{\bibinfo{volume}{72}}, \bibinfo{pages}{721}
  (\bibinfo{year}{2006}).

\bibitem[{\citenamefont{Mininni et~al.}(2006)\citenamefont{Mininni, Pouquet,
  and Montgomery}}]{mininni_06b}
\bibinfo{author}{\bibfnamefont{P.}~\bibnamefont{Mininni}},
  \bibinfo{author}{\bibfnamefont{A.}~\bibnamefont{Pouquet}}, \bibnamefont{and}
  \bibinfo{author}{\bibfnamefont{D.}~\bibnamefont{Montgomery}},
  \bibinfo{journal}{Phys. Rev. Lett.} \textbf{\bibinfo{volume}{97}},
  \bibinfo{pages}{244503} (\bibinfo{year}{2006}).

\bibitem[{\citenamefont{Borovsky and
  Funsten}(2003{\natexlab{a}})}]{borovsky_03a}
\bibinfo{author}{\bibfnamefont{J.~E.} \bibnamefont{Borovsky}} \bibnamefont{and}
  \bibinfo{author}{\bibfnamefont{H.~O.} \bibnamefont{Funsten}},
  \bibinfo{journal}{J. Geophys. Res.} \textbf{\bibinfo{volume}{108}},
  \bibinfo{pages}{1246} (\bibinfo{year}{2003}{\natexlab{a}}).

\bibitem[{\citenamefont{Borovsky and
  Funsten}(2003{\natexlab{b}})}]{borovsky_03b}
\bibinfo{author}{\bibfnamefont{J.~E.} \bibnamefont{Borovsky}} \bibnamefont{and}
  \bibinfo{author}{\bibfnamefont{H.~O.} \bibnamefont{Funsten}},
  \bibinfo{journal}{J. Geophys. Res.} \textbf{\bibinfo{volume}{108}},
  \bibinfo{pages}{1284} (\bibinfo{year}{2003}{\natexlab{b}}).

\bibitem[{\citenamefont{Franci et~al.}(2015)\citenamefont{Franci, Landi,
  Matteini, Verdini, and Hellinger}}]{franci_15}
\bibinfo{author}{\bibfnamefont{L.}~\bibnamefont{Franci}},
  \bibinfo{author}{\bibfnamefont{S.}~\bibnamefont{Landi}},
  \bibinfo{author}{\bibfnamefont{L.}~\bibnamefont{Matteini}},
  \bibinfo{author}{\bibfnamefont{A.}~\bibnamefont{Verdini}}, \bibnamefont{and}
  \bibinfo{author}{\bibfnamefont{P.}~\bibnamefont{Hellinger}},
  \bibinfo{journal}{Astrophys. J.} \textbf{\bibinfo{volume}{812}},
  \bibinfo{pages}{812:21} (\bibinfo{year}{2015}).

\bibitem[{\citenamefont{Gonz\`alez et~al.}(2019)\citenamefont{Gonz\`alez,
  Parashar, Gomez, Matthaeus, and Dmitruk}}]{gonzalez_19}
\bibinfo{author}{\bibfnamefont{C.~A.} \bibnamefont{Gonz\`alez}},
  \bibinfo{author}{\bibfnamefont{T.~N.} \bibnamefont{Parashar}},
  \bibinfo{author}{\bibfnamefont{D.}~\bibnamefont{Gomez}},
  \bibinfo{author}{\bibfnamefont{W.~H.} \bibnamefont{Matthaeus}},
  \bibnamefont{and} \bibinfo{author}{\bibfnamefont{P.}~\bibnamefont{Dmitruk}},
  \bibinfo{journal}{Phys. Plasmas} \textbf{\bibinfo{volume}{26}},
  \bibinfo{pages}{012306} (\bibinfo{year}{2019}).

\bibitem[{\citenamefont{Galtier and Meyrand}(2015)}]{galtier_15}
\bibinfo{author}{\bibfnamefont{S.}~\bibnamefont{Galtier}} \bibnamefont{and}
  \bibinfo{author}{\bibfnamefont{A.}~\bibnamefont{Meyrand}},
  \bibinfo{journal}{J. Plasma Phys.} \textbf{\bibinfo{volume}{81}},
  \bibinfo{pages}{325810106} (\bibinfo{year}{2015}).

\bibitem[{\citenamefont{Grappin et~al.}(1983)\citenamefont{Grappin, Pouquet,
  and L\'eorat}}]{grappin_83}
\bibinfo{author}{\bibfnamefont{R.}~\bibnamefont{Grappin}},
  \bibinfo{author}{\bibfnamefont{A.}~\bibnamefont{Pouquet}}, \bibnamefont{and}
  \bibinfo{author}{\bibfnamefont{J.}~\bibnamefont{L\'eorat}},
  \bibinfo{journal}{Astron. Astrophys.} \textbf{\bibinfo{volume}{126}},
  \bibinfo{pages}{51} (\bibinfo{year}{1983}).

\bibitem[{\citenamefont{Stawarz and Pouquet}(2015)}]{stawarz_15H}
\bibinfo{author}{\bibfnamefont{J.~E.} \bibnamefont{Stawarz}} \bibnamefont{and}
  \bibinfo{author}{\bibfnamefont{A.}~\bibnamefont{Pouquet}},
  \bibinfo{journal}{Phys. Rev. E} \textbf{\bibinfo{volume}{92}},
  \bibinfo{pages}{063102} (\bibinfo{year}{2015}).

\bibitem[{\citenamefont{Vasyliunas}(1975)}]{vasyliunas_75}
\bibinfo{author}{\bibfnamefont{V.~M.} \bibnamefont{Vasyliunas}},
  \bibinfo{journal}{Rev. Geophys. Space Phys.} \textbf{\bibinfo{volume}{13}},
  \bibinfo{pages}{303} (\bibinfo{year}{1975}).

\bibitem[{\citenamefont{Priest and Forbes}(2000)}]{priest_00}
\bibinfo{author}{\bibfnamefont{E.}~\bibnamefont{Priest}} \bibnamefont{and}
  \bibinfo{author}{\bibfnamefont{T.}~\bibnamefont{Forbes}},
  \emph{\bibinfo{title}{Magnetic Reconnection: MHD Theory and Applications}}
  (\bibinfo{publisher}{Cambridge University Press}, \bibinfo{year}{2000}).

\bibitem[{\citenamefont{Song et~al.}(2001)\citenamefont{Song, Gombosi, and
  Ridley}}]{song_01}
\bibinfo{author}{\bibfnamefont{P.}~\bibnamefont{Song}},
  \bibinfo{author}{\bibfnamefont{T.}~\bibnamefont{Gombosi}}, \bibnamefont{and}
  \bibinfo{author}{\bibfnamefont{A.}~\bibnamefont{Ridley}},
  \bibinfo{journal}{J. Geophys. Res.} \textbf{\bibinfo{volume}{106}},
  \bibinfo{pages}{8149} (\bibinfo{year}{2001}).

\bibitem[{\citenamefont{Cothran et~al.}(2005)\citenamefont{Cothran, Landreman,
  Brown, and Matthaeus}}]{cothran_05}
\bibinfo{author}{\bibfnamefont{C.~D.} \bibnamefont{Cothran}},
  \bibinfo{author}{\bibfnamefont{M.}~\bibnamefont{Landreman}},
  \bibinfo{author}{\bibfnamefont{M.~R.} \bibnamefont{Brown}}, \bibnamefont{and}
  \bibinfo{author}{\bibfnamefont{W.}~\bibnamefont{Matthaeus}},
  \bibinfo{journal}{Geophys. Res. Lett.} \textbf{\bibinfo{volume}{32}},
  \bibinfo{pages}{L03105} (\bibinfo{year}{2005}).

\bibitem[{\citenamefont{Mahajan and Yoshida}(1998)}]{mahajan_98}
\bibinfo{author}{\bibfnamefont{S.}~\bibnamefont{Mahajan}} \bibnamefont{and}
  \bibinfo{author}{\bibfnamefont{Z.}~\bibnamefont{Yoshida}},
  \bibinfo{journal}{Phys. Rev. Lett.} \textbf{\bibinfo{volume}{99}},
  \bibinfo{pages}{4863} (\bibinfo{year}{1998}).

\bibitem[{\citenamefont{Laveder et~al.}(2002)\citenamefont{Laveder, Passot, and
  Sulem}}]{laveder_02}
\bibinfo{author}{\bibfnamefont{D.}~\bibnamefont{Laveder}},
  \bibinfo{author}{\bibfnamefont{T.}~\bibnamefont{Passot}}, \bibnamefont{and}
  \bibinfo{author}{\bibfnamefont{P.}~\bibnamefont{Sulem}},
  \bibinfo{journal}{Phys. Plasmas} \textbf{\bibinfo{volume}{9}},
  \bibinfo{pages}{293} (\bibinfo{year}{2002}).

\bibitem[{\citenamefont{Mininni et~al.}(2003)\citenamefont{Mininni, G\'omez,
  and Mahajan}}]{mininni_03}
\bibinfo{author}{\bibfnamefont{P.}~\bibnamefont{Mininni}},
  \bibinfo{author}{\bibfnamefont{D.~O.} \bibnamefont{G\'omez}},
  \bibnamefont{and} \bibinfo{author}{\bibfnamefont{S.~M.}
  \bibnamefont{Mahajan}}, \bibinfo{journal}{Astrophys. J.}
  \textbf{\bibinfo{volume}{584}}, \bibinfo{pages}{1120} (\bibinfo{year}{2003}).

\bibitem[{\citenamefont{G\`omez et~al.}(2010)\citenamefont{G\`omez, Mininni,
  and Dmiturk}}]{gomez_10}
\bibinfo{author}{\bibfnamefont{D.}~\bibnamefont{G\`omez}},
  \bibinfo{author}{\bibfnamefont{P.}~\bibnamefont{Mininni}}, \bibnamefont{and}
  \bibinfo{author}{\bibfnamefont{P.}~\bibnamefont{Dmiturk}},
  \bibinfo{journal}{Phys. Rev. E} \textbf{\bibinfo{volume}{82}},
  \bibinfo{pages}{036406} (\bibinfo{year}{2010}).

\bibitem[{\citenamefont{Mininni et~al.}(2007)\citenamefont{Mininni, Alexakis,
  and Pouquet}}]{mininni_07}
\bibinfo{author}{\bibfnamefont{P.~D.} \bibnamefont{Mininni}},
  \bibinfo{author}{\bibfnamefont{A.}~\bibnamefont{Alexakis}}, \bibnamefont{and}
  \bibinfo{author}{\bibfnamefont{A.}~\bibnamefont{Pouquet}},
  \bibinfo{journal}{J. Plasma Phys.} \textbf{\bibinfo{volume}{73}},
  \bibinfo{pages}{377} (\bibinfo{year}{2007}).

\bibitem[{\citenamefont{Torbert et~al.}(2016)\citenamefont{Torbert, Burch,
  Giles, Gershman, Pollock, Dorelli, Avanov, Argall, Shuster, Strangeway
  et~al.}}]{torbert_16}
\bibinfo{author}{\bibfnamefont{R.~B.} \bibnamefont{Torbert}},
  \bibinfo{author}{\bibfnamefont{J.~L.} \bibnamefont{Burch}},
  \bibinfo{author}{\bibfnamefont{B.~L.} \bibnamefont{Giles}},
  \bibinfo{author}{\bibfnamefont{D.}~\bibnamefont{Gershman}},
  \bibinfo{author}{\bibfnamefont{C.~J.} \bibnamefont{Pollock}},
  \bibinfo{author}{\bibfnamefont{J.}~\bibnamefont{Dorelli}},
  \bibinfo{author}{\bibfnamefont{L.}~\bibnamefont{Avanov}},
  \bibinfo{author}{\bibfnamefont{M.~R.} \bibnamefont{Argall}},
  \bibinfo{author}{\bibfnamefont{J.}~\bibnamefont{Shuster}},
  \bibinfo{author}{\bibfnamefont{R.~J.} \bibnamefont{Strangeway}},
  \bibnamefont{et~al.}, \bibinfo{journal}{Geophys. Res. Lett.}
  \textbf{\bibinfo{volume}{43}}, \bibinfo{pages}{5918} (\bibinfo{year}{2016}).

\bibitem[{\citenamefont{Webster et~al.}(2018)\citenamefont{Webster, Burch,
  Reiff, Daou, Genestreti, Graham, Torbert, Ergun, Sazykin, Marshall
  et~al.}}]{webster_18}
\bibinfo{author}{\bibfnamefont{J.}~\bibnamefont{Webster}},
  \bibinfo{author}{\bibfnamefont{J.~L.} \bibnamefont{Burch}},
  \bibinfo{author}{\bibfnamefont{P.~H.} \bibnamefont{Reiff}},
  \bibinfo{author}{\bibfnamefont{A.~G.} \bibnamefont{Daou}},
  \bibinfo{author}{\bibfnamefont{K.~J.} \bibnamefont{Genestreti}},
  \bibinfo{author}{\bibfnamefont{D.~B.} \bibnamefont{Graham}},
  \bibinfo{author}{\bibfnamefont{R.~B.} \bibnamefont{Torbert}},
  \bibinfo{author}{\bibfnamefont{R.~E.} \bibnamefont{Ergun}},
  \bibinfo{author}{\bibfnamefont{S.~Y.} \bibnamefont{Sazykin}},
  \bibinfo{author}{\bibfnamefont{A.}~\bibnamefont{Marshall}},
  \bibnamefont{et~al.}, \bibinfo{journal}{J. Geophys. Res.}
  \textbf{\bibinfo{volume}{123}}, \bibinfo{pages}{4858} (\bibinfo{year}{2018}).

\bibitem[{\citenamefont{Shuster et~al.}(2019)\citenamefont{Shuster, Gershman,
  Chen, Wang, Bessho, Dorelli, da~Silva, Giles, Paterson, Denton
  et~al.}}]{shuster_19}
\bibinfo{author}{\bibfnamefont{J.~R.} \bibnamefont{Shuster}},
  \bibinfo{author}{\bibfnamefont{D.~J.} \bibnamefont{Gershman}},
  \bibinfo{author}{\bibfnamefont{L.-J.} \bibnamefont{Chen}},
  \bibinfo{author}{\bibfnamefont{S.}~\bibnamefont{Wang}},
  \bibinfo{author}{\bibfnamefont{N.}~\bibnamefont{Bessho}},
  \bibinfo{author}{\bibfnamefont{J.~C.} \bibnamefont{Dorelli}},
  \bibinfo{author}{\bibfnamefont{D.~E.} \bibnamefont{da~Silva}},
  \bibinfo{author}{\bibfnamefont{B.~L.} \bibnamefont{Giles}},
  \bibinfo{author}{\bibfnamefont{W.~R.} \bibnamefont{Paterson}},
  \bibinfo{author}{\bibfnamefont{R.~E.} \bibnamefont{Denton}},
  \bibnamefont{et~al.}, \bibinfo{journal}{Geophys. Res. Lett.}
  \textbf{\bibinfo{volume}{46}}, \bibinfo{pages}{7862} (\bibinfo{year}{2019}).

\bibitem[{\citenamefont{Ergun et~al.}(2018)\citenamefont{Ergun, Goodrich,
  Wilder, Ahmadi, Holmes, Eriksson, Stawarz, Nakamura, Genestreti, Hesse
  et~al.}}]{ergun_18}
\bibinfo{author}{\bibfnamefont{R.~E.} \bibnamefont{Ergun}},
  \bibinfo{author}{\bibfnamefont{K.~A.} \bibnamefont{Goodrich}},
  \bibinfo{author}{\bibfnamefont{F.~D.} \bibnamefont{Wilder}},
  \bibinfo{author}{\bibfnamefont{N.}~\bibnamefont{Ahmadi}},
  \bibinfo{author}{\bibfnamefont{J.~C.} \bibnamefont{Holmes}},
  \bibinfo{author}{\bibfnamefont{S.}~\bibnamefont{Eriksson}},
  \bibinfo{author}{\bibfnamefont{J.~E.} \bibnamefont{Stawarz}},
  \bibinfo{author}{\bibfnamefont{R.}~\bibnamefont{Nakamura}},
  \bibinfo{author}{\bibfnamefont{K.~J.} \bibnamefont{Genestreti}},
  \bibinfo{author}{\bibfnamefont{M.}~\bibnamefont{Hesse}},
  \bibnamefont{et~al.}, \bibinfo{journal}{Geophys. Res. Lett.}
  \textbf{\bibinfo{volume}{45}}, \bibinfo{pages}{3338} (\bibinfo{year}{2018}).

\bibitem[{\citenamefont{Mininni
  et~al.}(2011{\natexlab{a}})\citenamefont{Mininni, Dmitruk, Matthaeus, and
  Pouquet}}]{mininni_11}
\bibinfo{author}{\bibfnamefont{P.}~\bibnamefont{Mininni}},
  \bibinfo{author}{\bibfnamefont{P.}~\bibnamefont{Dmitruk}},
  \bibinfo{author}{\bibfnamefont{W.~H.} \bibnamefont{Matthaeus}},
  \bibnamefont{and} \bibinfo{author}{\bibfnamefont{A.}~\bibnamefont{Pouquet}},
  \bibinfo{journal}{Phys. Rev. E} \textbf{\bibinfo{volume}{83}},
  \bibinfo{pages}{016309} (\bibinfo{year}{2011}{\natexlab{a}}).

\bibitem[{\citenamefont{Mininni
  et~al.}(2011{\natexlab{b}})\citenamefont{Mininni, Rosenberg, Reddy, and
  Pouquet}}]{hybrid_11}
\bibinfo{author}{\bibfnamefont{P.}~\bibnamefont{Mininni}},
  \bibinfo{author}{\bibfnamefont{D.}~\bibnamefont{Rosenberg}},
  \bibinfo{author}{\bibfnamefont{R.}~\bibnamefont{Reddy}}, \bibnamefont{and}
  \bibinfo{author}{\bibfnamefont{A.}~\bibnamefont{Pouquet}},
  \bibinfo{journal}{Parallel Computing} \textbf{\bibinfo{volume}{37}},
  \bibinfo{pages}{316} (\bibinfo{year}{2011}{\natexlab{b}}).

\bibitem[{\citenamefont{Turner}(1986)}]{turner_86}
\bibinfo{author}{\bibfnamefont{L.}~\bibnamefont{Turner}},
  \bibinfo{journal}{IEEE Transactions on Plasma Science}
  \textbf{\bibinfo{volume}{PS-14}}, \bibinfo{pages}{849}
  (\bibinfo{year}{1986}).

\bibitem[{\citenamefont{Servidio
  et~al.}(2008{\natexlab{a}})\citenamefont{Servidio, Matthaeus, and
  Carbone}}]{servidio_08}
\bibinfo{author}{\bibfnamefont{S.}~\bibnamefont{Servidio}},
  \bibinfo{author}{\bibfnamefont{W.~H.} \bibnamefont{Matthaeus}},
  \bibnamefont{and} \bibinfo{author}{\bibfnamefont{V.}~\bibnamefont{Carbone}},
  \bibinfo{journal}{Phys. Plasmas} \textbf{\bibinfo{volume}{15}},
  \bibinfo{pages}{042314} (\bibinfo{year}{2008}{\natexlab{a}}).

\bibitem[{\citenamefont{Banerjee and Galtier}(2016)}]{banerjee_16}
\bibinfo{author}{\bibfnamefont{S.}~\bibnamefont{Banerjee}} \bibnamefont{and}
  \bibinfo{author}{\bibfnamefont{S.}~\bibnamefont{Galtier}},
  \bibinfo{journal}{Phys. Rev. E} \textbf{\bibinfo{volume}{93}},
  \bibinfo{pages}{033120} (\bibinfo{year}{2016}).

\bibitem[{\citenamefont{Ohsaki and Yoshida}(2005)}]{ohsaki_05}
\bibinfo{author}{\bibfnamefont{S.}~\bibnamefont{Ohsaki}} \bibnamefont{and}
  \bibinfo{author}{\bibfnamefont{Z.}~\bibnamefont{Yoshida}},
  \bibinfo{journal}{Phys. Plasmas} \textbf{\bibinfo{volume}{12}},
  \bibinfo{pages}{064505} (\bibinfo{year}{2005}).

\bibitem[{\citenamefont{Sorriso-Valvo et~al.}(2019)\citenamefont{Sorriso-Valvo,
  Catapano, Retin\`o, Le~Contel, Perrone, Roberts, Coburn, Panebianco,
  Valentini, Perri et~al.}}]{sorriso_19}
\bibinfo{author}{\bibfnamefont{L.}~\bibnamefont{Sorriso-Valvo}},
  \bibinfo{author}{\bibfnamefont{F.}~\bibnamefont{Catapano}},
  \bibinfo{author}{\bibfnamefont{A.}~\bibnamefont{Retin\`o}},
  \bibinfo{author}{\bibfnamefont{O.}~\bibnamefont{Le~Contel}},
  \bibinfo{author}{\bibfnamefont{D.}~\bibnamefont{Perrone}},
  \bibinfo{author}{\bibfnamefont{O.~W.} \bibnamefont{Roberts}},
  \bibinfo{author}{\bibfnamefont{J.~T.} \bibnamefont{Coburn}},
  \bibinfo{author}{\bibfnamefont{V.}~\bibnamefont{Panebianco}},
  \bibinfo{author}{\bibfnamefont{F.}~\bibnamefont{Valentini}},
  \bibinfo{author}{\bibfnamefont{S.}~\bibnamefont{Perri}},
  \bibnamefont{et~al.}, \bibinfo{journal}{Phys. Rev. Lett.}
  \textbf{\bibinfo{volume}{122}}, \bibinfo{pages}{035102}
  (\bibinfo{year}{2019}).

\bibitem[{\citenamefont{Politano and Pouquet}(1998)}]{politano_98g}
\bibinfo{author}{\bibfnamefont{H.}~\bibnamefont{Politano}} \bibnamefont{and}
  \bibinfo{author}{\bibfnamefont{A.}~\bibnamefont{Pouquet}},
  \bibinfo{journal}{Geophys. Res. Lett.} \textbf{\bibinfo{volume}{25}},
  \bibinfo{pages}{273} (\bibinfo{year}{1998}).

\bibitem[{\citenamefont{R\"udiger et~al.}(2011)\citenamefont{R\"udiger,
  Kitchatinov, and Brandenburg}}]{rudiger_11}
\bibinfo{author}{\bibfnamefont{G.}~\bibnamefont{R\"udiger}},
  \bibinfo{author}{\bibfnamefont{L.}~\bibnamefont{Kitchatinov}},
  \bibnamefont{and}
  \bibinfo{author}{\bibfnamefont{A.}~\bibnamefont{Brandenburg}},
  \bibinfo{journal}{Sol. Phys.} \textbf{\bibinfo{volume}{269}},
  \bibinfo{pages}{3} (\bibinfo{year}{2011}).

\bibitem[{\citenamefont{Pouquet et~al.}(1986)\citenamefont{Pouquet, Meneguzzi,
  and Frisch}}]{pouquet_86}
\bibinfo{author}{\bibfnamefont{A.}~\bibnamefont{Pouquet}},
  \bibinfo{author}{\bibfnamefont{M.}~\bibnamefont{Meneguzzi}},
  \bibnamefont{and} \bibinfo{author}{\bibfnamefont{U.}~\bibnamefont{Frisch}},
  \bibinfo{journal}{Phys. Rev. A} \textbf{\bibinfo{volume}{33}},
  \bibinfo{pages}{4266} (\bibinfo{year}{1986}).

\bibitem[{\citenamefont{Grappin et~al.}(1982)\citenamefont{Grappin, Frisch,
  L\'eorat, and Pouquet}}]{grappin_82}
\bibinfo{author}{\bibfnamefont{R.}~\bibnamefont{Grappin}},
  \bibinfo{author}{\bibfnamefont{U.}~\bibnamefont{Frisch}},
  \bibinfo{author}{\bibfnamefont{J.}~\bibnamefont{L\'eorat}}, \bibnamefont{and}
  \bibinfo{author}{\bibfnamefont{A.}~\bibnamefont{Pouquet}},
  \bibinfo{journal}{Astron. Astrophys.} \textbf{\bibinfo{volume}{102}},
  \bibinfo{pages}{6} (\bibinfo{year}{1982}).

\bibitem[{\citenamefont{Passot and Sulem}(2019)}]{passot_19}
\bibinfo{author}{\bibfnamefont{T.}~\bibnamefont{Passot}} \bibnamefont{and}
  \bibinfo{author}{\bibfnamefont{P.~L.} \bibnamefont{Sulem}},
  \bibinfo{journal}{J. Plasma Phys.} \textbf{\bibinfo{volume}{85}},
  \bibinfo{pages}{905850301} (\bibinfo{year}{2019}).

\bibitem[{\citenamefont{Perez and Boldyrev}(2009)}]{perez_09}
\bibinfo{author}{\bibfnamefont{J.-C.} \bibnamefont{Perez}} \bibnamefont{and}
  \bibinfo{author}{\bibfnamefont{S.}~\bibnamefont{Boldyrev}},
  \bibinfo{journal}{Phys. Rev. Lett.} \textbf{\bibinfo{volume}{102}},
  \bibinfo{pages}{025003} (\bibinfo{year}{2009}).

\bibitem[{\citenamefont{Meneguzzi et~al.}(1996)\citenamefont{Meneguzzi,
  Politano, Pouquet, and Zolver}}]{meneguzzi_96}
\bibinfo{author}{\bibfnamefont{M.}~\bibnamefont{Meneguzzi}},
  \bibinfo{author}{\bibfnamefont{H.}~\bibnamefont{Politano}},
  \bibinfo{author}{\bibfnamefont{A.}~\bibnamefont{Pouquet}}, \bibnamefont{and}
  \bibinfo{author}{\bibfnamefont{M.}~\bibnamefont{Zolver}},
  \bibinfo{journal}{J. Comp. Phys.} \textbf{\bibinfo{volume}{123}},
  \bibinfo{pages}{32} (\bibinfo{year}{1996}).

\bibitem[{\citenamefont{Yokoi}(2013)}]{yokoi_13}
\bibinfo{author}{\bibfnamefont{N.}~\bibnamefont{Yokoi}},
  \bibinfo{journal}{Geophys. Astrophys. Fluid Dyn.}
  \textbf{\bibinfo{volume}{107}}, \bibinfo{pages}{114} (\bibinfo{year}{2013}).

\bibitem[{\citenamefont{Yokoi et~al.}(2013)\citenamefont{Yokoi, Higashimori,
  and Hoshino}}]{yokoi_13c}
\bibinfo{author}{\bibfnamefont{N.}~\bibnamefont{Yokoi}},
  \bibinfo{author}{\bibfnamefont{K.}~\bibnamefont{Higashimori}},
  \bibnamefont{and} \bibinfo{author}{\bibfnamefont{M.}~\bibnamefont{Hoshino}},
  \bibinfo{journal}{Phys. Plasmas} \textbf{\bibinfo{volume}{20}},
  \bibinfo{pages}{122310} (\bibinfo{year}{2013}).

\bibitem[{\citenamefont{Titov et~al.}(2019)\citenamefont{Titov, Stepanov,
  Yokoi, Verma, and Samtaney}}]{titov_19}
\bibinfo{author}{\bibfnamefont{V.}~\bibnamefont{Titov}},
  \bibinfo{author}{\bibfnamefont{R.}~\bibnamefont{Stepanov}},
  \bibinfo{author}{\bibfnamefont{N.}~\bibnamefont{Yokoi}},
  \bibinfo{author}{\bibfnamefont{M.}~\bibnamefont{Verma}}, \bibnamefont{and}
  \bibinfo{author}{\bibfnamefont{R.}~\bibnamefont{Samtaney}},
  \bibinfo{journal}{Magnetohydrodynamics} \textbf{\bibinfo{volume}{55}},
  \bibinfo{pages}{225} (\bibinfo{year}{2019}).

\bibitem[{\citenamefont{Milosevich et~al.}(2017)\citenamefont{Milosevich,
  Lingam, and Morrison}}]{miloshevich_17}
\bibinfo{author}{\bibfnamefont{G.}~\bibnamefont{Milosevich}},
  \bibinfo{author}{\bibfnamefont{M.}~\bibnamefont{Lingam}}, \bibnamefont{and}
  \bibinfo{author}{\bibfnamefont{P.~J.} \bibnamefont{Morrison}},
  \bibinfo{journal}{Astrophys. J. Lett.} \textbf{\bibinfo{volume}{19}},
  \bibinfo{pages}{015007} (\bibinfo{year}{2017}).

\bibitem[{\citenamefont{Milosevich et~al.}(2020)\citenamefont{Milosevich,
  Passot, and Sulem}}]{miloshevich_20}
\bibinfo{author}{\bibfnamefont{G.}~\bibnamefont{Milosevich}},
  \bibinfo{author}{\bibfnamefont{T.}~\bibnamefont{Passot}}, \bibnamefont{and}
  \bibinfo{author}{\bibfnamefont{P.}~\bibnamefont{Sulem}},
  \bibinfo{journal}{Astrophys. J. Lett.} \textbf{\bibinfo{volume}{888}},
  \bibinfo{pages}{L7} (\bibinfo{year}{2020}).

\bibitem[{\citenamefont{Politano et~al.}(1989)\citenamefont{Politano, Pouquet,
  and Sulem}}]{politano_89}
\bibinfo{author}{\bibfnamefont{H.}~\bibnamefont{Politano}},
  \bibinfo{author}{\bibfnamefont{A.}~\bibnamefont{Pouquet}}, \bibnamefont{and}
  \bibinfo{author}{\bibfnamefont{P.}~\bibnamefont{Sulem}},
  \bibinfo{journal}{Phys. Fluids B} \textbf{\bibinfo{volume}{1}},
  \bibinfo{pages}{2330} (\bibinfo{year}{1989}).

\bibitem[{\citenamefont{Sahraoui et~al.}(2006)\citenamefont{Sahraoui, Galtier,
  and Belmont}}]{sahraoui_06}
\bibinfo{author}{\bibfnamefont{F.}~\bibnamefont{Sahraoui}},
  \bibinfo{author}{\bibfnamefont{S.}~\bibnamefont{Galtier}}, \bibnamefont{and}
  \bibinfo{author}{\bibfnamefont{G.}~\bibnamefont{Belmont}},
  \bibinfo{journal}{J. Plasma Phys.} \textbf{\bibinfo{volume}{73}},
  \bibinfo{pages}{723} (\bibinfo{year}{2006}).

\bibitem[{\citenamefont{Meyrand and Galtier}(2012)}]{meyrand_12}
\bibinfo{author}{\bibfnamefont{R.}~\bibnamefont{Meyrand}} \bibnamefont{and}
  \bibinfo{author}{\bibfnamefont{S.}~\bibnamefont{Galtier}},
  \bibinfo{journal}{Phys. Rev. Lett.} \textbf{\bibinfo{volume}{109}}
  (\bibinfo{year}{2012}).

\bibitem[{\citenamefont{Lee}(1952)}]{lee_52}
\bibinfo{author}{\bibfnamefont{T.~D.} \bibnamefont{Lee}},
  \bibinfo{journal}{Quart. Appl. Math.} \textbf{\bibinfo{volume}{10}},
  \bibinfo{pages}{69} (\bibinfo{year}{1952}).

\bibitem[{\citenamefont{Kraichnan}(1967)}]{kraichnan_67}
\bibinfo{author}{\bibfnamefont{R.}~\bibnamefont{Kraichnan}},
  \bibinfo{journal}{Phys. Fluids} \textbf{\bibinfo{volume}{10}},
  \bibinfo{pages}{1417} (\bibinfo{year}{1967}).

\bibitem[{\citenamefont{Kraichnan}(1973)}]{kraichnan_73}
\bibinfo{author}{\bibfnamefont{R.}~\bibnamefont{Kraichnan}},
  \bibinfo{journal}{J. Fluid Mech.} \textbf{\bibinfo{volume}{59}},
  \bibinfo{pages}{745} (\bibinfo{year}{1973}).

\bibitem[{\citenamefont{Cichowlas et~al.}(2005)\citenamefont{Cichowlas,
  Bona\"iti, Debbasch, and Brachet}}]{cichowlas_05}
\bibinfo{author}{\bibfnamefont{C.}~\bibnamefont{Cichowlas}},
  \bibinfo{author}{\bibfnamefont{P.}~\bibnamefont{Bona\"iti}},
  \bibinfo{author}{\bibfnamefont{F.}~\bibnamefont{Debbasch}}, \bibnamefont{and}
  \bibinfo{author}{\bibfnamefont{M.}~\bibnamefont{Brachet}},
  \bibinfo{journal}{Phys Rev. Lett.} \textbf{\bibinfo{volume}{95}}
  (\bibinfo{year}{2005}).

\bibitem[{\citenamefont{Nastrom and Gage}(1985)}]{nastrom_85}
\bibinfo{author}{\bibfnamefont{G.~D.} \bibnamefont{Nastrom}} \bibnamefont{and}
  \bibinfo{author}{\bibfnamefont{K.}~\bibnamefont{Gage}}, \bibinfo{journal}{J.
  Atmos. Sci.} \textbf{\bibinfo{volume}{42}}, \bibinfo{pages}{950}
  (\bibinfo{year}{1985}).

\bibitem[{\citenamefont{Koprov et~al.}(2005)\citenamefont{Koprov, Koprov,
  Ponomarev, and Chkhetiani}}]{koprov_05}
\bibinfo{author}{\bibfnamefont{B.}~\bibnamefont{Koprov}},
  \bibinfo{author}{\bibfnamefont{V.}~\bibnamefont{Koprov}},
  \bibinfo{author}{\bibfnamefont{V.}~\bibnamefont{Ponomarev}},
  \bibnamefont{and}
  \bibinfo{author}{\bibfnamefont{O.}~\bibnamefont{Chkhetiani}},
  \bibinfo{journal}{Dokl. Phys.} \textbf{\bibinfo{volume}{50}},
  \bibinfo{pages}{419} (\bibinfo{year}{2005}).

\bibitem[{\citenamefont{Krstulovic et~al.}(2011)\citenamefont{Krstulovic,
  Brachet, and Pouquet}}]{krstulovic_11}
\bibinfo{author}{\bibfnamefont{G.}~\bibnamefont{Krstulovic}},
  \bibinfo{author}{\bibfnamefont{M.}~\bibnamefont{Brachet}}, \bibnamefont{and}
  \bibinfo{author}{\bibfnamefont{A.}~\bibnamefont{Pouquet}},
  \bibinfo{journal}{Phys. Rev. E} \textbf{\bibinfo{volume}{84}},
  \bibinfo{pages}{016410} (\bibinfo{year}{2011}).

\bibitem[{\citenamefont{Mininni and Pouquet}(2007)}]{mininni_07b}
\bibinfo{author}{\bibfnamefont{P.~D.} \bibnamefont{Mininni}} \bibnamefont{and}
  \bibinfo{author}{\bibfnamefont{A.}~\bibnamefont{Pouquet}},
  \bibinfo{journal}{Phys. Rev. Lett.} \textbf{\bibinfo{volume}{99}},
  \bibinfo{pages}{254502} (\bibinfo{year}{2007}).

\bibitem[{\citenamefont{Zhu et~al.}(2014)\citenamefont{Zhu, Yang, and
  Zhu}}]{zhuJZ_14}
\bibinfo{author}{\bibfnamefont{J.-Z.} \bibnamefont{Zhu}},
  \bibinfo{author}{\bibfnamefont{W.}~\bibnamefont{Yang}}, \bibnamefont{and}
  \bibinfo{author}{\bibfnamefont{G.-Y.} \bibnamefont{Zhu}},
  \bibinfo{journal}{J. Fluid Mech.} \textbf{\bibinfo{volume}{739}},
  \bibinfo{pages}{479} (\bibinfo{year}{2014}).

\bibitem[{\citenamefont{Servidio
  et~al.}(2008{\natexlab{b}})\citenamefont{Servidio, Matthaeus, and
  Dmitruk}}]{servidio_08b}
\bibinfo{author}{\bibfnamefont{S.}~\bibnamefont{Servidio}},
  \bibinfo{author}{\bibfnamefont{W.}~\bibnamefont{Matthaeus}},
  \bibnamefont{and} \bibinfo{author}{\bibfnamefont{P.}~\bibnamefont{Dmitruk}},
  \bibinfo{journal}{Phys. Rev. Lett} \textbf{\bibinfo{volume}{100}},
  \bibinfo{pages}{095005} (\bibinfo{year}{2008}{\natexlab{b}}).

\bibitem[{\citenamefont{Mininni and Pouquet}(2009{\natexlab{a}})}]{mininni_09c}
\bibinfo{author}{\bibfnamefont{P.}~\bibnamefont{Mininni}} \bibnamefont{and}
  \bibinfo{author}{\bibfnamefont{A.}~\bibnamefont{Pouquet}},
  \bibinfo{journal}{Phys Rev E} \textbf{\bibinfo{volume}{79}},
  \bibinfo{pages}{026304} (\bibinfo{year}{2009}{\natexlab{a}}).

\bibitem[{\citenamefont{Pouquet et~al.}(1976)\citenamefont{Pouquet, Frisch, and
  L\'eorat}}]{pouquet_76}
\bibinfo{author}{\bibfnamefont{A.}~\bibnamefont{Pouquet}},
  \bibinfo{author}{\bibfnamefont{U.}~\bibnamefont{Frisch}}, \bibnamefont{and}
  \bibinfo{author}{\bibfnamefont{J.}~\bibnamefont{L\'eorat}},
  \bibinfo{journal}{J. Fluid Mech.} \textbf{\bibinfo{volume}{77}},
  \bibinfo{pages}{321} (\bibinfo{year}{1976}).

\bibitem[{\citenamefont{Brissaud et~al.}(1973)\citenamefont{Brissaud, Frisch,
  L\'eorat, Lesieur, and Mazure}}]{brissaud_73}
\bibinfo{author}{\bibfnamefont{A.}~\bibnamefont{Brissaud}},
  \bibinfo{author}{\bibfnamefont{U.}~\bibnamefont{Frisch}},
  \bibinfo{author}{\bibfnamefont{J.}~\bibnamefont{L\'eorat}},
  \bibinfo{author}{\bibfnamefont{M.}~\bibnamefont{Lesieur}}, \bibnamefont{and}
  \bibinfo{author}{\bibfnamefont{A.}~\bibnamefont{Mazure}},
  \bibinfo{journal}{Phys. Fluids} \textbf{\bibinfo{volume}{16}},
  \bibinfo{pages}{1366} (\bibinfo{year}{1973}).

\bibitem[{\citenamefont{Baerenzung et~al.}(2011)\citenamefont{Baerenzung,
  Mininni, Pouquet, and Rosenberg}}]{baerenzung_11}
\bibinfo{author}{\bibfnamefont{J.}~\bibnamefont{Baerenzung}},
  \bibinfo{author}{\bibfnamefont{P.}~\bibnamefont{Mininni}},
  \bibinfo{author}{\bibfnamefont{A.}~\bibnamefont{Pouquet}}, \bibnamefont{and}
  \bibinfo{author}{\bibfnamefont{D.}~\bibnamefont{Rosenberg}},
  \bibinfo{journal}{J. Atmos. Sci.} \textbf{\bibinfo{volume}{68}},
  \bibinfo{pages}{2757} (\bibinfo{year}{2011}).

\bibitem[{\citenamefont{Mininni and Pouquet}(2009{\natexlab{b}})}]{mininni_09}
\bibinfo{author}{\bibfnamefont{P.}~\bibnamefont{Mininni}} \bibnamefont{and}
  \bibinfo{author}{\bibfnamefont{A.}~\bibnamefont{Pouquet}},
  \bibinfo{journal}{Phys. Rev. E} \textbf{\bibinfo{volume}{80}},
  \bibinfo{pages}{025401} (\bibinfo{year}{2009}{\natexlab{b}}).

\bibitem[{\citenamefont{M\"uller et~al.}(2012)\citenamefont{M\"uller, Malapaka,
  and Busse}}]{mueller_12}
\bibinfo{author}{\bibfnamefont{W.}~\bibnamefont{M\"uller}},
  \bibinfo{author}{\bibfnamefont{S.}~\bibnamefont{Malapaka}}, \bibnamefont{and}
  \bibinfo{author}{\bibfnamefont{A.}~\bibnamefont{Busse}},
  \bibinfo{journal}{Phys. Rev. E} \textbf{\bibinfo{volume}{85}},
  \bibinfo{pages}{015302} (\bibinfo{year}{2012}).

\bibitem[{\citenamefont{M\"uller and Malapaka}(2013)}]{mueller_13}
\bibinfo{author}{\bibfnamefont{W.}~\bibnamefont{M\"uller}} \bibnamefont{and}
  \bibinfo{author}{\bibfnamefont{S.}~\bibnamefont{Malapaka}},
  \bibinfo{journal}{Geophys. Astrophys. Fluid Dyn.}
  \textbf{\bibinfo{volume}{107}}, \bibinfo{pages}{93} (\bibinfo{year}{2013}).

\bibitem[{\citenamefont{Frishman and Herbert}(2018)}]{frishman_18}
\bibinfo{author}{\bibfnamefont{A.}~\bibnamefont{Frishman}} \bibnamefont{and}
  \bibinfo{author}{\bibfnamefont{C.}~\bibnamefont{Herbert}},
  \bibinfo{journal}{Phys. Rev. Lett.} \textbf{\bibinfo{volume}{120}}
  (\bibinfo{year}{2018}).

\bibitem[{\citenamefont{Alexandrova et~al.}(2008)\citenamefont{Alexandrova,
  Lacombe, and Mangeney}}]{alexandrova_08}
\bibinfo{author}{\bibfnamefont{O.}~\bibnamefont{Alexandrova}},
  \bibinfo{author}{\bibfnamefont{C.}~\bibnamefont{Lacombe}}, \bibnamefont{and}
  \bibinfo{author}{\bibfnamefont{A.}~\bibnamefont{Mangeney}},
  \bibinfo{journal}{Ann. Geophys.} \textbf{\bibinfo{volume}{26}},
  \bibinfo{pages}{3585} (\bibinfo{year}{2008}).

\bibitem[{\citenamefont{Huang et~al.}(2017)\citenamefont{Huang, Hadid,
  Sahraoui, Yuan, and Deng}}]{huang_17}
\bibinfo{author}{\bibfnamefont{S.~Y.} \bibnamefont{Huang}},
  \bibinfo{author}{\bibfnamefont{L.~Z.} \bibnamefont{Hadid}},
  \bibinfo{author}{\bibfnamefont{F.}~\bibnamefont{Sahraoui}},
  \bibinfo{author}{\bibfnamefont{Z.~G.} \bibnamefont{Yuan}}, \bibnamefont{and}
  \bibinfo{author}{\bibfnamefont{X.~H.} \bibnamefont{Deng}},
  \bibinfo{journal}{Astrophys. J. Lett.} \textbf{\bibinfo{volume}{863}},
  \bibinfo{pages}{L10} (\bibinfo{year}{2017}).

\bibitem[{\citenamefont{Chasapis et~al.}(2018)\citenamefont{Chasapis,
  Matthaeus, Parashar, Wan, Haggerty, Pollock, Giles, Paterson, Dorelli,
  Gershman et~al.}}]{chasapis_18}
\bibinfo{author}{\bibfnamefont{A.}~\bibnamefont{Chasapis}},
  \bibinfo{author}{\bibfnamefont{W.~H.} \bibnamefont{Matthaeus}},
  \bibinfo{author}{\bibfnamefont{T.~N.} \bibnamefont{Parashar}},
  \bibinfo{author}{\bibfnamefont{M.}~\bibnamefont{Wan}},
  \bibinfo{author}{\bibfnamefont{C.~C.} \bibnamefont{Haggerty}},
  \bibinfo{author}{\bibfnamefont{C.~J.} \bibnamefont{Pollock}},
  \bibinfo{author}{\bibfnamefont{B.~L.} \bibnamefont{Giles}},
  \bibinfo{author}{\bibfnamefont{W.~R.} \bibnamefont{Paterson}},
  \bibinfo{author}{\bibfnamefont{J.}~\bibnamefont{Dorelli}},
  \bibinfo{author}{\bibfnamefont{D.~J.} \bibnamefont{Gershman}},
  \bibnamefont{et~al.}, \bibinfo{journal}{Astrophys. J. Lett.}
  \textbf{\bibinfo{volume}{856}}, \bibinfo{pages}{L19} (\bibinfo{year}{2018}).

\bibitem[{\citenamefont{Phan et~al.}(2018)\citenamefont{Phan, Eastwood, Shay,
  Drake, Sonnerup, Fujimoto, Cassak, {\O }ieroset, Burch, Torbert
  et~al.}}]{phan_18}
\bibinfo{author}{\bibfnamefont{T.~D.} \bibnamefont{Phan}},
  \bibinfo{author}{\bibfnamefont{J.~P.} \bibnamefont{Eastwood}},
  \bibinfo{author}{\bibfnamefont{M.~A.} \bibnamefont{Shay}},
  \bibinfo{author}{\bibfnamefont{J.~F.} \bibnamefont{Drake}},
  \bibinfo{author}{\bibfnamefont{B.~U.~{\"O}.} \bibnamefont{Sonnerup}},
  \bibinfo{author}{\bibfnamefont{M.}~\bibnamefont{Fujimoto}},
  \bibinfo{author}{\bibfnamefont{P.~A.} \bibnamefont{Cassak}},
  \bibinfo{author}{\bibfnamefont{M.}~\bibnamefont{{\O }ieroset}},
  \bibinfo{author}{\bibfnamefont{J.~L.} \bibnamefont{Burch}},
  \bibinfo{author}{\bibfnamefont{R.~B.} \bibnamefont{Torbert}},
  \bibnamefont{et~al.}, \bibinfo{journal}{Nature}
  \textbf{\bibinfo{volume}{557}}, \bibinfo{pages}{202} (\bibinfo{year}{2018}).

\bibitem[{\citenamefont{Bandyopadhyay et~al.}(2019)\citenamefont{Bandyopadhyay,
  Sorriso-Valvo, Chasapis, Hellinger, Matthaeus, Verdini, Landi, Franci,
  Matteini, Giles et~al.}}]{bandyopadhyay_19}
\bibinfo{author}{\bibfnamefont{R.}~\bibnamefont{Bandyopadhyay}},
  \bibinfo{author}{\bibfnamefont{L.}~\bibnamefont{Sorriso-Valvo}},
  \bibinfo{author}{\bibfnamefont{A.}~\bibnamefont{Chasapis}},
  \bibinfo{author}{\bibfnamefont{P.}~\bibnamefont{Hellinger}},
  \bibinfo{author}{\bibfnamefont{W.~H.} \bibnamefont{Matthaeus}},
  \bibinfo{author}{\bibfnamefont{A.}~\bibnamefont{Verdini}},
  \bibinfo{author}{\bibfnamefont{S.}~\bibnamefont{Landi}},
  \bibinfo{author}{\bibfnamefont{L.}~\bibnamefont{Franci}},
  \bibinfo{author}{\bibfnamefont{L.}~\bibnamefont{Matteini}},
  \bibinfo{author}{\bibfnamefont{B.~L.} \bibnamefont{Giles}},
  \bibnamefont{et~al.}, \bibinfo{journal}{ArXiv:1907.06802}
  (\bibinfo{year}{2019}).

\bibitem[{\citenamefont{Alexakis et~al.}(2006)\citenamefont{Alexakis, Mininni,
  and Pouquet}}]{alexakis_06}
\bibinfo{author}{\bibfnamefont{A.}~\bibnamefont{Alexakis}},
  \bibinfo{author}{\bibfnamefont{P.}~\bibnamefont{Mininni}}, \bibnamefont{and}
  \bibinfo{author}{\bibfnamefont{A.}~\bibnamefont{Pouquet}},
  \bibinfo{journal}{Astrophys. J.} \textbf{\bibinfo{volume}{640}},
  \bibinfo{pages}{335} (\bibinfo{year}{2006}).

\bibitem[{\citenamefont{Zhdankin et~al.}(2016)\citenamefont{Zhdankin, Boldyrev,
  and Uzdensky}}]{zhdankin_16}
\bibinfo{author}{\bibfnamefont{V.}~\bibnamefont{Zhdankin}},
  \bibinfo{author}{\bibfnamefont{S.}~\bibnamefont{Boldyrev}}, \bibnamefont{and}
  \bibinfo{author}{\bibfnamefont{D.~A.} \bibnamefont{Uzdensky}},
  \bibinfo{journal}{Phys. Plasmas} \textbf{\bibinfo{volume}{23}},
  \bibinfo{pages}{055705} (\bibinfo{year}{2016}).

\bibitem[{\citenamefont{Mininni and Pouquet}(2013)}]{mininni_13}
\bibinfo{author}{\bibfnamefont{P.}~\bibnamefont{Mininni}} \bibnamefont{and}
  \bibinfo{author}{\bibfnamefont{A.}~\bibnamefont{Pouquet}},
  \bibinfo{journal}{Phys. Rev. E} \textbf{\bibinfo{volume}{87}},
  \bibinfo{pages}{033002} (\bibinfo{year}{2013}).

\bibitem[{\citenamefont{Pouquet et~al.}(1988)\citenamefont{Pouquet, Sulem, and
  Meneguzzi}}]{pouquet_88}
\bibinfo{author}{\bibfnamefont{A.}~\bibnamefont{Pouquet}},
  \bibinfo{author}{\bibfnamefont{P.~L.} \bibnamefont{Sulem}}, \bibnamefont{and}
  \bibinfo{author}{\bibfnamefont{M.}~\bibnamefont{Meneguzzi}},
  \bibinfo{journal}{Phys. Fluids} \textbf{\bibinfo{volume}{31}},
  \bibinfo{pages}{2635} (\bibinfo{year}{1988}).

\bibitem[{\citenamefont{Scott and Wang}(2005)}]{scott_05}
\bibinfo{author}{\bibfnamefont{R.}~\bibnamefont{Scott}} \bibnamefont{and}
  \bibinfo{author}{\bibfnamefont{F.}~\bibnamefont{Wang}}, \bibinfo{journal}{J.
  Phys. Oceano.} \textbf{\bibinfo{volume}{35}}, \bibinfo{pages}{1650}
  (\bibinfo{year}{2005}).

\bibitem[{\citenamefont{Skamarock et~al.}(2014)\citenamefont{Skamarock, Park,
  Klemp, and Snyder}}]{skamarock_14}
\bibinfo{author}{\bibfnamefont{W.~C.} \bibnamefont{Skamarock}},
  \bibinfo{author}{\bibfnamefont{S.-H.} \bibnamefont{Park}},
  \bibinfo{author}{\bibfnamefont{J.~B.} \bibnamefont{Klemp}}, \bibnamefont{and}
  \bibinfo{author}{\bibfnamefont{C.}~\bibnamefont{Snyder}},
  \bibinfo{journal}{J. Atm. Sci.} \textbf{\bibinfo{volume}{71}},
  \bibinfo{pages}{4369} (\bibinfo{year}{2014}).

\bibitem[{\citenamefont{Passot et~al.}(2018)\citenamefont{Passot, Sulem, and
  Tassi}}]{passot_18}
\bibinfo{author}{\bibfnamefont{T.}~\bibnamefont{Passot}},
  \bibinfo{author}{\bibfnamefont{P.~L.} \bibnamefont{Sulem}}, \bibnamefont{and}
  \bibinfo{author}{\bibfnamefont{E.}~\bibnamefont{Tassi}},
  \bibinfo{journal}{Phys. Plasmas} \textbf{\bibinfo{volume}{25}},
  \bibinfo{pages}{041207} (\bibinfo{year}{2018}).

\bibitem[{\citenamefont{Schekochihin et~al.}(2009)\citenamefont{Schekochihin,
  Cowley, Dorland, Hammett, Howes, Quataert, and Tatsuno}}]{schekochihin_09}
\bibinfo{author}{\bibfnamefont{A.}~\bibnamefont{Schekochihin}},
  \bibinfo{author}{\bibfnamefont{S.}~\bibnamefont{Cowley}},
  \bibinfo{author}{\bibfnamefont{W.}~\bibnamefont{Dorland}},
  \bibinfo{author}{\bibfnamefont{G.}~\bibnamefont{Hammett}},
  \bibinfo{author}{\bibfnamefont{G.}~\bibnamefont{Howes}},
  \bibinfo{author}{\bibfnamefont{E.}~\bibnamefont{Quataert}}, \bibnamefont{and}
  \bibinfo{author}{\bibfnamefont{T.}~\bibnamefont{Tatsuno}},
  \bibinfo{journal}{Astrophys. J. Supp.} \textbf{\bibinfo{volume}{182}},
  \bibinfo{pages}{310} (\bibinfo{year}{2009}).

\bibitem[{\citenamefont{Cho}(2011)}]{cho_11}
\bibinfo{author}{\bibfnamefont{J.}~\bibnamefont{Cho}}, \bibinfo{journal}{Phys
  Rev. Lett.} \textbf{\bibinfo{volume}{106}}, \bibinfo{pages}{191104}
  (\bibinfo{year}{2011}).

\bibitem[{\citenamefont{Ji}(1999)}]{ji_99b}
\bibinfo{author}{\bibfnamefont{H.}~\bibnamefont{Ji}}, \bibinfo{journal}{Phys.
  Rev. Lett.} \textbf{\bibinfo{volume}{83}}, \bibinfo{pages}{3198}
  (\bibinfo{year}{1999}).

\bibitem[{\citenamefont{Galtier and Bhattacharjee}(2003)}]{galtier_03b}
\bibinfo{author}{\bibfnamefont{S.}~\bibnamefont{Galtier}} \bibnamefont{and}
  \bibinfo{author}{\bibfnamefont{A.}~\bibnamefont{Bhattacharjee}},
  \bibinfo{journal}{Phys. Plasmas} \textbf{\bibinfo{volume}{10}},
  \bibinfo{pages}{3065} (\bibinfo{year}{2003}).

\bibitem[{\citenamefont{Kim and Cho}(2015)}]{kim_15}
\bibinfo{author}{\bibfnamefont{H.}~\bibnamefont{Kim}} \bibnamefont{and}
  \bibinfo{author}{\bibfnamefont{J.}~\bibnamefont{Cho}},
  \bibinfo{journal}{Astrophys. J. Suppl.} \textbf{\bibinfo{volume}{801}},
  \bibinfo{pages}{75} (\bibinfo{year}{2015}).

\end{thebibliography}
  \end{document}